\documentclass[12pt,preprint2]{aastex}
\usepackage{emulateapj5}
\usepackage{onecolfloat5}

\newcommand{\cxo}{\textit{CXO}}
\newcommand{\rosat}{\textit{ROSAT}}

\newcommand{\xmm}{\textit{XMM}}
\newcommand{\chandra}{\textit{Chandra}}
\newcommand{\beq}{\begin{equation}}
\newcommand{\eeq}{\end{equation}}
\newcommand{\mc}{\multicolumn}
\newcommand{\expnt}[2]{\ensuremath{#1 \times 10^{#2}}}   % scientific notation
\newcommand{\gsim}{\gtrsim}
\newcommand{\lsim}{\lesssim}
\newcommand{\kms}{\ensuremath{\mbox{ km s}^{-1}}}

\newcommand{\nh}{\ensuremath{N_{\rm H}}}
\newcommand{\snr}{SNR~G093.3+6.9}
\newcommand{\snrb}{SNR~G315.4\ensuremath{-}2.3}
\newcommand{\snrc}{SNR~G084.2\ensuremath{-}0.8}
\newcommand{\snrd}{SNR~G127.1+0.5}
\newcommand{\Ga}{G093.3$+$6.9}
\newcommand{\Gb}{G315.4$-$2.3}
\newcommand{\Gc}{G084.2$+$0.8}
\newcommand{\Gd}{G127.1$+$0.5}

\newcommand{\scrhi}{H$\,${\tiny \sc I}}
\newcommand{\scrovi}{O$\,${\tiny \sc VI}}
\newcommand{\psrckl}{PSR~J1814$-$1744}
\newcommand{\psrmsk}{PSR~J1847$-$0130}

\begin{document}
\twocolumn[
\slugcomment{Accepted by ApJS}

\title{An X-ray Search for Compact Central Sources in Supernova
  Remnants I: SNRs \Ga, \Gb, \Gc, \& \Gd}
\shorttitle{A Search for Compact Central Sources in Four SNRs}
\shortauthors{Kaplan et al.}

\author{D.~L.~Kaplan\altaffilmark{1}, D.~A.~Frail\altaffilmark{2},
  B.~M.~Gaensler\altaffilmark{3}, E.~V.~Gotthelf\altaffilmark{4}, 
  S.~R.~Kulkarni\altaffilmark{1}, P.~O.~Slane\altaffilmark{3}, \and\
  A.~Nechita\altaffilmark{5}}

\begin{abstract}
Most astronomers now accept that stars more massive than about
$9\,M_{\odot}$ explode as supernovae and leave stellar remnants,
either neutron stars or black holes, with neutron stars being more
prevalent.  Recent modeling of the explosions suggests a significant
diversity in the key natal properties --- rotation rate, velocity, and magnetic
field strength --- of the resulting neutron stars that account for
the association of active radio pulsars, pulsar wind nebulae, and
magnetars with supernova remnants (SNRs). The discovery of a central
X-ray source in Cas~A, the youngest known Galactic SNR, dramatized the
expected diversity. However, less than half of the SNRs within 5 kpc
have identified central sources, and only three are identified as the
remnants of Type~Ia SNe.  Here, we report a systematic effort to
search for compact central sources in the remaining 23 SNRs of this
distance limited sample.  Our search was inspired, on empirical
considerations, by the enigmatic faint X-ray source in Cas~A;
motivated, on theoretical grounds, by the expectation that young
neutron stars emit cooling X-ray emission; and made possible by the
superb angular resolution offered by the {\em Chandra} X-ray mission
and the sensitivity of the \textit{XMM-Newton} mission.  

In this first paper we report {\em Chandra} observations of four SNRs
(\Ga, \Gb, \Gc, and \Gd).  We have undertaken a systematic optical/IR
identification program of the X-ray sources detected in the field
of each SNR. Foreground (flare stars,
active stars) and background (active galactic nuclei) sources have
identifiable IR/optical counterparts. In contrast, the counterparts of
neutron stars (or black holes) are expected to be very faint.  We are
able to account for all the well-detected X-ray sources and thus able
to state with some confidence that there are no associated central
sources down to a 
level of one tenth of that of the Cas~A central source, $L_{X}\lsim
10^{31}\mbox{ ergs s}^{-1}$.  We compare our limits with cooling
curves for neutron stars and find that any putative neutron stars in
these SNRs must be cooling faster than  expected for traditional
$1.35\,M_{\odot}$ neutron stars and that any putative pulsar must have
low spin-down luminosities $\dot E \lsim 10^{34}\mbox{ ergs
s}^{-1}$. However, our limits are unable to constrain the presence or
absence of more unusual options, such as relatively more massive
neutron stars with $M\gsim 1.45\,M_\odot$, neutron stars with exotic
interiors,  or quiescent black holes.
In subsequent papers, we will report on the X-ray and optical/IR
observations of the remaining members of the 5-kpc sample.
\end{abstract}

]

\altaffiltext{1}{Department of Astronomy, 105-24 California Institute
  of Technology, Pasadena, CA 91125; \texttt{dlk,srk@astro.caltech.edu} }
\altaffiltext{2}{National Radio Astronomy Observatory, P.O. Box O,
  Socorro, NM 87801; \texttt{dfrail@nrao.edu}.}
\altaffiltext{3}{Harvard-Smithsonian Center for Astrophysics, 60
  Garden Street, MS-6, Cambridge, MA 02138;
  \texttt{bgaensler,slane@cfa.harvard.edu}.} 
\altaffiltext{4}{Columbia Astrophysics Laboratory, Columbia
  University, 550 West 120th Street, New York, NY 10027; \texttt{evg@astro.columbia.edu}.}
\altaffiltext{5}{Harvard College, Cambridge, MA 02138;
  \texttt{nechita@fas.harvard.edu}.}

\section{Introduction}
Understanding the deaths of massive stars is one of the frontiers of
modern astrophysics.  Considerable observational evidence substantiates
the idea that stars below $M_{w}\sim 8\,M_\odot$ end their lives
as white dwarfs \citep{weidmann87}, while the detection of neutrinos
from SN~1987A dramatically illustrated that more massive stars undergo
core collapse \citep{hkk+87}.  The outcome of core collapse can either
be a neutron star or a black hole \citep*{whw02}.  However, there are
great uncertainties in the mapping between initial mass of the star
and the end product, and even more uncertainties in the natal
properties of the stellar remnant.  It is these uncertainties that
give observers opportunities to make new discoveries and theorists to
predict or ``postdict'' these discoveries.

The first issue --- the state of the star prior to collapse --- is
very dependent on the mass loss history of stars \citep{hfw+03}, a
phenomenon that is poorly understood and can easily be modified by the
presence of a binary companion.  For solar metallicity (the situation
relevant to this paper), stars with between $M_l \sim 9\,M_\odot$ and
$M_u\sim 25\,M_\odot$ are expected to form a neutron star
\citep{hfw+03}, while stars above $M_u$ are expected to form a black
hole either by fall-back of material which transmutes the neutron star
to a black hole or by direct collapse.  As an aside we note that
progenitors with masses between $M_{w}$ and $M_l$ may form an
O-Ne-Mg white dwarf that may collapse to neutron stars or simply
explode \citep{mnys80,nomoto84,nomoto87}.

%% The supernovae resulting from these explosions are paradoxically less
%% energetic for more massive stars given the increase in gravitational
%% binding energy for the core and a relatively fixed explosion energy
%% \citep{fryer99}.  Likewise the supernova remnants (SNRs) may also be
%% less energetic.  Unfortunately, an isolated black hole in the center
%% of a faint SNR is virtually impossible to detect.

The second issue --- the natal properties of the stellar remnant ---
apparently involves delicate physics but has strong observational
ramifications.  The gravitational binding energy of a neutron star is
$10^{53}\,$erg, of which only 1\% appears to be coupled to the ejecta
(which ultimately powers the SNR).  Even more minuscule fractions go
into rotational energy, kinetic energy (bulk motion) and magnetic
fields.  It is now generally agreed that three dimensional effects in
the explosion determine the natal properties \citep{burrows00,kpj+03}.

So far the discussion has assumed that the only gross parameter of
interest is the mass of the star. However, it is likely that rotation
of the progenitor can also profoundly affect the outcome. 

These two issues are now propelling two different areas of inquiry.
The relationship between the progenitor properties (mass, rotation)
and the gross outcome of core collapse (neutron star or black hole) is
observationally being determined by systematic studies of supernovae
and GRBs and their interconnection. The second area is in
understanding the natal properties of neutron stars, which is the main
focus of the paper.

The discovery of pulsars in the Vela SNR \citep*{lvm68} and the Crab
Nebula \citep{sr68} made concrete the suggestion that core collapse
results in neutron stars \citep{bz34}, some of which manifest
themselves as radio pulsars.  Young pulsars, in addition to pulsing in
the radio, can and usually do power synchrotron nebulae \citep{wp78}
that are indirect markers of pulsars.  These synchrotron nebulae are
commonly called pulsar wind nebulae, or PWNe.  Over the following two
decades, the notion that neutron stars resemble the Crab pulsar
guided the search for central objects as well as intensive radio
mapping of SNRs. As a result of these efforts, the term ``composite''
SNR (PWN + shell) was added to the SNR lexicon (see \citealt{mgh+79}).

However, over the last five years there have been three developments
that have severely revised our picture of young neutron stars.  First,
astronomers have come to accept of tremendous diversity in the natal
properties of young neutron stars.  Anomalous X-ray pulsars (AXPs),
soft $\gamma$-ray repeaters (SGRs), nearby thermal and radio quiet
neutron stars, long period radio pulsars with high inferred magnetic
fields (HBPSR) are now routinely found in astronomical literature.
These new classes of neutron stars have primarily come from high
energy (X-ray and $\gamma$-ray) observations.  Second, there is
appreciation that the radio luminosities of PWNe is poorly dependent
on the spin-down luminosity of their central pulsars. For example,
energetic pulsars may have  faint PWNe (e.g., PSR~J1119$-$6127;
\citealt{gs03}), and very sensitive imaging of the regions around
identified central sources has frequently yielded only upper limits to
the surface brightness of putative PWNe (e.g., \citealt*{gbs00}).
Third, seven nearby cooling NSs have been identified
\citep{haberl03} through \rosat\ observations. Many of these neutron
stars do not appear to be evolved versions of standard radio pulsars;
e.g., RX~J0720.4$-$3125 has a period longer than almost any known
radio pulsar but has a typical $B$-field strength
\citep{kkvkm02,zhc+02}.

While this diversity is clearly demonstrated observationally, theory
and simulation cannot yet constrain the fundamental birth properties
of neutron stars \citep[e.g.,][]{bom+03}.  Models still have
difficulties achieving explosions, much less following the activity in
the post-collapse object in any detail.

Three years ago we began a program of observationally investigating the
stellar remnants in a volume-limited census of Galactic supernova remnants.
The approach we took was inspired by the first light picture of \chandra,
the discovery of a central X-ray source in the well-studied and
youngest known supernova remnant in our Galaxy, Cassiopeia~A
\citep{t99}.  The nature of the object continues to be debated
\citep{pza+00,cph+01,mrj+02}.  However, one conclusion is crystal clear:
the X-ray source is not a standard radio pulsar (unbeamed or otherwise).

The basis of our effort is that observationally, all central sources
in SNRs known to date, regardless of the band of their initial
identification ($\gamma$-ray, X-ray, or radio)
appear to possess detectable X-ray emission. Theoretically, we expect
thermal X-ray emission from young neutron stars. Thus, on both counts
the search for central sources in young remnants is very well motivated.
However, a follow-up program is essential since many other foreground
sources such as flare stars, young stars, and accreting sources and
background sources such as AGN dominate the source counts \citep{hg88,mcba00}.
Fortunately, the sub-arcsecond spatial resolution of \chandra\ allow
efficient filtering of such contaminating objects.

To this end, we have identified a sample of SNRs within 5\,kpc of the
Sun which do not have known radio pulsars or PWNe and have not been
associated with Type~Ia explosions (Table~\ref{tab:snrs}.  We
successfully proposed for a ``large'' \chandra\ effort in AO-3 to
image nine SNRs. This initial allocation has been supplemented with
additional time in AO-4 of \chandra\ and AO-2 of \xmm.  Followup of
the X-ray sources has been undertaken with a plethora of ground based
telescopes in the optical and near-IR bands (Palomar 60-inch, Palomar
200-inch, Las Campanas 40-inch, ESO 3.5-m, Magellan 6.5-m and Keck
10-m). Here, we report the first analysis of four SNRs for which the
followup is now complete. The analysis for the remaining remnants will
be reported in future papers.

The organization of the paper is as follows. In \S~\ref{sec:class} we
summarize the empirical X-ray properties of the known sample of young
neutron stars. Such a summary is essential since the guiding principle
of our effort is to place our search for central objects against the
framework of existing classes of sources. Specifically, we are not
entirely guided by the relatively poorly understood cooling of neutron
stars. Our search has been designed to find objects as faint as one
tenth of the central X-ray source in Cassiopeia~A.  In
\S~\ref{sec:sample} we present a complete sample of cataloged SNRs
within 5\,kpc. Of these, 18 have an identified central source or are
known to be a composite remnant, while three are thought the be the
results of Type~Ia SNe. The remaining 23 seemingly ``hollow'' SNRs
form our primary sample.  By ``hollow'', we refer to SNRs that have
distinct shells but no obvious indication of central neutron stars ---
see \citet{vg97}.  Section~\ref{sec:obs} has a general overview of our
observations and analysis techniques: in
\S~\ref{sec:Chandra-survey}--\ref{sec:timing} we present a summary of
the details of our \chandra\ observations and data reduction, and a
likewise global description of the extensive multi-wavelength followup
is given in \S~\ref{sec:cpt}.  We follow this by detailed descriptions
of each of the first four SNRs, its observations, and identification
of counterparts to its X-ray sources in \S\S~\ref{sec:snra},
\ref{sec:snrb}, \ref{sec:snrc}, and \ref{sec:snrd} for SNRs \Ga, \Gb,
\Gc, and \Gd, respectively.  It must be appreciated that complete
identification of all X-ray sources is essential, given the small
sample size.  We hope that our detailed cataloging will be of help to
efforts such as {\em ChaMPlane} \citep{gzh+03}.  Finally, in
\S~\ref{sec:limits} we discuss what limits our data can place on the
existence of central sources in the four SNRs, and we conclude in
\S~\ref{sec:conclusions}.

%\begin{landscape}
\begin{deluxetable}{l c c l r c c c c l}
\setlength{\tabcolsep}{0.03in}
\tablecaption{SNRs Within 5 kpc\label{tab:snrs}}
\tablewidth{0pt}
\tabletypesize{\scriptsize}
\tablehead{
\colhead{SNR G} & \colhead{Other Name} & \colhead{Dist.} & \colhead{Distance Method} & \colhead{Size\tablenotemark{a}} &
\colhead{Type\tablenotemark{b}} & \colhead{Central Source/} & \colhead{X-ray?\tablenotemark{d}} & \colhead{Sample\tablenotemark{e}} & \colhead{Refs\tablenotemark{f}} \\
 &  & \colhead{(kpc)} & & \colhead{(arcmin)} & & \colhead{Ia} \\
}
\startdata
004.5+6.8   &Kepler\tablenotemark{g} & 4.5 & Optical expansion/\scrhi & 3 & S & Ia?\tablenotemark{c}
&\nodata &\nodata & 1\\
005.4$-$1.2 & Duck & 5 & \scrhi\ absorption & 35 & C? & pulsar?  & yes
& \nodata & 2,3 \\
006.4$-$0.1 &W 28\tablenotemark{g} & 3.0 & OH masers & 42 & S & \nodata &\nodata 
 &\nodata & 4\\ 
011.2$-$0.3 & & 5 & \scrhi\ absorption & 4 & C & pulsar  & yes &\nodata& 5 \\
013.3$-$1.3 & & 2.4 & CO absorption & 70 & S & \nodata  & \nodata & \xmm-AO2 \\[0.1in]
034.7$-$0.4 & W44 & 2.5 & \scrhi\ absorption & 35 & C & pulsar  & yes& \nodata& 6,7,8\\
039.7$-$2.0 & W50 & 5& VLBI & 120 & ? & SS 433  & yes &\nodata& 9,10 \\
053.6$-$2.2 &3C 400.2 & 2.8 & \scrhi\ association & 33& S & \nodata
 &\nodata & \cxo-AO3 \\ 
054.4$-$0.3 &HC 40 & 3.3 & CO association & 40 & S & \nodata  &\nodata & \cxo-AO3 \\ 
065.3+5.7   & & 0.8 & Optical velocity & 310 & S & \nodata  &\nodata & \cxo-AO4
\\[0.1in]
069.0+2.7 & CTB 80 & 2 & \scrhi\ absorption & 80 & ? & pulsar  & yes &\nodata& 11,12\\
074.0$-$8.5 & Cygnus Loop & 0.44 & Optical proper motion & 230 & S &
X-ray source?  & \nodata & \cxo-AO4 & 13\\
078.2+2.1 & $\gamma$ Cygni & 1.2 & OH/CO association & 60 & S &
\nodata  &\nodata & \xmm-AO2 & 14 \\
084.2$-$0.8 & & 4.5 & \scrhi/CO association & 20 & S & \nodata  &\nodata & \cxo-AO3
\\ 	     
089.0+4.7 & HB 21 & 0.8 & OB association & 120 & S & \nodata  &\nodata & \nodata\\[0.1in]
093.3+6.9   &DA 530 & 3.5 & \scrhi\ absorption/X-ray fitting & 27 & S & \nodata &\nodata  & \cxo-AO3 \\ 
111.7$-$2.1 &Cas A & 3.4 & Optical expansion & 5 & S & CCO & yes &
\nodata & 15 \\
114.3+0.3 & & 3.5 & \scrhi\ association & 90 & S & pulsar & yes &\nodata & 16,17,18 \\
116.5+1.1 & & 4 & \scrhi\ association & 80 & S & \nodata &\nodata  & \nodata
\\
116.9+0.2   &CTB 1 & 3.1 & Optical lines & 34 & S & \nodata &\nodata& \cxo-AO3& \\[0.1in] 
119.5+10.2 & CTA 1 & 1.4 &\scrhi\ association  & 90 & S &
$\gamma$-/X-ray neutron star &\nodata  & \nodata & 19,20,21  \\
120.1+1.4 & Tycho & 2.4 & Proper motion/shock vel. & 8 & S & Ia &
\nodata & \nodata & 22 \\
127.1+0.5   &R5 & 1.3 & Assoc.\ with NGC 559 & 45 & S & \nodata &\nodata  & \cxo-AO3 \\
130.7+3.1 & 3C 58 & 3.2 & \scrhi\ absorption & 9 & F & pulsar &yes &\nodata& 23,24\\
132.7+1.3 & HB 3 & 2.2 & Interactions with ISM & 80 & S & \nodata &\nodata  &
\xmm-AO2 \\[0.1in]
156.2+5.7 & & 1.3 & NEI fits & 110 & S & \nodata &\nodata  & \cxo-AO4& \\
160.8+2.6 & HB9 & $<4$ & \scrhi\, optical vel. & 140 & S & \nodata &\nodata  &
\cxo-AO4 & \\
166.0+4.3   &VRO 42.05.01 & 4.5 & \scrhi\ association & 55 & S & \nodata &\nodata  & \cxo-AO3 \\ 
166.2+2.5 & & 4 & \scrhi\ interaction & 90 & S & \nodata &\nodata  & \nodata &
\\
180.0$-$1.7 &S147 &1.2 & pulsar dispersion measure & 180 & S & pulsar & yes &\nodata& 25,26,27\\[0.1in]
184.6$-$5.8 & Crab& 2 & proper motion/radial vel.\ & 7 & F & pulsar & yes &\nodata& 28 \\
189.1+3.0   &IC 443 & 1.5 & Opt.\ vel./assoc.\ with S249 & 45 & S &
neutron star & yes & \nodata & 29 \\
203.0+12.0 & Monogem Ring & 0.3 & \scrovi\ absorption/modeling & 1500
 & S & pulsar & yes & \nodata & 30\\
205.5+0.5 & Monoceros & 1.2 & Optical velocity & 220 & S & \nodata &\nodata  &
\cxo-AO4\\
260.4$-$3.4 & Puppis A & 2.2 & \scrhi\ association  & 60 & S &
CCO & yes & \nodata & 31,32 \\ 
263.9$-$3.3 & Vela & 0.3 & pulsar parallax & 255 & C & pulsar & yes
&\nodata& 33 \\[0.1in]
296.5+10.0 & PKS 1209$-$51/52 & 2.1 & \scrhi\ association  & 90 & S
& CCO/X-ray pulsar & yes & \nodata & 34,35\\
309.8+0.0   & & 3.6 & \scrhi\ absorption & 25 & S & \nodata &\nodata  & \cxo-AO3 \\ 
315.4$-$2.3 &RCW 86 & 2.8 & Optical lines & 42 & S & \nodata &\nodata  &
\cxo-AO3 & 36\\
327.4+0.4 & Kes 27 & 4.3 & \scrhi\ absorption/interact. & 30 & S
& \nodata &\nodata  & \nodata & \\
327.6+14.6 & SN~1006 & 2.2 & Spectra/proper motion & 30 & S & Ia &
\nodata & \nodata &  37,38\\[0.1in]
330.0+15.0 & Lupus Loop & 0.8 & \scrhi\ & 180 & S & \nodata &\nodata  &
\cxo-AO4&  \\
332.4$-$0.4 & RCW 103 & 3.3 & \scrhi\ absorption & 10 & S & CCO
& yes &\nodata& 39 \\
343.1$-$2.3 & & 2.5 &\scrhi\ absorption  & 32 & C & pulsar &yes& \nodata& 40,41\\
354.1+0.1 & & 5 & recombination lines & 15 & C & pulsar? & no & \nodata& 42,43\\
\enddata
\tablenotetext{a}{Major axis.}
\tablenotetext{b}{Types are: S (shell), C (composite), and F (filled,
  or PWN), as
  determined by \citet{green01}.  A
  question mark indicates that the type is poorly determined.}
\tablenotetext{c}{Indicates if a central source is known, or if the
  SNR is thought to be of type Ia.  If a central source is known and
  it falls into one of the classes from \S~\ref{sec:class}, it is
  labeled accordingly.}
\tablenotetext{d}{Indicates whether the central source has been
  detected in X-rays.  See Table~\ref{tab:psrs}.}
\tablenotetext{e}{Refers to X-ray samples of shell SNRs described in \S~\ref{sec:Chandra-survey}.}
\tablenotetext{f}{References deal only with the central source or Ia classification.
General SNR properties were taken from \citet{green01} and references therein.}
\tablenotetext{g}{Has already been observed with \chandra.}

\tablerefs{1: \citet{kt99}; 2: \citet{fk91}; 3: \citet{mjk+91}; 4:
\citet{yswk00}; 5: \citet{ttdm97}; 6: \citet*{wcd91}; 7:
\citet{hhs+97}; 8: \citet{pks02}; 9: \citet*{cgc75}; 10:
\citet{wwgs83}; 11: \citet{kcb+88}; 12: \citet{mgb+02}; 13:
\citet{mtt+98}; 14: \citet{bkc+96}; 15: \citet{t99}; 16:
\citet{kpha93}; 17: \citet*{bbt96}; 18: \citet*{frs93}; 19:
\citet*{sss95}; 20: \citet{ssb+97}; 21: \citet{szh+04}; 22:
\citet{baade45}; 23: \citet{mss+02}; 24: \citet{csl+02}; 25:
\citet{acj+96}; 26: \citet{rn03}; 27: \citet{klh+03}; 28:
\citet{ccl+69}; 29: \citet{ocw+01}; 30: \citet{tbb+03}; 31:
\citet*{pzt99}; 32: \citet*{ztp99}; 33: \citet{lvm68}; 34: \citet{hb84};
35: \citet{zpst00}; 36: \citet{vbdk00}; 37: \citet{fwlh88}; 38:
\citet{apg01}; 39: \citet*{gph97}; 40: \citet*{mop93}; 41: \citet*{bbt95}; 42:
\citet{fgw94}; 43: \citet{bt97}.
%36: \citet{gcm+95};
%18: \citet{sh82}; 
%19: \citet{mtd82}; 
} 
\end{deluxetable}
%\end{landscape}

\section{X-ray Properties of Young Neutron Stars}
\label{sec:class}

The first manifestations of neutron stars associated with SNRs were
traditional rotation-powered pulsars (such as those in the Crab and
Vela SNRs).  With the advent of high-energy missions a number of new
classes were discovered.  These include Soft Gamma Repeaters
\citep[SGRs; for a review see][]{hur00}, Anomalous X-ray Pulsars
\citep[AXPs; for a review see][]{mcis02}, and Compact Central Objects
\citep[CCOs; for a review see][]{psgz02}.  Finally, recent radio surveys have uncovered
central Crab-like pulsars with field strengths beyond $10^{13}$~G ---
the so-called high-B pulsars \citep[HBPSRs;][]{ckl+00,gvbt00}.  SGRs
and AXPs have been suggested to be magnetars \citep{dt92,td95}:
neutron stars with extremely strong field strengths, $B\gsim
10^{15}$~G.  CCO is the generic name for a heterogeneous group of
X-ray sources emitting largely unpulsed soft (thermal) emission and
lacking detectable radio emission.  Another possibly related class are
the Isolated Neutron Stars \citep[INSs;][]{haberl03}, also called
Radio-Quiet Neutron Stars (RQNSs) or Dim Thermal Neutron Stars
(DTNSs).  Below we summarize the general properties of each of these classes in
turn.

\begin{deluxetable}{c c c c c c r l}
\tablecaption{X-ray Properties of Central Sources\label{tab:psrs} from
Table~\ref{tab:snrs}}
\tablewidth{0pt}
\tabletypesize{\footnotesize}
\tablehead{
\colhead{Source} & \colhead{SNR G} & \colhead{$kT_{\infty}^{\rm BB}$} &
\colhead{$\Gamma$} &
\colhead{$\log_{10} L_{X}$\tablenotemark{a}} &
\colhead{X-ray} &
\colhead{Age\tablenotemark{b}}  & \colhead{Refs} \\
 & & \colhead{(keV)} & & \colhead{$(\mbox{ergs s}^{-1})$}
& \colhead{PWN?}&
\colhead{(kyr)}\\
}
\startdata
PSR~B1757$-$24 & 005.4$-$1.2\phn & \nodata & 1.6 & 33.0 &
Yes\tablenotemark{e}& $>39$ &1,2 \\
PSR~J1811$-$1925 & 011.2$-$0.3\phn & \nodata & 1.7 & 33.7 & Yes &1.6 & 3,4,5 \\
PSR~B1853+01 & 034.7$-$0.4\phn & \nodata & 1.3 & 31.2 &Yes &20 & 6 \\
SS 433\tablenotemark{c} & 039.7$-$2.0\phn & \nodata & 0.7 & 35\phm{.0} &\nodata&5--40 & 7 \\
PSR~B1951+32 & 069.0+2.7\phn & \nodata & 1.6 & 32.9 &Yes &64 & 8 \\[0.1in]
CXO J232327.8+584842 & 111.7$-$2.1\phn & 0.7\phn & 3.0 & 33.5/34.7\tablenotemark{h} &No& 0.3 & 9\\
PSR~B2334+61 & 114.3+0.3\phn & \nodata & 2.0\tablenotemark{d}&  31.7&\nodata\tablenotemark{g} & 41 & 10 \\
RX J0007.0+7302 & 119.5+10.2 & 0.14 & 1.5 & 31.2 & Yes & 13 & 11 \\
PSR~J0205+6449 & 130.7+3.1\phn & \nodata & 1.7 & 32.2 &Yes & 0.8 & 12,13\\
PSR~J0538+2817 & 180.0$-$1.7\phn & 0.16 &\nodata & 32.9 &Yes & 30& 14,15 \\[0.1in]
PSR~B0531+21 & 184.6$-$5.8\phn & \nodata & 1.6 & 35.8 &Yes  & 1.0 &16 \\
CXO J061705.3+222127 & 189.1+3.0\phn  & 0.7 & \nodata & 31.3 &Yes  &30 & 17 \\
PSR~B0656+14 & 203.0+12.0 &0.07+0.14 & 1.5 & 31.1 & No & 100 & 18,19,20\\
RX~J0822$-$4300 & 260.4$-$3.4\phn & 0.4\phn & \nodata & 33.5 &No &3.7 & 21,22\\
PSR~B0833$-$45 & 263.9$-$3.3\phn & \nodata & 2.5 & 32.5 &Yes &11 & 23\\[0.1in]
1E~1207.4$-$5209 & 296.5+10.0 & 0.26 & \nodata & 33.1 &No &7 & 24\\
%PSR~B1509$-$58 & 320.4$-$1.2 & \nodata & 1.4 & 34.9 & 1.6 &15\\[0.1in]
1E~161348$-$5055\tablenotemark{c} & 332.4$-$0.4\phn & 0.6\phn & \nodata & 32.1--33.9 & No& 2 & 25
\\
PSR~B1706$-$44 & 343.1$-$2.3\phn & 0.14 & 2.0 & 32.6 & Yes\tablenotemark{e} &18 & 26\\
PSR~B1727$-$33 & 354.1+0.1\phn & \nodata & \nodata & $<32.6$/$<32.5$ &\nodata\tablenotemark{g} &15
& 16 \\ 
\enddata 
\tablerefs{1: \citet{kggl01}; 2: \citet{gf00}; 3: \citet{ttdm97}; 4:
\citet{rtk+03}; 5: \citet{sg02}; 6: \citet{pks02}; 7: \citet{kkmb96};
8: \citet{mgb+02}; 9: \citet*{mti02}; 10: \citet{bbt96}; 11:
\citet{szh+04}; 12: \citet{shm02}; 13: \citet{stephenson71}; 14:
\citet{rn03}; 15: \citet{klh+03}; 16: \citet{bt97}; 17:
\citet{ocw+01}; 18: \citet{ms02}; 19: \citet{btgg03}; 20:
\citet{gcf+96}; 21: \citet{ztp99}; 22: \citet{psgz02}; 23:
\citet{pzs+01}; 24: \citet{spzt02}; 25: \citet*{gpv99}; 26:
\citet*{ghd02}. In addition, general pulsar data have been taken from
\citet{hm03}.}

\tablenotetext{a}{Luminosity for
only the point-source in the 0.5--2.0~keV band, assuming  the distance
from Table~\ref{tab:snrs}.  Upper limits to the luminosity are given for
a blackbody with $kT_\infty=0.25$~keV and for a power-law with $\Gamma=2.2$.}
\tablenotetext{b}{The best estimate of the age of the SNR if known,
  otherwise the pulsar spin-down age $P/2\dot P$.}
\tablenotetext{c}{Possibly not an isolated neutron star.}
\tablenotetext{d}{Assumed.}  
\tablenotetext{e}{The X-ray PWNe here are significantly fainter compared
  to the pulsars than for other sources \citep{kggl01,ghd02}.}
\tablenotetext{g}{The current X-ray data do not
  sufficiently constrain the existence of a nebula.}
\tablenotetext{h}{The current X-ray data do not constrain the
  spectrum: either a blackbody or power-law model is possible.}
\end{deluxetable}

\subsection{Radio Pulsars}
\label{sec:psrdesc}
Radio pulsars observed in the X-rays often have two-component spectra.
Here, the thermal component tends to be softer ($\lsim 0.3$~keV) and
the power-law component harder ($\Gamma=1.5$--2.5) than those of AXPs;
see \citet{bt97}, \citet{pccm02}, and Tables~\ref{tab:psrs} and
\ref{tab:radpsrs}.  A rough relation was originally discovered by
\citet{sw88} between the X-ray luminosity and the rotational energy
loss rate $\dot E$: $L_{{\rm X},0.1-2.4\,{\rm keV}} \approx 10^{-3}
\dot E$ \citep{bt97}.  While this relation has been updated for
specific classes of neutron stars and different energy bands
\citep[e.g.,][]{pccm02} and has considerable scatter, it still holds
on average .  The observed X-ray luminosities of radio pulsars then
vary between $10^{31}\mbox{ ergs s}^{-1}$ and $10^{37}\mbox{ ergs
s}^{-1}$, depending on their values of $\dot E$ (and through that,
their values of $P$ and $\dot P$).

While only millisecond pulsars and ``old'' ($>10^6$~yr) pulsars from
the sample of \citet{pccm02}, have values of $\dot E$ less than
$10^{34}\mbox{ ergs s}^{-1}$, there have been three young pulsars
discovered recently that are more extreme.  These are the HBPSRs
\psrckl\ ($P=4.0$~s and $\dot E=\expnt{4.7}{32}\mbox{ ergs s}^{-1}$;
\citealt{ckl+00}), \psrmsk\ ($P=6.7$~s and $\dot
E=\expnt{1.7}{32}\mbox{ ergs s}^{-1}$; \citealt{msk+03}), and
PSR~J1718-37184 ($P=3.4$~s and $\dot E=\expnt{1.6}{33}\mbox{ ergs
s}^{-1}$; \citealt{mks+03}).  The first two do not have detected X-ray
emission (\citealt*{pkc00}, \citealt{msk+03}), while the third does
have a very faint ($L_{2-10\,{\rm keV}}\approx \expnt{9}{29}\mbox{
ergs s}^{-1}$) X-ray counterpart [the energetic ($\dot E \gsim
10^{36}\mbox{ ergs s}^{-1}$) HBPSRs have brighter X-ray counterparts
(\citealt{gak+02,gs03}, \citealt*{hcg03})], consistent with their
values of $\dot E$.

\begin{deluxetable}{c c c c c c c c}
\tablecaption{Properties of Rotation-Powered Pulsars Associated with SNRs\label{tab:radpsrs}}
\tablewidth{0pt}
\tablehead{\colhead{PSR} & \colhead{SNR} & \colhead{$P$} &
  \colhead{$\tau$\tablenotemark{a}} & \colhead{$\log_{10}\dot E$} & \colhead{$D$} &
  \colhead{$\log_{10} L_{X}$\tablenotemark{b}} & \colhead{Refs.} \\
 & & \colhead{(ms)} & \colhead{(kyr)} & \colhead{$(\mbox{ergs
      s}^{-1})$} & \colhead{(kpc)} & \colhead{$(\mbox{ergs
      s}^{-1})$} & \\
}
\startdata
J0205+6449 &	G130.7+3.1&	\phn66	&5.4 & 37.4	& 3.2  & 32.2      & \\
J0537$-$6910 &	N157B\tablenotemark{c}	 &	\phn16	&5.0 & 38.7	& 49.4 & 38.3& 	1,2\\
J0538+2817 &	G180.0$-$1.7&	143	&620 & 34.7	& 1.8  & 32.9	   & \\
B0531+21&	G184.6$-$5.8&	\phn33	&1.2 & 38.7	& 2.0  & 35.8	   & \\
B0540$-$69&	0540$-$69.3\tablenotemark{c}	 &	\phn50	&1.7 & 38.2	& 49.4 & 36.3 & 	2,3 \\[0.1in]
B0833$-$45&	G263.9$-$3.3&	\phn89	&11  & 36.8	& 0.3  & 32.5	   & \\
J1016$-$5857 &  G284.3$-$1.8 &107 & 21 & 36.5 & 3 &32.5 &  4\\
J1119$-$6127 &	G292.2$-$0.5&	408	&1.6 & 36.4	& 6    & 32.5 & 	5\\
J1124$-$5916 &	G292.0+1.8&	135	&2.9 & 37.1	& $>6$  & 33.0 & 	6,7\\
B1338$-$62 &	G308.8$-$0.1&	193	&12  & 36.1	& 9    &32.8	   & 8\\[0.1in]
B1509$-$58&	G320.4$-$1.2&	151	&1.6 & 37.3	& 5.2  & 34.1 & 	9,10\\
B1643$-$43 &    G341.2+0.9 & 232 & 33 & 35.5 & 7 & \nodata &\\
B1706$-$44&	G343.1$-$2.3&	102	&18  & 36.5	& 2.5  & 32.6	   & \\
B1727$-$33&	G354.1+0.1&	139	&26  & 36.1	& 5    & $<$32.6	   & \\
B1757$-$24&	G005.4$-$1.2&	125	&16  & 36.4	& 5    & 33.0	   & \\[0.1in]
J1811$-$1925 &	G011.2$-$0.3&	\phn65	&23  & 36.8	& 5    & 33.7	   & \\
J1846$-$0258 &	G029.7$-$0.3&	324	& 0.72 & 36.9	& 19   & 35.4 & 	11\\
B1853+01&	G034.7$-$0.4&	267	&20  & 35.6	& 2.5  & 31.2	   & \\
J1930+1852 &    G054.1+0.3 &136 & 2.9 & 37.1 & 5 & 33.3 & 12\\
B1951+32&	G069.0+2.7&	\phn40	&110 & 36.6	& 2    & 35	   & \\[0.1in]
J2229+6114 &	G106.6+2.9&	\phn52	&11  & 37.3	& 3    & 32.8 & 	13,14\\
B2334+61&	G114.3+0.3&	495	&41  & 34.8	& 3.5  & 31.7      & \\
\enddata
\tablecomments{Pulsar-SNR associations are largely drawn from \citet{kh02}.}
\tablenotetext{a}{Characteristic age $\tau\equiv P/2\dot P$.}
\tablenotetext{b}{X-ray luminosity  in the 0.5--2.0~keV band.  Upper limits to the luminosity are given for
a power-law with $\Gamma=2.2$.}
\tablenotetext{c}{In the Large Magellanic Cloud.}
\tablerefs{1: \citet{wgcd01}; 2: \citet{gotthelf03}; 3: \citet{gw00b};
  4: \citet{cbm+01}; 5: \citet{gs03}; 6: \citet{hsb+01}; 7:
  \citet{gw03}; 8: \citet*{gkm03}; 9: \citet{gcm+95}; 10:
  \citet{gak+02}; 11: \citet{hcg03}; 12: \citet{clb+02}; 13:
  \citet*{kup01}; 14: \citet{hcg+01} .  Also see references from
  Tabs.~\ref{tab:snrs} and \ref{tab:psrs}.}
\end{deluxetable}

\subsection{Pulsar Wind Nebulae}
\label{sec:pwndesc}
Pulsar wind nebulae (PWNe), are bright, centrally condensed nebulae
with non-thermal (power-law) X-ray and radio spectra often associated
with young, energetic pulsars and SNRs (here we refer only to
``bubble'' PWNe, as differentiated by the bow-shock PWNe produced by
the motion of the pulsars through the ambient medium; for reviews, see
\citealt{che98} or \citealt{gaensler03}).  The photon indices range
from 1.3--2.3 \citep{gotthelf03}, similar to those of pulsars, but
they are roughly $\sim 10$ times as luminous for a given $\dot E$
\citep{gotthelf04} and the sizes range from a few
arcseconds to several arcminutes.  PWNe, both X-ray and radio, offer a
great advantage over bare radio pulsars for inferring the existence of
neutron stars: they are unbeamed.  This fact has historically been
used in a number of cases to infer the existence of energetic pulsars
where the pulsar had not been seen itself,  such as 3C~58
(\citealt*{bhs82}, \citealt{fm93}), N157B \citep{wg98}, and Kes~75 \citep{bh84}.

The evolution of a PWN (see \citealt*{vdsdk03} for a recent review)
begins as it expands supersonically through the shocked ejecta of the
SNR.  Here, the radius of the PWN is $\propto \dot E^{1/5} t^{6/5}$
\citep[assuming that the total luminosity of the neutron star is
$\propto \dot E$;][]{vdsagt01,vds03}, so it depends only weakly on
$\dot E$ but more strongly on the remnant age.  Eventually, though,
the reverse shock of the SNR crashes back on the PWN at a time $t_{\rm
col} \sim 10^4$~yr \citep[][although this depends on the ejecta mass]{vdsdk03}.  After this occurs , the spherical
portion of the PWN continues to expand subsonically with its radius
$\propto t^{11/15}$ \citep{vdsagt01}, while at the same time the
pulsar moves away from the center of the SNR and begins to form a
bow-shock nebula (when its motion through the SNR ejecta becomes
supersonic at ages of $\gsim 3000$~yr; \citealt{vdsdk03}).  The
passage of the reverse shock should re-heat and energize the PWN,
causing it to brighten.

\subsection{AXPs \& SGRs}
\label{sec:axpdesc}
In the X-ray band, traditional AXPs are characterized by a
two-component spectrum: a power-law with index 3--4, and a soft
blackbody with temperature 0.3--0.7~keV \citep{mcis02}.  While the
distances are quite uncertain, especially to those not in SNRs, the
luminosities are relatively constant and $>10^{34}\mbox{ ergs s}^{-1}$
(in the 1--10~keV band; \citealt{mcis02}), substantially greater than
their spin-down energy loss rates.  These sources are characterized by
slow spin periods (6--12~s) with substantial period derivatives that
indicate the presence of very strong magnetic fields ($B>10^{14}$~G).

There are two objects, however, that may indicate AXPs can have
substantial X-ray variability.  First, AX~J184453$-$025640
\citep{vgtg00} is a 7~s X-ray pulsar which varied by a factor of
$\gsim 10$ in flux but whose properties are otherwise found to be
fully consistent with an AXP.  More recently, the AXP-candidate
XTE~J1810$-$197 was discovered in \textit{RXTE} data \citep{ims+03};
\chandra\ and \xmm\ data confirm the pulsations present in the
\textit{RXTE} data and allow for comparison with archival
\textit{Einstein}, \textit{ROSAT}, and \textit{ASCA} data where it is
a factor of $\approx 100$ fainter \citep{ghbb03}.  In the bright
(current) state, XTE~J1810$-$197 has an absorbed X-ray flux
(0.5--10~keV) of $\approx \expnt{4}{-11}\mbox{ ergs cm}^{-2}\mbox{
s}^{-1}$, while \rosat\ data from 1993 has a flux of $\approx
\expnt{5}{-13}\mbox{ ergs cm}^{-2}\mbox{ s}^{-1}$.  Converting the
fluxes to luminosities is uncertain due to the largely unconstrained
distance, but \citet{ghbb03} determine an upper limit of 5~kpc.  With
this, the current luminosity in the 0.5--2~keV range is $\lsim
\expnt{3}{35}\mbox{ ergs s}^{-1}$, similar to other AXPs, but the
``quiescent'' luminosity is $\sim 10^{33}\mbox{ ergs s}^{-1}$.  While
less certain, AX~J184453$-$025640 shows a roughly similar range of
luminosities.

SGRs have roughly similar quiescent luminosities \citep{hur00},
although their spectra are somewhat harder.  However, none of the SGRs
is firmly associated with a SNR \citep{gsgv01}.  There is now mounting
observational evidence that SGRs and AXPs are related objects
(\citealt*{gkw02}, \citealt{kkm+03,kgw+03}), confirming the hypothesis of
\citet{td96}.  In what follows, we do not treat the SGRs as separate
objects since their quiescent X-ray properties are sufficiently
similar to those of AXPs.

\subsection{Cooling Radio-Quiet Neutron Stars/Compact Central Objects}
\label{sec:rqnsdesc}
Regardless of emission at other wavelengths, young neutron stars
should have thermal X-ray emission from their surfaces as they cool.
Indeed, it is likely that \rosat\ has detected a
number of nearby, $\sim 10^6$~yr (too old to be in SNRs) neutron stars
\citep{ttzc00,haberl03} with no detectable radio emission
(\citealt*{kkvk03}, \citealt{kvkm+03,johnston03}) --- hence the name RQNS.  The
\rosat\ sources have temperatures of $\lsim 100$~eV.  This thermal
emission is almost certain to be present at some level, regardless of
whatever other processes are occurring (radio emission, accretion,
etc.), but the exact level of emission depends on the mass of the
neutron star and on the presence or absence of exotic particles
(pions, kaons, hyperons, free quarks, etc.) and/or processes (i.e.\
direct Urca cooling) in the core (\citealt*{kyg02}, \citealt{ttt+02}).

A number of superficially similar sources known as CCOs have been
discovered in SNRs, such as the sources in Cas~A and Puppis~A.  The
temperatures of CCOs should be higher than those of the field RQNSs by
a factor of 5 or so, depending on cooling physics
\citep[e.g.,][]{page98}.  For some of these sources, it is likely that
there is little if any radio emission due to the low values of $\dot
E$ \citep{kkvkm02}, but for others the lack of such emission may just
be beaming effect (i.e., as with Geminga), implying beaming fractions
of $\sim 50$\% \citep{bj99}.  The radii of the blackbody fits to the
CCOs in Table~\ref{tab:psrs} are typically less than the 10--15~km
expected for a neutron star (presumably since a blackbody, while
providing an adequate spectral fit, does not actually represent the
emission of the surface) --- values of $R_{\infty}\approx 1$--3~km are
common.

\section{Survey Design}
\subsection{A Volume-Limited Sample of Shell SNRs}
\label{sec:sample}
\begin{figure}
\plotone{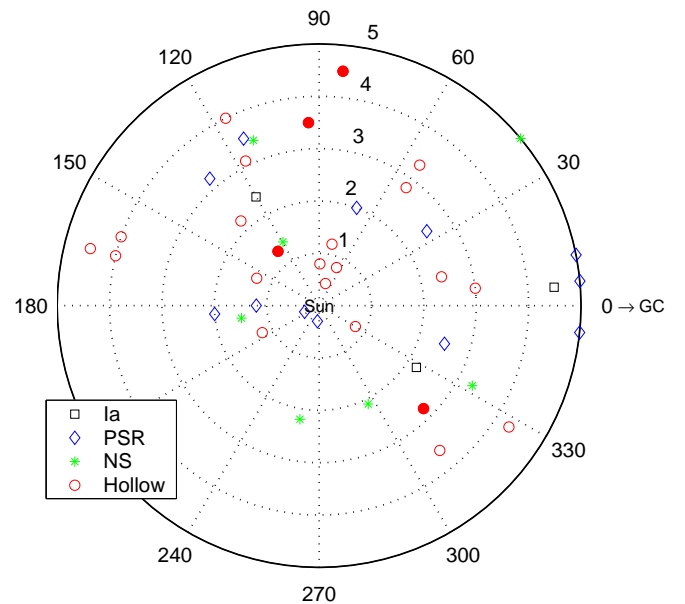}
\caption{Galactic distribution of SNRs from Table~\ref{tab:snrs}.
  Galactic longitude $l$ (degrees) is plotted against distance (kpc),
  with the Sun at the origin and the Galactic center to the right.
  Black squares are the Ia SNRs, blue diamonds are associated with
  radio pulsars, green asterisks are associated with other types of
  neutron stars (RQNS, etc.), and the red circles are the hollow SNRs.
  The four filled circles are the SNRs from this paper with detailed
  analyses.  }
\label{fig:snrs}
\end{figure}

The success of this effort hinges upon defining an objectively
constructed sample so that strong conclusions can be drawn not only
from detections but also non-detections. Bearing this in mind, we
identified all of the SNRs\footnote{Drawn from the Galactic SNR
Catalog \citep{g00}.  Since we constructed the table, the catalog has
been updated \citep{green01}, and it is this list and more recent
references that we use to determine the properties of the SNRs.} that
are at a distance of less than 5 kpc as determined from a reliable
distance determination (e.g.\ derived from the kinematic velocity of
associated line emission or a parallax, rather than from \nh\ or the
$\Sigma-D$ relation). This sample is comprised of the 45 SNRs listed
in Table~\ref{tab:snrs}.  This is not an entirely complete sample, as
illustrated by the void in the third quadrant in Figure~\ref{fig:snrs}
and the relative paucity of distant SNRs toward the Galactic center,
but the criteria for inclusion in the sample (detected SNR with a
robust distance) should not be correlated with the properties of the
central objects.  We estimate that we cover $\sim 15$\% of the
Galactic molecular gas, and hence sites of massive star formation,
with our distance criteria (based on  \citealt{dame93}).
For the X-ray observations we eliminated all SNRs that were of Type Ia
or those that are already associated with a NS and/or central
synchrotron nebula\footnote{We assume that sources like IC~443, which
has a synchrotron nebula but where pulsations have not yet been
detected, still do have central neutron stars.} --- i.e.\ we only
include those that are type S in Table~\ref{tab:snrs}.

The shell remnants are the major remaining sample where the neutron
star population has not been systematically assessed.  For the
SNRs with PWNe, some estimate as to the neutron star's properties
can be made even if the source itself has not been detected, but the
shell remnants permit no such estimation.   Therefore, these
shell-type SNRs are the subject of our survey.  Below we first the
status of neutron star velocities, an issue that affect our survey
design.  We then continue to describe the survey itself.

\subsection{Neutron Star Velocities}
Radio pulsars have high space velocities, among the highest in the
Galaxy, as measured from timing, scintillation, and interferometry.
Various authors \citep[e.g.,][]{hp97,lbh97,cc98,acc02} have
modeled the distribution slightly differently, but they all agree that a
substantial number of pulsars move with speeds $>300\mbox{ km
s}^{-1}$, while $\approx 90\%$ have speeds $<700\mbox{ km
s}^{-1}$.

The space velocities of neutron stars can also be inferred from their
offsets from the centers of associated supernova remnants.  Such an
approach demands reliable estimates for the distance and age of the
system under consideration. It also assumes that the geometric center
of a SNR is easily identifiable, and that this center corresponds to
the site of the supernova. With these caveats in mind, such an analysis
potentially provides a direct measurement of the neutron star velocity
distribution, free from the many selection effects associated with proper
motion studies.

\citet*{fgw94} carried out a detailed study of the offsets
of radio pulsars from the centers of SNRs. From a sample of 15
pulsar/SNR associations, \citet{fgw94} concluded that young radio pulsars
are born with projected space velocities ranging between $50$~\kms
and $1000$~\kms, with a mean of $\sim$500~\kms. While at that time
this distribution represented a somewhat more rapidly-moving population
than that which had been inferred from proper motion studies, it agrees
well with the more recent determinations discussed above.

However, the preceding discussion assumes that pulsars were born near
the centers of what are now SNRs.  Some authors dispute that this is
always the case \citep{gvar02}, suggesting that if the SN progenitors
have large space velocities they could evacuate a bubble with their
winds, move toward the edges of those bubbles, and then explode (in a
so-called off-center cavity explosion).
This would make the origins of neutron stars closer to the edges of
SNRs.  While an interesting possibility for a few sources, the large
number of associations where the neutron star is close to the center
of the SNR mean that this hypothesis cannot work for the majority of sources.

It is not yet known whether other populations of neutron star have
different velocity distributions from that seen for radio pulsars.
The location of several SGRs on or beyond the perimeters of SNRs
originally led to the suggestion that SGRs had very high space
velocities, $v_\perp \sim 1000-2000$~\kms, as might result from the
anisotropic neutrino emission associated with the formation of such
highly-magnetized objects \citep[e.g.,][]{dt92}. However, it has since
been argued that many of these SGR/SNR associations are spurious
\citep{lx00,gsgv01}, in which case these inferred velocities are not
valid. On the other hand, several AXPs have convincing associations
with young SNRs. In all these cases, the AXP lies close to the
geometric center of its SNR, implying projected space velocities for
this population $\la500$~\kms\ (\citealt*{ggv99}, \citealt{gsgv01}),
consistent with the velocities of radio pulsars. The emerging and
still enigmatic class of central compact objects (CCOs) are also
centrally located in young SNRs (see \citealt{psgz02} for a review),
and most likely also have relatively low space velocities.

What little is known about the velocities of older neutron stars that
are not radio pulsars roughly agrees with the situation for radio
pulsars: the velocities are high, $\gsim 100\mbox{ km
s}^{-1}$. Specifically, the velocities of the INSs
RX~J1856.5$-$3754 and RX~J0720.4$-$3125 are both $\approx 200\mbox{ km
s}^{-1}$ (\citealt*{kvka02}, \citealt*{mzh03}).

\section{Observations and Data Analysis}
\label{sec:obs}
In this section we give an overview of the analysis procedure that we
used for the different SNRs.  We start by describing the splitting of
the SNRs into observationally-based sub-samples, of which the
\chandra\ ACIS sub-sample is the major component discussed here
(\S~\ref{sec:Chandra-survey}).  We then describe the analysis of the
\chandra\ data that were used to identify potential compact objects
(\S~\ref{sec:proc}).  Finally, we describe the motivation for and
basic analysis of the optical and infrared followup observations that
were used to reject contaminating foreground and background X-ray
sources (\S~\ref{sec:cpt}).  Following this section we present the
actual detailed analyses of the four SNRs in this paper
(\S~\ref{sec:results}).

\subsection{\chandra\ ACIS Survey}
\label{sec:Chandra-survey}
\begin{figure}
\plotone{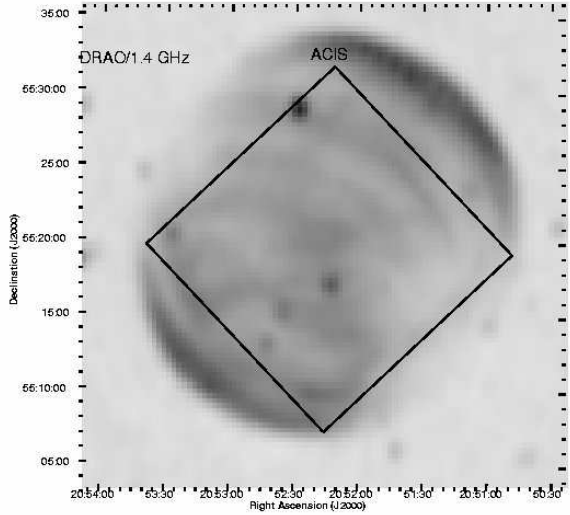}
\caption{DRAO 1.4~GHz radio image of \snr\ \citep{lrr+99}, showing the
  placement of the ACIS-I detector.}
\label{fig:radio}
\end{figure}

\begin{figure}
\plotone{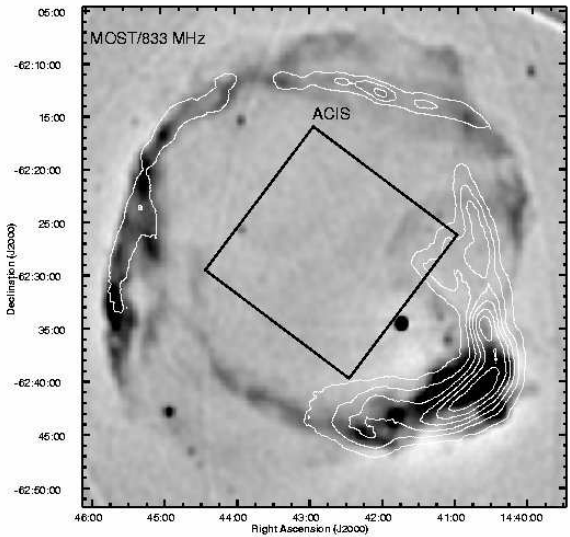}
\caption{MOST \citep{wg96} 833~MHz radio image of \snrb, showing the placement of
  the ACIS-I detector.  The bright region toward the south-west is
  the nebula RCW~86 \citep*{rcw60}.  The contours are from \rosat\ PSPC
  data (sequence RP500078).}
\label{fig:radiob}
\end{figure}

\begin{figure}
\plotone{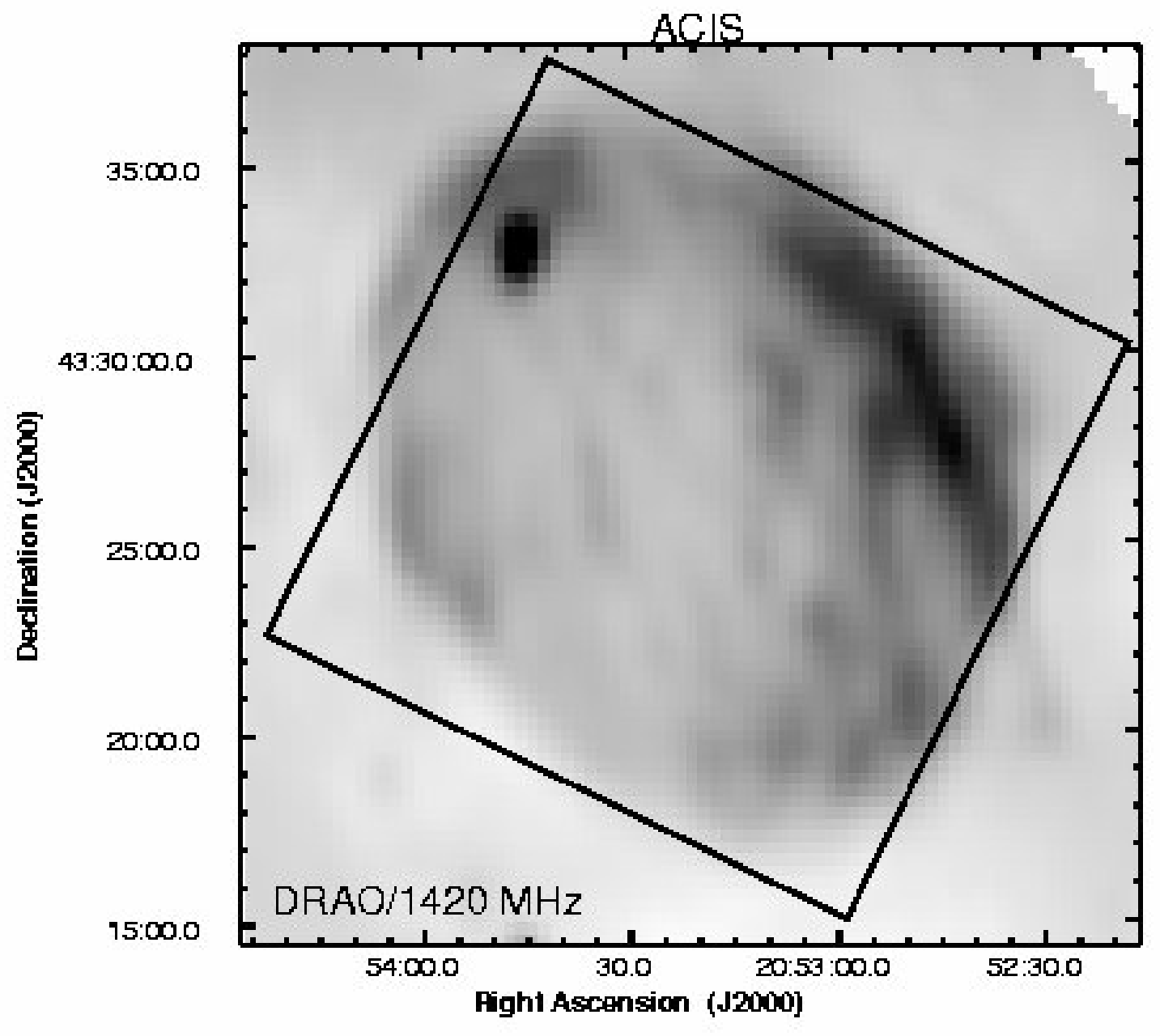}
\caption{DRAO 1.4~GHz radio image of \snrc\ \citep{tgp+03}, showing the
  placement of the ACIS-I detector.}
\label{fig:radioc}
\end{figure}

\begin{figure}
\plotone{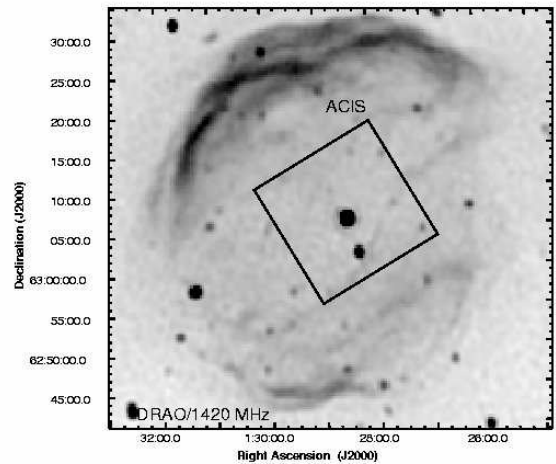}
\caption{DRAO 1.4~GHz radio image of \snrd, showing the
  placement of the ACIS-I detector.}
\label{fig:radiod}
\end{figure}

We defined three sub-samples among the 23 SNRs that had no central
sources from Table~\ref{tab:snrs}.  The sub-samples were defined
largely by size so that the X-ray observations have a good chance of
identifying the central compact source.  Our primary sub-sample was
designed for the \chandra\ ACIS-I detector, with its $16\arcmin$
field-of-view.  We parameterize as follows: a neutron star has a
typical transverse velocity of $100v_{100}$~km~s$^{-1}$, distance $d$
in kpc, and an age of $10^4t_4$~yr.  To ensure that a NS lies within
$8\arcmin$ of its SNRs geometric center (and so will fall on the
ACIS-I array; see Figs.~\ref{fig:radio}--\ref{fig:radiod} for
illustrations of the ACIS field-of-view), we require $t_{4} \leq
2.27d/v_{100}$. For a Sedov-phase remnant, we have $t_{4}=7\times
10^{-3} \theta d T_{7}^{-1/2}$, where $\theta$~arcmin is the SNR
diameter and $10^7\,T_{7}$~K is the shock temperature. Our limit is
thus $\theta \leq 324 T_{7}^{1/2}v_{100}^{-1}$.  We expect $T_{7}
\approx 1$ for a broad range of SNR ages; given the somewhat weak
dependence on $T$ we adopt $T_{7}=1$ in this calculation.  A
conservative constraint on pulsar velocities is $v_{100} \leq 7$.
This then yields the condition $\theta \lsim 46~{\rm arcmin}$.  There
are 14 SNRs that meet this criterion, of which 3 (Cas~A, IC~443, and
Kepler) have already been observed by \chandra.  This then leaves 11
SNRs for further \chandra\ observations, although SNR~G327.4+0.4 only
recently had a distance determination and was not included in the
original \chandra\ sample.  We therefore proposed for ACIS-I
observations of the 10 remaining SNRs (identified by \cxo-AO3 in
Tab.~\ref{tab:snrs}), and were awarded observations of 9 (the tenth,
G006.4$-$0.1, was awarded to J.~Rho in another AO-3 proposal).

We then constructed a sample of the larger SNRs for \xmm, and one of
the largest SNRs using the literature and the \rosat\ All-Sky Survey
Bright Source Catalog (BSC) \citep{rbs}, coupled with \chandra\
snap-shot images.  These samples are identified by \xmm-AO2 and
\cxo-AO4 in Table~\ref{tab:snrs} (our \xmm\ proposal was
for seven SNRs, of which we were only allocated time for the three
that are identified as ``\xmm-AO2').  We defer detailed discussion of
these samples to later papers.

To determine exposure times for these sources, we examined the types
of neutron stars found in the SNRs from Table~\ref{tab:snrs} (see also
\citealt{cph+01} and \S~\ref{sec:class}).  These neutron stars are listed in
Table~\ref{tab:psrs}.  We see two groups among them: those with
non-thermal spectra (characterized by a power-law with photon index
$\Gamma \sim 1.6$) and those with thermal spectra (characterized by a
blackbody with temperature $kT_{\infty}\approx 0.5$~keV).  Among the
shell-type SNRs in Table~\ref{tab:snrs}, the thermal sources
predominate, and these also produce the lowest X-ray count-rates.
Therefore we computed exposure times for a thermal source with
$kT_{\infty}=0.25$~keV (toward the low-end of those in
Tab.~\ref{tab:psrs}) and a bolometric luminosity $L_{\rm
bol}=10^{32}\mbox{ ergs s}^{-1}$, a factor of $\sim 10$ lower than
those of most of the thermal sources.  For the column densities \nh\
we used the best available estimates from the literature.  For these
shell SNRs, there is very little contribution from the SNR itself in
the interior, especially with the resolution of \chandra, so this was
not an issue, although we calculate the expected background
contribution, using surface-brightnesses compiled from the literature.
The exposure times are those that should detect a source with a
prototypical spectrum but a factor of 10 less luminous than normal
with a significance of at least $5\,\sigma$ above the background.

\subsection{X-ray Point-Source Detection}
\label{sec:proc}
\begin{figure}
\plotone{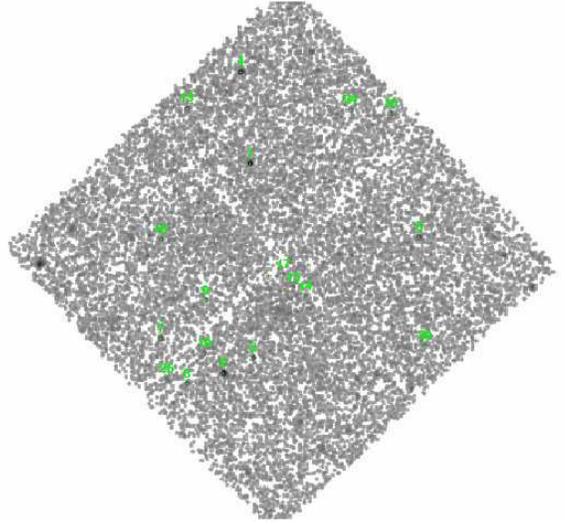}
\caption{Smoothed ACIS-I image (0.3--8.0~keV) of \snr.  North is up, and East
  is to the left.  The brightness is scaled  proportional to the 
  logarithm of the counts in $2\arcsec$ bins, smoothed with a Gaussian
filter.  The sources from Table~\ref{tab:srcs} are labeled.}
\label{fig:acis}
\end{figure}

\begin{figure}
\plotone{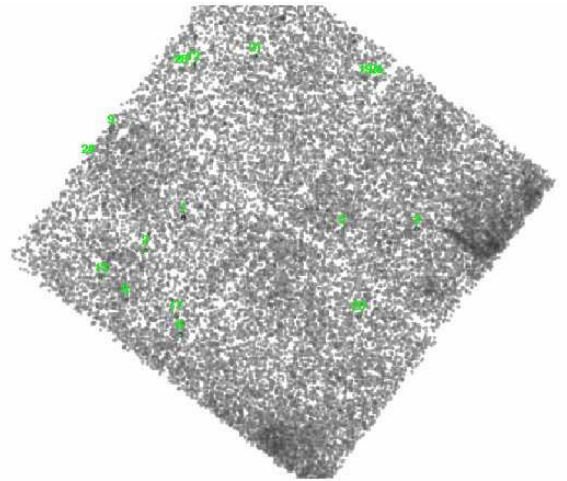}
\caption{Smoothed ACIS-I image (0.3--8.0~keV) of \snrb.  North is up, and East
  is to the left.  The brightness is scaled  proportional to the 
  logarithm of the counts in $2\arcsec$ bins, smoothed with a Gaussian
filter.  The sources from Table~\ref{tab:srcsb} are labeled.  The
  diffuse emission to the West and South-West is the RCW~86 complex.}
\label{fig:acisb}
\end{figure}

\begin{figure}
\plotone{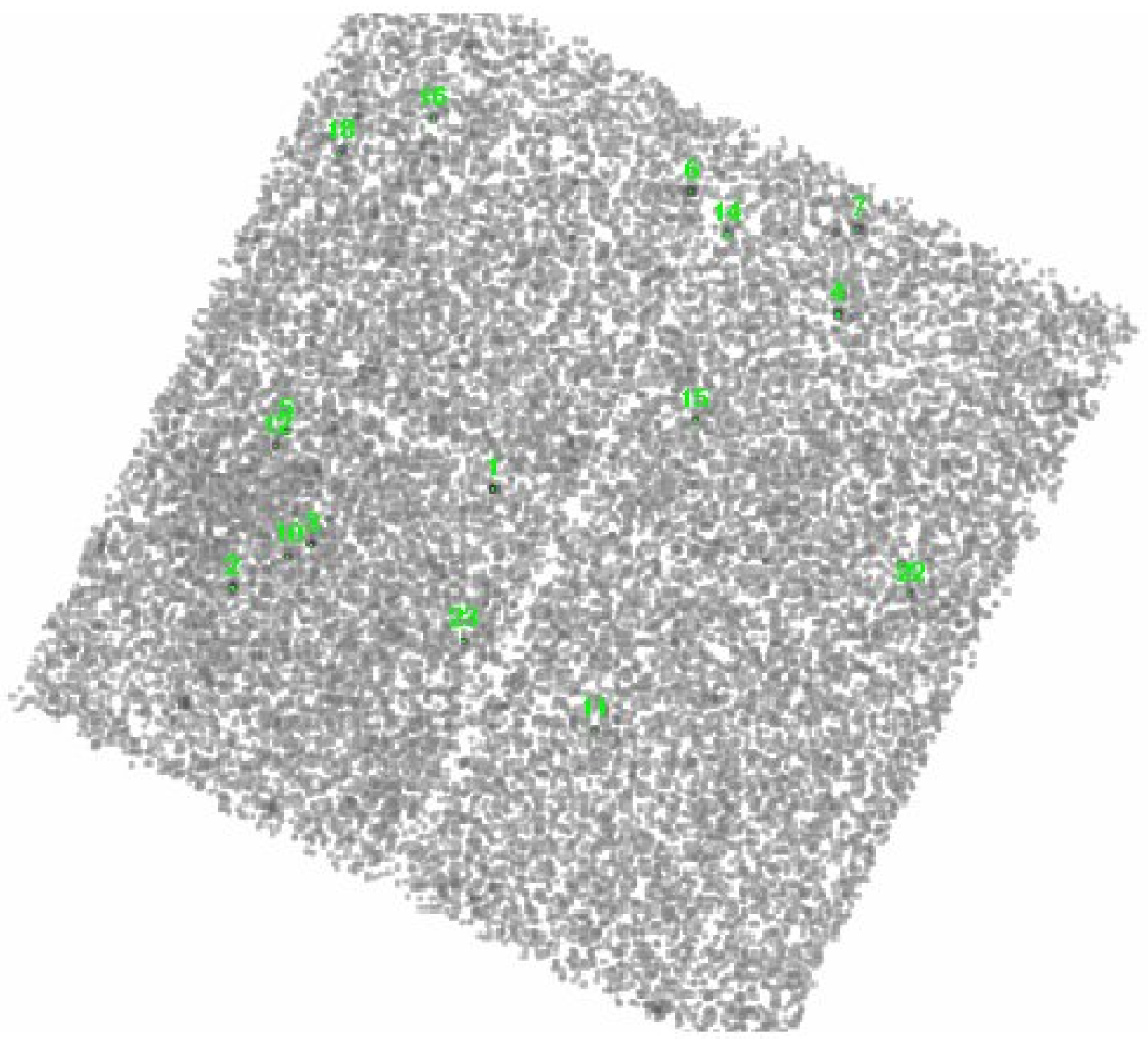}
\caption{Smoothed ACIS-I image (0.3--8.0~keV) of \snrc.  North is up, and East
  is to the left.  The brightness is scaled  proportional to the 
  logarithm of the counts in $2\arcsec$ bins, smoothed with a Gaussian
filter.  The sources from Table~\ref{tab:srcsc} are labeled.}
\label{fig:acisc}
\end{figure}

\begin{figure}
\plotone{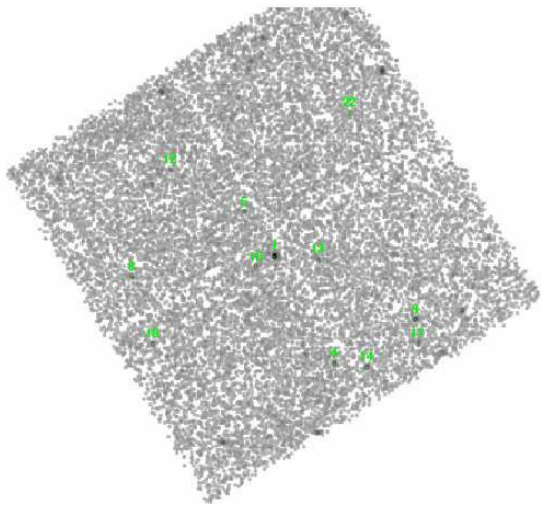}
\caption{Smoothed ACIS-I image (0.3--8.0~keV) of \snrd.  North is up, and East
  is to the left.  The brightness is scaled  proportional to the 
  logarithm of the counts in $2\arcsec$ bins, smoothed with a Gaussian
filter.  The sources from Table~\ref{tab:srcsd} are labeled.}
\label{fig:acisd}
\end{figure}

To analyze the data, we first reprocessed the raw (level 1) event data
to take advantage of updates in the \chandra\ calibration since the
data were first taken.  Specifically, the reprocessing included a
correction for charge transfer inefficiency (CTI;
\citealt{tbng02})\footnote{Following
\url{http://asc.harvard.edu/ciao/threads/acisapplycti/}.}, and we
removed the $\pm0.5$~pixel randomization added to the events.  We did
not include any correction for the degradation of the quantum
efficiency (QE) of the ACIS detectors\footnote{See
\url{http://cxc.harvard.edu/cal/Acis/Cal\_prods/qeDeg/}.}, as at no
point did we do a complete spectral analysis that would have used the
available correction techniques\footnote{See
\url{http://cxc.harvard.edu/ciao/threads/apply\_acisabs/}.}.  These
corrections are minor ($\approx 10$\%~year$^{-1}$) and are beyond our
level of accuracy.  We selected only events that have the ``standard''
\textit{ASCA} grades (0, 2, 3, 4, and 6).  After generating a new
level 2 event file, we corrected the data for \chandra\ aspect
errors\footnote{Following
\url{http://asc.harvard.edu/cal/ASPECT/fix\_offset/fix\_offset.cgi}.}:
for example, the change for \snr\ was $-0\farcs13$ in Right Ascension
and $0\farcs32$ in Declination.  Smoothed images of the data are shown
in Figure~\ref{fig:acis}--\ref{fig:acisd}.

Starting from this level 2 event file, we processed the data and
extracted sources much as was done in the \chandra\ Deep Field
\citep[CDF;][]{bah+01} and the 82-ksec observation of the Orion Nebula
\citep{fbg+02}.  First we separated the level 2 events by energy into
three bands:  0.3--2.0~keV (L), 2.0--8.0~keV (H), and 0.3--8.0~keV
(A).  These bands are slightly modified from those of the CDF and
Orion, as our potential sources are softer.  
We then separated the
data by CCD, using only the four ACIS-I CCDs.  This then gave twelve
separate event files ($3\mbox{ bands} \times 4\mbox{ CCDs}$).  

For each of these event files, we generated instrument and exposure
maps, following standard CIAO
threads\footnote{\url{http://asc.harvard.edu/ciao/threads/expmap\_acis\_single/.}}.
Creating an exposure map requires a source model: for this model, we
assumed a blackbody with a temperature of $kT_\infty=0.25$~keV
(similar to that of the Cas~A central point source) and a column
density appropriate for each SNR (\S\S~\ref{sec:snra}--\ref{sec:snrd}).

Then, for each of the event files (now with exposure maps), we ran
\texttt{wavdetect} \citep{fkrl02}, using wavelet scales of $1,
\sqrt{2}, 2, \sqrt{8},\ldots,16$~pixels and a nominal energy of
1.5~keV.  The significance threshold was $10^{-6}$.

We then merged the \texttt{wavdetect} source files for further
analysis.  First we merged by CCD, creating a source list for each
band.  Then, to merge the data from different bands (which are
spatially coincident), we considered sources identical if the
positions (as determined by \texttt{wavdetect}) differed by less than
$2\farcs5$, for sources at off-axis angles of $<6\arcmin$, or differed
by less than $4\arcsec$, for sources at off-axis angles of $\geq
6\arcmin$.  While these are large tolerances given the typical
astrometric uncertainty of \texttt{wavdetect} ($0\farcs6$ for on-axis
sources, approaching $1\farcs5$ for off-axis sources), the source
density is so low ($\sim 0.2\mbox{ arcmin}^{-2}$ for this observation)
that the number of false matches is negligible.  Even so, only 5
sources had match-radii of $> 1\farcs5$, and most were $<0\farcs5$.
After manually examining the list of sources, we then removed those
that appeared spurious.  We then limited the source list to those
sources with $\geq 10$~counts.  This limit allows us to have enough
photons that the position is well determined (critical when trying to
identify counterparts) and that an estimate can be made of the
hardness ratio.  Given that our exposure times were calculated such
that a low-luminosity neutron star would be detected with
$>25$~counts, the 10-count limit is conservative.  The final merged
source list contains 12--18 sources, depending on the SNR.  The
sources are all consistent with being point-sources.

\begin{figure}
% produced by hardness.m
\plotone{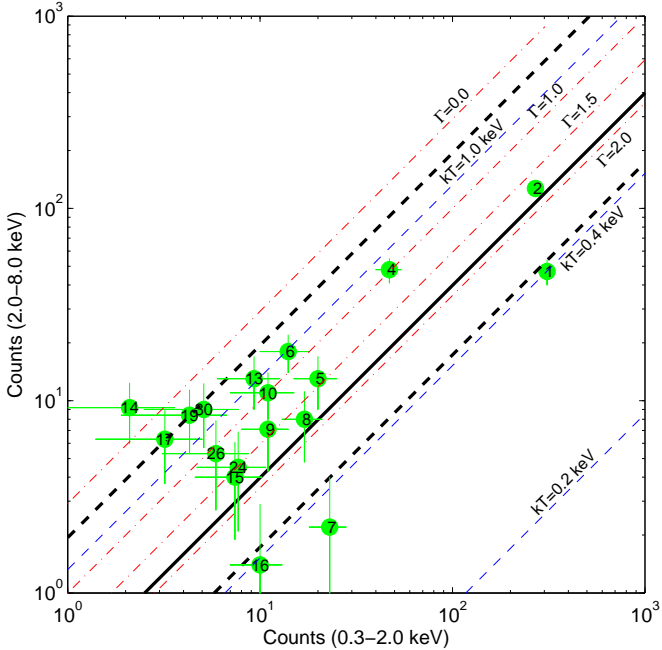}
\caption{L-band count vs.\ H-band counts for \snr.  The sources from
  Table~\ref{tab:srcs} are shown in green and are numbered.  Model
  spectra (computed using \texttt{PIMMS}) with
  $\nh=\expnt{2}{21}\mbox{ cm}^{-2}$ are plotted as diagonal lines:
  power-law models as red dash-dotted lines ($\Gamma$ is as indicated)
  and blackbody models as blue dashed lines ($kT_\infty$ is as
  indicated).  The solid thick black line represents the median
  spectrum from the CDF/Orion studies, with the dashed thick black
  lines showing the 25- and 75-percentile spectra.  }
\label{fig:hard}
\end{figure}

\begin{figure}
\plotone{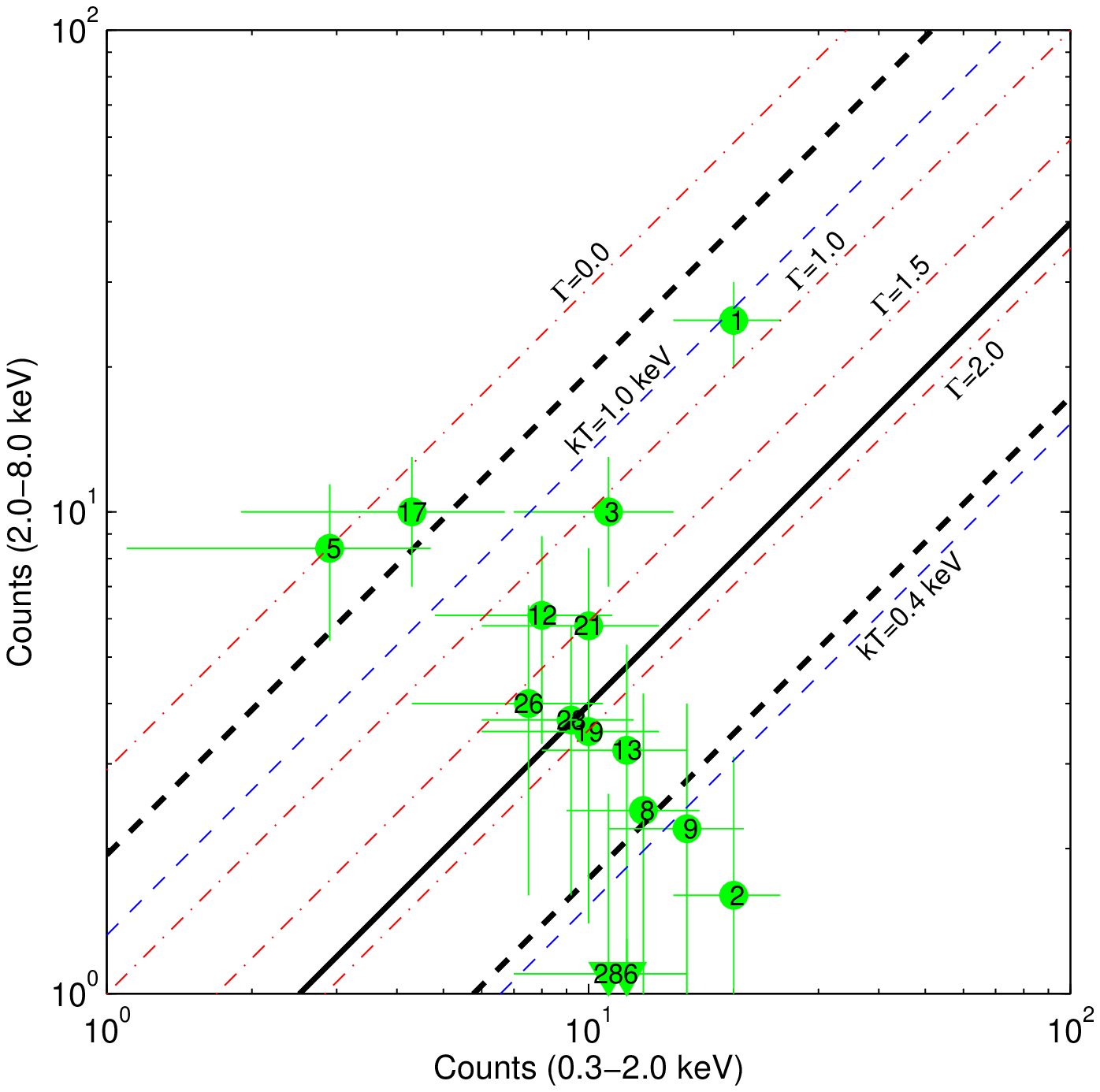}
\caption{L-band count vs.\ H-band counts for \snrb.  The sources from
  Table~\ref{tab:srcsb} are shown in green and are numbered.  Model
  spectra (computed using \texttt{PIMMS}) with
  $\nh=\expnt{2}{21}\mbox{ cm}^{-2}$ are plotted as diagonal lines:
  power-law models as red dash-dotted lines ($\Gamma$ is as indicated)
  and blackbody models as blue dashed lines ($kT_\infty$ is as
  indicated).  The solid thick black line represents the median
  spectrum from the CDF/Orion studies, with the dashed thick black
  lines showing the 25- and 75-percentile spectra. }
\label{fig:hardb}
\end{figure}

\begin{figure}
% produced by hardness.m
\plotone{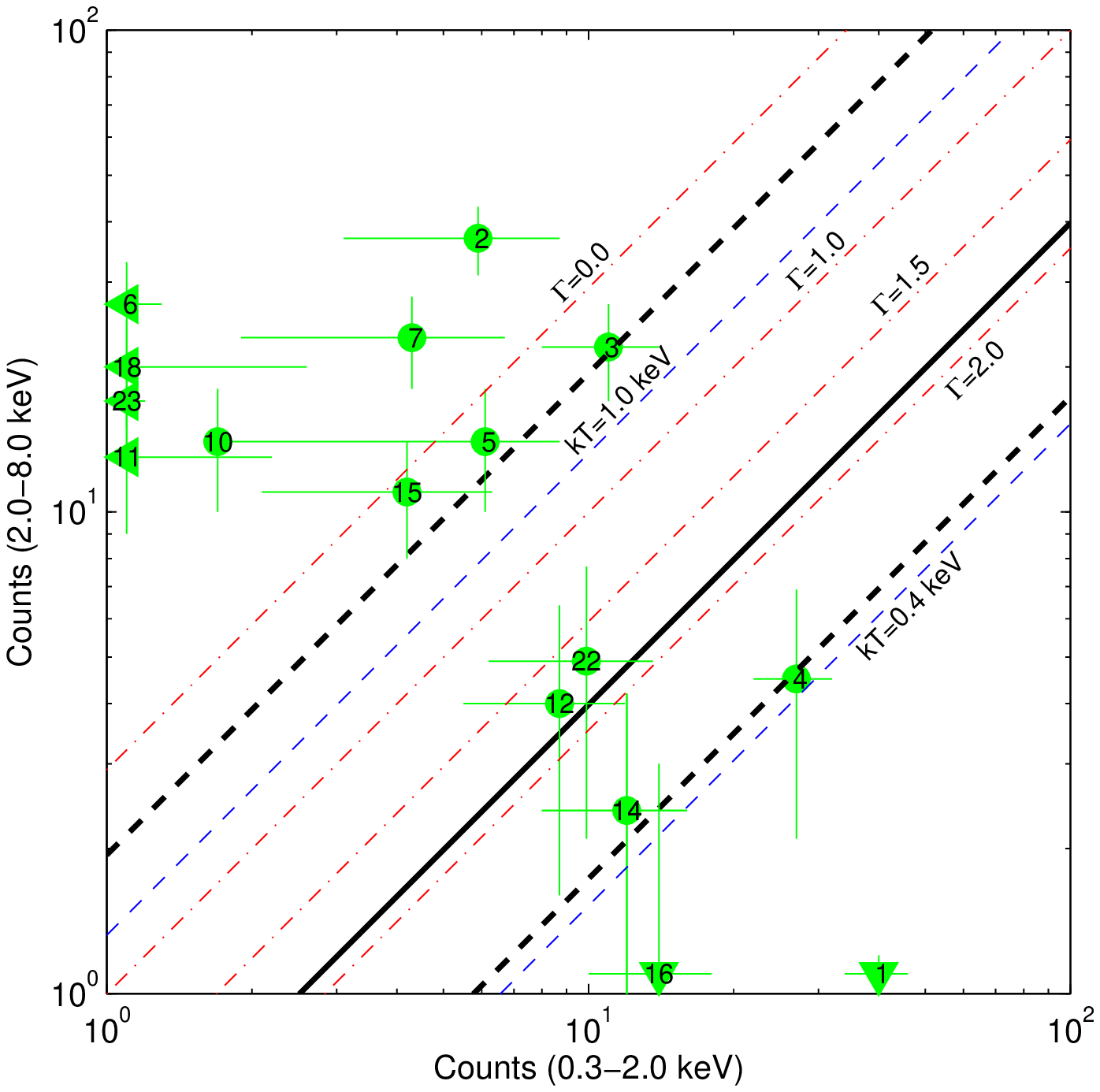}
\caption{L-band count vs.\ H-band counts for \snrc.  The sources from
  Table~\ref{tab:srcsc} are shown in green and are numbered.  Model spectra (computed
  using \texttt{PIMMS}) with $\nh=\expnt{2}{21}\mbox{ cm}^{-2}$ are
  plotted as  diagonal lines: power-law models as red dash-dotted lines
  ($\Gamma$ is as indicated) and blackbody models as blue dashed
  lines ($kT_\infty$ is as indicated).  The solid thick black line represents the
  median spectrum from the
CDF/Orion studies, with the dashed thick black lines showing the 25-
  and 75-percentile spectra.   }
\label{fig:hardc}
\end{figure}

\begin{figure}
% produced by hardness.m
\plotone{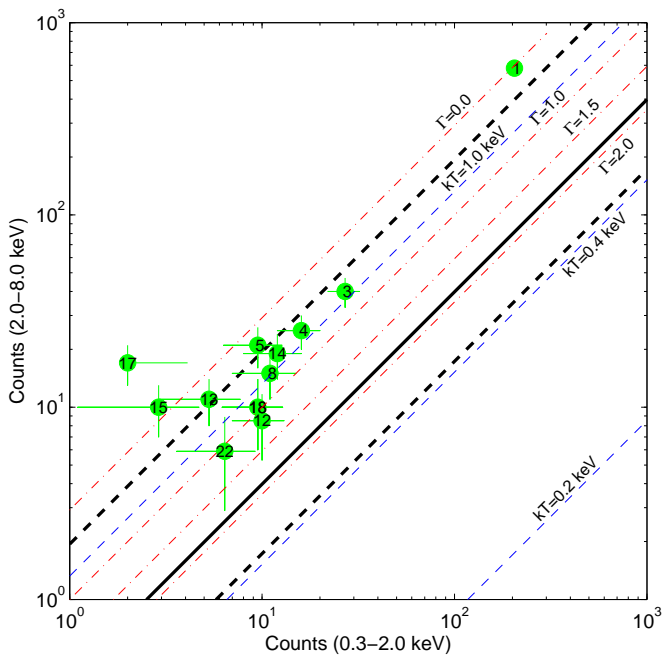}
\caption{L-band count vs.\ H-band counts for \snrd.  The sources from
  Table~\ref{tab:srcsd} are shown in green and are numbered.  Model spectra (computed
  using \texttt{PIMMS}) with $\nh=\expnt{2}{21}\mbox{ cm}^{-2}$ are
  plotted as  diagonal lines: power-law models as red dash-dotted lines
  ($\Gamma$ is as indicated) and blackbody models as blue dashed
  lines ($kT_\infty$ is as indicated).  The solid thick black line represents the
  median spectrum from the
CDF/Orion studies, with the dashed thick black lines showing the 25-
  and 75-percentile spectra.  }
\label{fig:hardd}
\end{figure}

For each source, we then performed additional aperture photometry.
This allowed us to (1) use knowledge of the psf size in determination
of the source counts and (2) obtain source counts for sources that
were detected in only one or two bands.  The radii of the photometric
apertures were determined from the analytic fit to the 95\% encircled
energy radius given in \citet{fbg+02}.  We measured the number of
counts in these apertures for each band and subtracted the number of
counts in a background annulus extending from 2--3 times the 95\%
encircled energy radius to determine the net number of counts.  To aid
in comparison with the CDF/Orion data sets, we also extracted photons
in the more standard 0.5--2.0~keV band.  The final source data are
presented in Tables~\ref{tab:srcs}--\ref{tab:srcsd}.  We plot the
L-band counts vs.\ the H-band counts for the detected sources in
Figures~\ref{fig:hard}--\ref{fig:hardd}.

%\begin{landscape}
\begin{deluxetable}{r r r c c r c c c c c c}
%\rotate
\tablecaption{X-ray Sources in \snr\label{tab:srcs}}
\tablewidth{0pt}
\tabletypesize{\scriptsize}
\tablehead{ \colhead{ID\tablenotemark{a}} & \colhead{CXOU~J\tablenotemark{b}} & \colhead{$\alpha$} & \colhead{$\delta$} & \colhead{$r_{90}$\tablenotemark{c}} & \colhead{$\Delta R$\tablenotemark{d}} & \colhead{Counts$_{\rm L}$} & \colhead{Counts$_{\rm H}$} & \colhead{Counts$_{\rm A}$} & \colhead{Counts} & \colhead{HR$_{\rm L,H}$\tablenotemark{e}} \\
&  & \mc{2}{c}{(J2000)} & \colhead{($\arcsec$)} & \colhead{($\arcmin$)} &  & & & \colhead{(${0.5-2.0\mbox{ keV}}$)} \\
}
\startdata
1 & 205222.8+552343 & 20:52:22.89 & +55:23:43.7 & 0.5 & 3.5 &  310(20) &47(7) &360(20) &290(20) &$-$0.73(0.04) \\
2 & 205230.9+551437 &  20:52:30.98 & +55:14:37.5 & 0.6 & 6.4 &  270(20) &127(12) &390(20) &260(20) &$-$0.35(0.05) \\
4 & 205225.8+552741 &  20:52:25.82 & +55:27:41.6 & 1.3 & 7.4 &  47(7) &48(7) &93.9(10.2) &44(7) &\phs0.01(0.11) \\
5 & 205222.0+551516 &  20:52:22.01 & +55:15:16.9 & 0.8 & 5.3 &  20(5) &13(4) &33(6) &19(4) &$-$0.20(0.18) \\
6 & 205231.3+552031 &  20:51:31.37 & +55:20:31.6 & 1.4 & 6.1 &  14(4) &18(4) &32(6) &13(4) &\phs0.13(0.19) \\[0.1in]
7 & 205250.2+551606 &  20:52:50.20 & +55:16:06.6 & 1.5 & 6.8 &  23(5) &2.2(1.8) &25(5) &22(5) &$-$0.83(0.14) \\
8 & 205242.1+551409 &  20:52:42.19 & +55:14:09.8 & 1.6 & 7.5 &  17(4) &8.0(3.2) &25(5) &16(4) &$-$0.36(0.21) \\
9 & 205226.4+551746 &  20:52:36.48 & +55:17:46.0 & 0.7 & 4.2 &  11(3) &7.1(2.8) &18(4) &11(3) &$-$0.19(0.24) \\
10 & 205139.8+552553 &  20:51:39.85 & +55:25:53.4 & 1.7 & 7.3 &  11(4) &11(3) &22(5) &9.3(3.5) &$-$0.02(0.23) \\
13 & 205242.2+552607 &  20:52:42.26 & +55:26:07.1 & 1.8 & 6.9 &  9.3(3.3) &13(4) &22(5) &9.5(3.3) &\phs0.16(0.23) \\[0.1in]
14 & 205205.9+551758 &  20:52:05.96 & +55:17:58.5 & 0.7 & 2.8 &  2.1(1.5) &9.2(3.2) &11(3) &2.1(1.5) &\phs0.63(0.24) \\
15 & 205209.8+551821 &  20:52:09.87 & +55:18:21.2 & 0.8 & 2.2 &  7.4(2.8) &4.0(2.1) &11(3) &7.4(2.8) &$-$0.30(0.30) \\
16 & 205250.0+552025 &  20:52:50.09 & +55:20:25.6 & 1.1 & 5.1 &  10(3) &1.4(1.5) &12(4) &10(3) &$-$0.77(0.24) \\
17 & 205212.6+551854 & 20:52:12.65 & +55:18:54.4 & 0.6 & 1.6 & 3.2(1.8) &6.3(2.6) &10(3) &3.2(1.8) &\phs0.33(0.31) \\
19 & 205236.8+551528 &  20:52:36.89 & +55:15:28.8 & 1.4 & 6.0 &  4.3(2.4) &8.4(3.0) &12(4) &3.2(2.1) &\phs0.33(0.29) \\[0.1in]
24 & 205152.2+552602 &  20:51:52.23 & +55:26:02.7 & 1.8 & 6.4 &  7.7(3.0) &4.5(2.4) &12(4) &7.1(2.8) &$-$0.26(0.30) \\
26 & 205248.3+551422 &  20:52:48.32 & +55:14:22.3 & 3.1 & 7.8 &  5.9(2.8) &5.3(2.6) &12(4) &5.9(2.8) &$-$0.05(0.34) \\
30 & 205129.7+551548 &  20:51:29.72 & +55:15:48.5 & 3.0 & 7.9 &  5.1(2.6) &9.0(3.3) &14(4) &4.0(2.4) &\phs0.28(0.29) \\
\enddata
\tablenotetext{a}{Internal identifier of the form \snr:$N$.}
\tablenotetext{b}{Official IAU name.}
\tablenotetext{c}{Approximate 90\% confidence radius.}
\tablenotetext{d}{Angle from the center of the SNR.}
\tablenotetext{e}{Hardness ratio, computed according to $(C_{\rm
    H}-C_{\rm L})/(C_{\rm H}+C_{\rm L})$, where $C$ is the
  number of counts in a band.}
\tablecomments{Quantities in parentheses are 1-$\sigma$
  uncertainties.  Here and in Tables~\ref{tab:srcsb}--\ref{tab:srcsd},
  ${\rm Counts}_{A}={\rm Counts}_L + {\rm Counts}_H$, but the three
  columns have been rounded separately.
}
\end{deluxetable}
%\end{landscape}

%\begin{landscape}
\begin{deluxetable}{r r r c c r c c c c c c}
%\rotate
\tablecaption{X-ray Sources in \snrb\label{tab:srcsb}}
\tablewidth{0pt}
\tabletypesize{\scriptsize}
\tablehead{  \colhead{ID\tablenotemark{a}} & \colhead{CXOU~J\tablenotemark{b}} & \colhead{$\alpha$} & \colhead{$\delta$} & \colhead{$r_{90}$\tablenotemark{c}} & \colhead{$\Delta R$\tablenotemark{d}} & \colhead{Counts$_{\rm L}$} & \colhead{Counts$_{\rm H}$} & \colhead{Counts$_{\rm A}$} & \colhead{Counts} & \colhead{HR$_{\rm L,H}$\tablenotemark{e}} \\
&  & \mc{2}{c}{(J2000)} & \colhead{($\arcsec$)} & \colhead{($\arcmin$)} &  & & & \colhead{(${0.5-2.0\mbox{ keV}}$)} \\
}
\startdata
1 & 144319.3$-$622804 & 14:43:19.31 & $-$62:28:04.2 & 0.7 & 3.4 &  20(5) &25(5) &45(7)
&19(4) &\phs0.11(0.15)  \\
2 & 144333.7$-$622928 &  14:43:33.74 & $-$62:29:28.2 & 1.1 & 5.2 &  20(5) &1.6(1.5) &22(5) &20(5) &$-$0.85(0.13) \\
3 & 144151.5$-$622833 &  14:41:51.59 & $-$62:28:33.5 & 2.0 & 6.8 &
11(4) &10(3) &22(5) &12(4) &$-$0.05(0.23) \\
5 & 144219.5$-$622834 &  14:42:19.54 & $-$62:28:34.5 & 1.1 & 3.5 &  2.9(1.8) &8.4(3) &11(3) &2.9(1.8) &\phs0.49(0.28) \\
6 & 144320.6$-$623308 &  14:43:20.63 & $-$62:33:08.6 & 2.0 & 6.0 &  12(4) &$-$0.5(0.2) &11(4) &12(4) &$-$1.09(0.04) \\[0.1in]
8 & 144341.0$-$623138 &  14:43:41.05 & $-$62:31:38.3 & 1.9 & 6.8 &  13(4) &2.4(1.8) &16(4) &14(4) &$-$0.69(0.21) \\
9 & 144346.3$-$622413 &  14:43:46.37 & $-$62:24:13.1 & 3.4 & 7.7 &  16(5) &2.2(1.8) &16(5) &14(4) &$-$0.77(0.19) \\
12 & 144315.3$-$622128 &  14:43:15.36 & $-$62:21:28.8 & 1.6 & 7.5 &  8.0(3.2) &6.1(2.8) &14(4) &7.2(3) &$-$0.13(0.30) \\
13 & 144350.4$-$623040 &  14:43:50.45 & $-$62:30:40.0 & 2.2 & 7.4 &  12(4) &3.2(2.1) &15(4) &12(4) &$-$0.57(0.25) \\
17 & 144322.1$-$623219 &  14:43:22.12 & $-$62:32:19.3 & 1.7 & 5.4 &  4.3(2.4) &10(3) &15(4) &4.3(2.4) &\phs0.41(0.27) \\[0.1in]
19 & 144210.9$-$622202 &  14:42:10.94 & $-$62:22:02.8 & 2.7 & 7.7 &  10(4) &3.5(2.1) &14(4) &8.8(3.3) &$-$0.50(0.26) \\
21 & 144252.1$-$622107 &  14:42:52.16 & $-$62:21:07.1 & 2.4 & 7.2 &  10(4) &5.8(2.6) &17(4) &9.8(3.5) &$-$0.28(0.26) \\
23 & 144213.1$-$623220 &  14:42:13.16 & $-$62:32:20.4 & 1.5 & 5.8 &  9.2(3.2) &3.7(2.1) &12(4) &9.0(3.2) &$-$0.43(0.27) \\
26 & 144207.1$-$622204 &  14:42:07.16 & $-$62:22:04.3 & 3.8 & 8.0 &  7.5(3.2) &4.0(2.4) &12(4) &8.0(3.2) &$-$0.30(0.33) \\
28 & 144320.0$-$622138 &  14:43:20.06 & $-$62:21:38.7 & 1.9 & 7.5 &  11(4) &0.6(1.5) &12(4) &11(4) &$-$0.90(0.24)  \\
\enddata
\tablenotetext{a}{Internal identifier of the form \snrb:$N$.}
\tablenotetext{b}{Official IAU name.}
\tablenotetext{c}{Approximate 90\% confidence radius.}
\tablenotetext{d}{Angle from the center of the SNR.}
\tablenotetext{e}{Hardness ratio, computed according to $(C_{\rm
    H}-C_{\rm L})/(C_{\rm H}+C_{\rm L})$, where $C$ is the
  number of counts in a band.}
\tablecomments{Quantities in parentheses are 1-$\sigma$ uncertainties.}
\end{deluxetable}
%\end{landscape}

%\begin{landscape}
\begin{deluxetable}{r r r c c r c c c c c c}
\tablecaption{X-ray Sources in \snrc\label{tab:srcsc}}
\tablewidth{0pt}
\tabletypesize{\scriptsize}
\tablehead{  \colhead{ID\tablenotemark{a}} & \colhead{CXOU~J\tablenotemark{b}} & \colhead{$\alpha$} & \colhead{$\delta$} & \colhead{$r_{90}$\tablenotemark{c}} & \colhead{$\Delta R$\tablenotemark{d}} & \colhead{Counts$_{\rm L}$} & \colhead{Counts$_{\rm H}$} & \colhead{Counts$_{\rm A}$} & \colhead{Counts} & \colhead{HR$_{\rm L,H}$\tablenotemark{e}} \\
&  &  \mc{2}{c}{(J2000)}& \colhead{($\arcsec$)} & \colhead{($\arcmin$)} &  & & & \colhead{(${0.5-2.0\mbox{ keV}}$)} \\
}
\startdata
1 & 205328.9+432659 & 20:53:28.96 & +43:26:59.1 & 0.5 & 1.3 &  40(6) &$-$0.3(0.1)&40(6) &38(6) &$-$1.01(0.01)\\
2 & 205357.9+432459 & 20:53:57.92 & +43:24:59.3 & 1.3 & 6.9 &  5.9(2.8) &37(6) &44(7) &7.2(3) &\phs0.73(0.12) \\
3 & 205349.1+432551 & 20:53:49.18 & +43:25:51.2 & 0.9 & 5.1 &  11(3) &22(5) &33(6)&11(3) &\phs0.33(0.17) \\
4 & 205250.5+433030 & 20:52:50.55 & +43:30:30.0 & 1.2 & 6.5 &  27(5) &4.5(2.4) &31(6) &27(5) &$-$0.71(0.14)\\
5 & 205352.0+422809 & 20:53:52.03 & +43:28:09.7 & 1.2 & 5.5 &  6.1(2.6) &14(4) &20(5) &5.8(2.6) &\phs0.40(0.21)\\[0.1in]
6 & 205306.9+423259 & 20:53:06.89 & +43:32:59.7 & 1.2 & 6.3 &  $-$0.5(0.2) &27(6)&26(6) &$-$0.5(0.2) &\phs1.04(0.02) \\
7 & 205248.2+433214 & 20:52:48.25 & +43:32:14.0 & 2.0 & 7.9 &  4.3(2.4) &23(5) &26(6) &4.0(2.4) &\phs0.69(0.16) \\
10 &205351.7+432537 &  20:53:51.76 & +43:25:37.9 & 1.4 & 5.7 &  1.7(1.8) &14(4) &16(4) &1.7(1.8) &\phs0.79(0.21) \\
11 &205317.6+432206&  20:53:17.66 & +43:22:06.8 & 1.2 & 5.2 &  0.3(1.1) &13(4) &13(4)&0.3(1.1) &\phs0.95(0.16)\\
12 &205353.1+432751&  20:53:53.13 & +43:27:51.5 & 1.1 & 5.7 &  8.7(3.2) &4.0(2.4)&13(4) &8.7(3.2) &$-$0.37(0.30)  \\[0.1in]
14 &205302.8+433207&  20:53:02.87 & +43:32:07.6 & 1.5 & 5.9 &  12(4) &2.4(1.8) &13(4) &12(4) &$-$0.67(0.23) \\
15 &205306.5+432822&  20:53:06.50 & +43:28:22.2 & 0.6 & 3.0 &  4.2(2.1) &11(3) &16(4) &4.2(2.1) &\phs0.46(0.23) \\
16 &205335.6+433427&  20:53:35.66 & +43:34:27.6 & 1.9 & 7.6 &  14(4) &0.4(1.9) &12(4) &9.3(3.5) &$-$0.94(0.26) \\
18 &205346.0+433345&  20:53:46.04 & +43:33:45.7 & 2.4 & 7.8 &  $-$0.4(1.5) &20(5) &20(5) &$-$0.1(1.5) &\phs1.04(0.16) \\
22 &205242.6+432451&  20:52:42.61 & +43:24:51.0 & 1.2 & 7.6 &  9.9(3.7) &4.9(2.8)&15(5) &8.3(3.3) &$-$0.34(0.30)  \\[0.1in]
23 &205332.2+432355&  20:53:32.27 & +43:23:55.7 & 0.7 & 3.9 &  $-$0.3(0.1) &17(4)&17(4) &$-$0.3(0.1) &\phs1.03(0.02) \\
\enddata
\tablenotetext{a}{Internal identifier of the form \snrc:$N$.}
\tablenotetext{b}{Official IAU name.}
\tablenotetext{c}{Approximate 90\% confidence radius.}
\tablenotetext{d}{Angle from the center of the SNR.}
\tablenotetext{e}{Hardness ratio, computed according to $(C_{\rm
    H}-C_{\rm L})/(C_{\rm H}+C_{\rm L})$, where $C$ is the
  number of counts in a band.}
\tablecomments{Quantities in parentheses are 1-$\sigma$ uncertainties.}
\end{deluxetable}
%\end{landscape}

%\begin{landscape}
\begin{deluxetable}{r r r c c r c c c c c c}
\tablecaption{X-ray Sources in \snrd\label{tab:srcsd}}
\tablewidth{0pt}
\tabletypesize{\scriptsize}
\tablehead{ \colhead{ID\tablenotemark{a}}  &
  \colhead{CXOU~J\tablenotemark{b}} &
  \colhead{$\alpha$} & \colhead{$\delta$} &
  \colhead{$r_{90}$\tablenotemark{c}} & \colhead{$\Delta R$\tablenotemark{d}} &
  \colhead{Counts$_{\rm L}$} & \colhead{Counts$_{\rm H}$} &
  \colhead{Counts$_{\rm A}$} & \colhead{Counts$_{0.5-2.0\mbox{
	keV}}$} & \colhead{HR$_{\rm L,H}$\tablenotemark{e}}  \\ 
&  & \mc{2}{c}{(J2000)} & \colhead{($\arcsec$)} & \colhead{($\arcmin$)} &  & & &  \\
}
\startdata
1 & 012830.6+630629  &01:28:30.64 & +63:06:29.9 & 0.4 & 0.2 &  205(15) &580(20)&780(30) &182(14) &\phs0.48(0.03) \\
3 & 012736.6+630345  &01:27:36.60 & +63:03:45.5 & 1.3 & 6.9 &  27(5) &40(7) &68(9) &25(5) &\phs0.19(0.13)\\
4 & 012807.7+630150  &01:28:07.72 & +63:01:50.5 & 1.3 & 5.5 &  16(4) &25(5) &41(7) &17(4) &\phs0.21(0.16) \\
5 & 012842.3+630825  &01:28:42.31 & +63:08:25.6 & 0.6 & 2.2 &  9.5(3.2) &21(5) &32(6)&8.4(3) &\phs0.38(0.17)\\
8 & 012925.4+630535  &01:29:25.43 & +63:05:35.3 & 1.7 & 6.1 &  11(4) &15(4) &26(5) &9.3(3.3) &\phs0.17(0.21) \\[0.1in]
12 &012910.8+631014  & 01:29:10.86 & +63:10:14.8 & 1.6 & 5.7 &  10(3) &8.5(3.2) &20(5) &11(3) &$-$0.10(0.25) \\
13 &012813.9+630621  & 01:28:13.91 & +63:06:21.1 & 0.9 & 2.1 &  5.3(2.4) &11(3) &17(4)&6.3(2.6) &\phs0.33(0.24) \\
14 &012755.2+630141  & 01:27:55.23 & +63:01:41.6 & 1.5 & 6.4 &  12(4) &19(5) &33(6) &7.9(3) &\phs0.23(0.18) \\
15 &012837.8+630602  & 01:28:37.83 & +63:06:02.8 & 0.8 & 0.8 &  2.9(1.8) &10(3) &13(4) &2.9(1.8) &\phs0.56(0.24) \\
17 &012735.7+630241  & 01:27:35.75 & +63:02:41.6 & 1.7 & 7.4 &  2.0(2.1) &17(4) &19(5) &0.1(1.5) &\phs0.79(0.21) \\[0.1in]
18 &012917.4+630242  & 01:29:17.42 & +63:02:42.3 & 2.1 & 6.4 &  9.5(3.3) &10(4) &20(5) &7.7(3) &\phs0.03(0.25)\\
22 &012801.8+631244  & 01:28:01.83 & +63:12:44.5 & 2.0 & 7.1 &  6.4(2.8) &5.9(3) &12(4) &5.3(2.6) &$-$0.04(0.33) \\
%57 &  01:28:45.31 & +63:03:50.0 & 0.4 & 3.1 &  0.00(1e$-$05) &10(3) &3.2(1.8) &0.00(1e$-$05) &\phs1.00(0.00) \\
\enddata
\tablenotetext{a}{Internal identifier of the form \snrd:$N$.}
\tablenotetext{b}{Official IAU name.}
\tablenotetext{c}{Approximate 90\% confidence radius.}
\tablenotetext{d}{Angle from the center of the SNR.}
\tablenotetext{e}{Hardness ratio, computed according to $(C_{\rm
    H}-C_{\rm L})/(C_{\rm H}+C_{\rm L})$, where $C$ is the
  number of counts in a band.}
\tablecomments{Quantities in parentheses are 1-$\sigma$ uncertainties.}
\end{deluxetable}
%\end{landscape}

\subsubsection{Nomenclature}
Herein, for convenience, we label the \chandra-detected X-ray sources
by their field identification --- for example, \snr:5 refers to the
5th source in the \snr\ field.  This is not meant to imply that all of
these sources are associated with the SNRs --- the vast majority are
not.  These are meant to be internal designations only and do not
replace the official IAU names of the form CXOU~JHHMMSS.s$\pm$DDMMSS,
where HHMMSS.s represents the Right Ascension and $\pm$DDMMSS
represents the Declination.  When detailing the source identifications
(Tabs.~\ref{tab:srcs}--\ref{tab:srcsd}) we give both the internal and
official names, but in the rest of this paper we use only the internal
names.  When initially identifying X-ray sources we numbered then
consecutively, but the numbers in
Tables~\ref{tab:srcs}--\ref{tab:srcsd} are no longer consecutive as we
have removed sources at radii $>8\arcmin$ and with fewer than
10~counts (\S~\ref{sec:proc}).

\subsection{X-ray Extended-Source Detection}
\label{sec:ext}
% limits A*theta^2+B*theta
% A =pi*counts, B=sqrt(26*counts)
No extended sources were detected during the \texttt{wavdetect} runs,
with a maximum wavelet scale of 16 pixels or $8\arcsec$.  We also
manually examined the X-ray images for sources with larger sizes and
did not find any (there is some extended emission in \snrb\ toward
the south and west, but that is almost certainly due to diffuse thermal
emission from the SNR/RCW~86 complex as it has the same general
spectrum; see Fig.~\ref{fig:radiob}).

To quantify the limits on extended emission, we determine the
average background counts in a region free from sources.  These counts
are presented in Table~\ref{tab:pwn}, where we also find the
$3\,\sigma$ limits to extended emission.

\begin{deluxetable}{l c c c c}
\tablecaption{Limits to Extended X-ray Emission in SNRs \Ga, \Gb, \Gc,
  \& \Gd\label{tab:pwn}}
\tablewidth{0pt}
\tablehead{
\colhead{SNR} & \colhead{BG Counts} & \colhead{Count Limits} &
\colhead{$L$ Limits} & \colhead{$L(\theta=1\arcmin)$ Limits} \\
 & \colhead{$(\times 10^{-2}\mbox{ arcsec}^{-2})$} & \colhead{$(\times
  10^{-2})$} &\colhead{$(\times \theta^210^{28}\mbox{ ergs s}^{-1})$} & \colhead{$(\times 10^{32}\mbox{ ergs s}^{-1})$} \\
}
\startdata
\Ga & 1.75(1) & $6\theta^2+70\theta$ & $\phn8$ &$3$ \\
\Gb & 2.10(2) & $7\theta^2+70\theta$ & $10$ &$4$ \\
\Gc & 2.34(2) & $6\theta^2+70\theta$ &$10$ &$4$ \\
\Gd & 2.03(2) & $7\theta^2+80\theta$ &$\phn1$ &$0.3$ \\
\enddata
\tablecomments{$\theta$ is the radius of the extended region in
  arcseconds.  The limits assume uniform weighting over $\theta$, and
  are at the $3\,\sigma$ level.  The
  counts are in the 0.3--8.0~keV range.  The luminosity limits assume
  a power-law with $\Gamma=1.5$.}
\end{deluxetable}

% G093: 16178 good
%       area=3.6951e+06 pix^2=9.2377e+05 asec^2

% G315: 16130 good
%       area=3.0704e+06 pix^2=7.6760e+05 asec^2

% G127: 15704 good
%       area=3.215159e6 pix^2=7.7196e5 asec^2

\subsection{X-ray Timing}
\label{sec:timing}
\begin{figure}
% ../../G093/sources/plot_lc.m
\plottwo{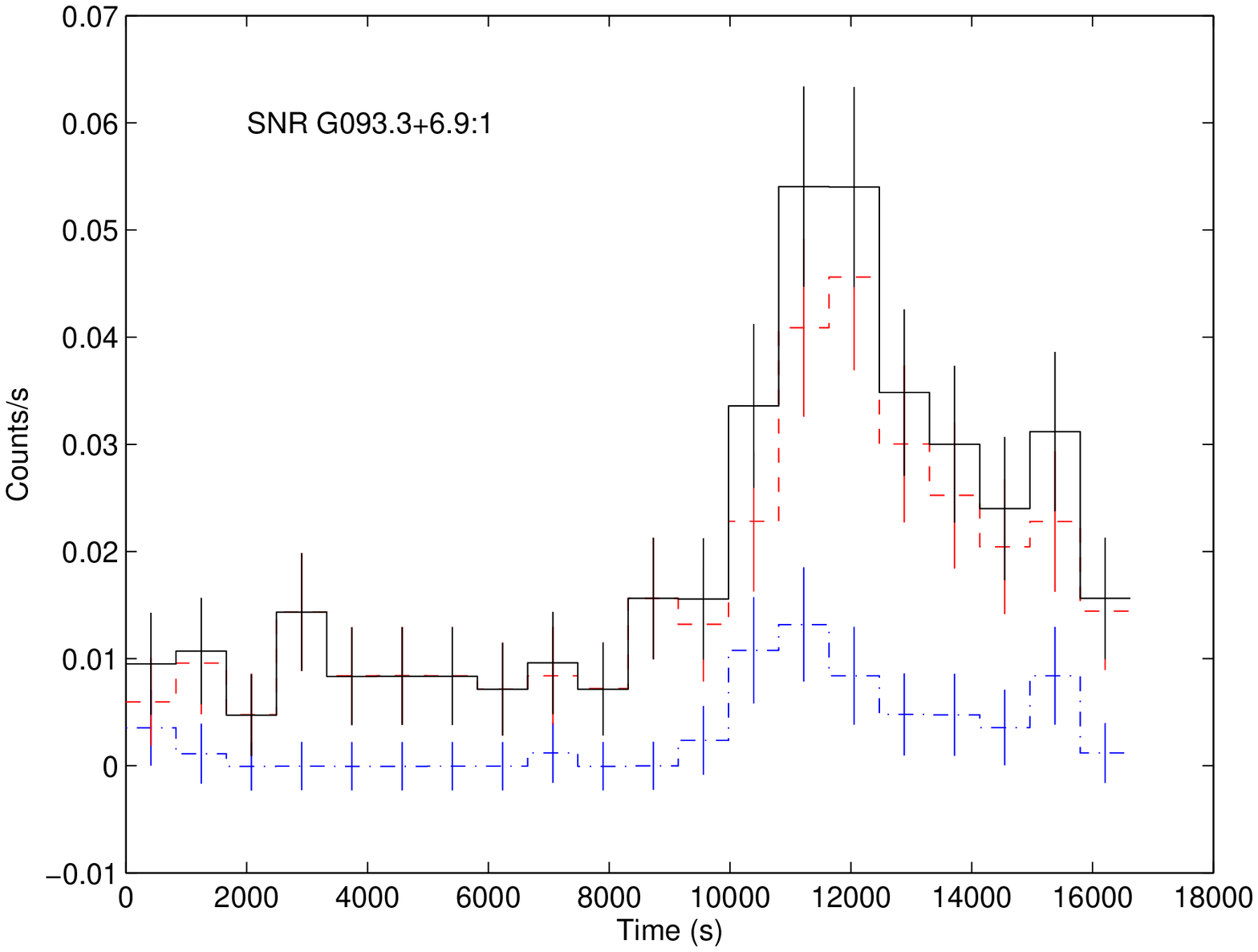}{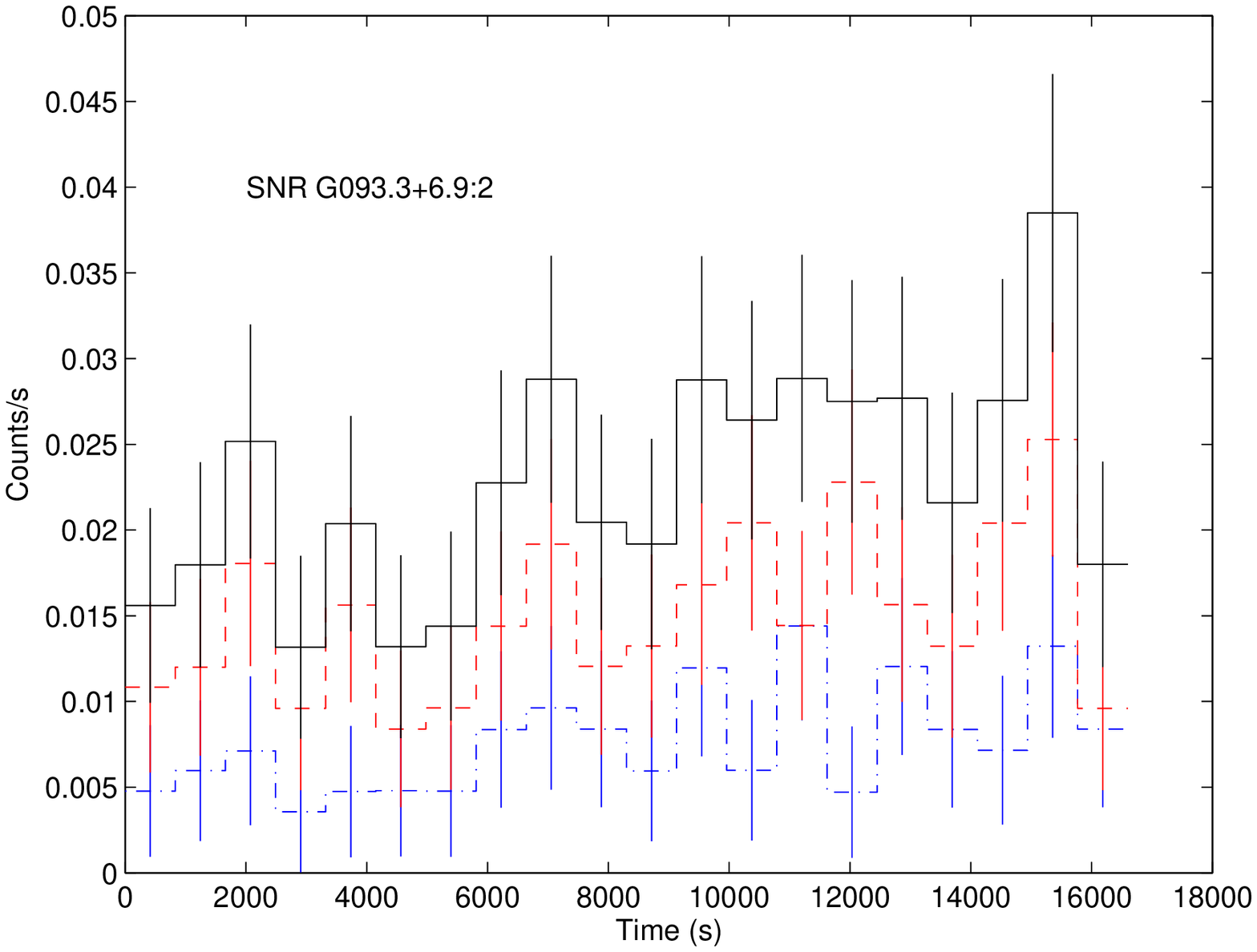}
\caption{Background-subtracted X-ray lightcurves of sources in \snr\ with
  $>100$~counts.  Left: \snr:1; right: \snr:2.  The dashed red line is
  0.3--2.0~keV (L-band), the dot-dashed blue line is 2.0--8.0~keV
  (H-band) and the solid black line is 0.3--8.0~keV (A-band).  A flare
  lasting $\approx 6000$~s and with an amplitude change of $\approx
  500$\% is present in the lightcurve of \snr:1.  }
\label{fig:lc}
\end{figure}

The majority of the detected sources have too few photons for
meaningful analyses of their variability.  We can perform some
analysis, though, with the brightest sources (we set a limit of
100~counts for a lightcurve, which eliminated all of the sources in
SNRs \Gb\ and \Gc).  In Figure~\ref{fig:lc} we show X-ray lightcurves
for the two sources in \snr\ with $>100$~counts, \snr:1 and \snr:2.
The lightcurve for \snr:2 is quite constant, but that for \snr:1 has a
significant flare lasting $\approx 1.5$~hr.  Searches for periodic
variation, however, showed nothing.  With the 3.2-s sampling of ACIS-I
we were unable to search for rapid variability, but the low
count-rates made that impossible anyway.  We took the barycentered
arrival-time data for \snr:1 and \snr:2 and performed FFTs with the
data binned into 4-s intervals.  No significant peaks in the
periodogram were found over the frequency range 0.01--0.12~Hz.

\begin{figure}
% ../../G127/sources/plot_lc.m
\plotone{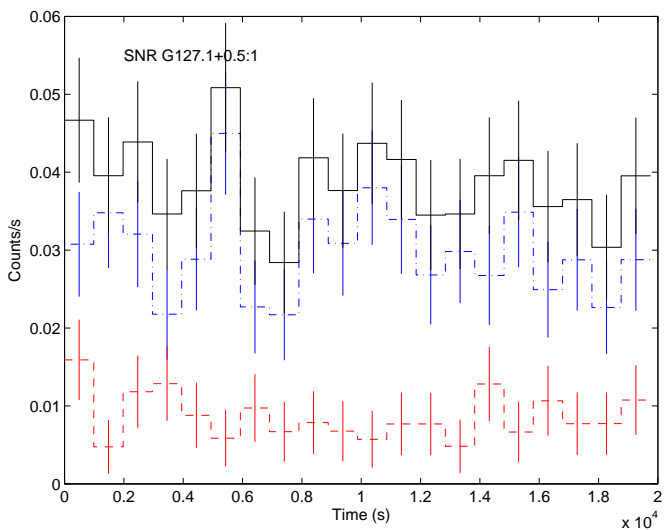}
\caption{Background-subtracted X-ray lightcurve of \snrd:1.  The dashed red line is
  0.3--2.0~keV (L-band), the dot-dashed blue line is 2.0--8.0~keV
  (H-band) and the solid black line is 0.3--8.0~keV (A-band).}
\label{fig:lcd}
\end{figure}

In Figure~\ref{fig:lcd} we show the X-ray lightcurve for the only
source in \snrd\ that has $>100$~counts: \snrd:1.  The lightcurve is
consistent with a constant flux.

\subsection{Multi-wavelength Followup \& Counterpart Identification}
\label{sec:cpt}
\subsubsection{Motivation}
After identifying X-ray sources with \chandra\ (or \xmm), the question
is then to determine which, if any, are the compact remnants of the
SNRs.  We have used source-count statistics
\citep{bah+01,mcba00,fbg+02} to estimate the number of
foreground/background sources given the $\nh$ and diffuse SNR
background toward each target --- we expect to detect 10--50 field
sources toward each SNR in the \cxo-AO3 sample.  Most of these sources
will be detected with few counts (10--30), and will therefore not
be suitable for X-ray spectroscopy that could, of itself, determine
which are neutron stars.  Because of their small X-ray count-rates,
weeding out interlopers requires multi-wavelength observations.

Isolated neutron stars have high X-ray to optical flux ratios
\citep*{hvkk00} --- deep optical/IR imaging is therefore an efficient
way to identify background sources.  We thus follow our X-ray
observations with successively deeper optical and IR observations,
identifying progressively fainter counterparts as we go.

Interloper sources, on the other hand, typically have much brighter optical/IR
counterparts \citep{hg88,gzh+03}.  In the Galactic plane, the majority
of 
sources are either nearby bright stars or active late-type stars.
Other possible sources are RS~CVn binaries, X-ray binaries, or
cataclysmic variables.  The extragalactic sources are usually AGN or
star-forming galaxies, although some nearby spiral galaxies are also
detectable \citep[e.g.,][]{bcc+03}.  Stars with spectral types A--M can
have some detectable X-ray emission (largely dependent on rotation and magnetic
fields) that appears as a hot (5--$\expnt{7}{6}$~K) plasma
\citep{kc00}.  X-ray binaries usually have harder spectra
characterized by power-laws with indices $\Gamma\approx 1.5$--2
\citep{gzh+03}.  Stars are unresolved optical/IR sources, but for
nearby binaries the X-ray emission may not lie exactly coincident with the
optical/IR emission (if, for example, the optical emission is from the
stars but the X-ray emission is from interacting coronae between the
stars) and the binary members may be merged to give the appearance of
an extended sources.  The extragalactic sources also have hard spectra
with power-law indices $\Gamma\approx 1.5$--2 \citep{bva+02}, and are
optically fainter ($R\gsim 18$).  While the nuclei of these galaxies
are unresolved, the closer galaxies can have resolved optical/IR
emission.

There are a small number of galaxies that have somewhat extreme
X-ray-to-optical flux ratios.  In a sample of 503 X-ray sources over
$448\mbox{ arcmin}^2$, \citet{kab+04} have identified a few sources
with X-ray-to-optical (850~nm) flux ratios exceeding 100, including 7
sources without 850~nm detections. While the identities are unclear,
they suggest that these galaxies have such faint optical counterparts
through a combination of intrinsic reddening and high redshift.
However, these galaxies should not significantly impact our sample.
We might expect 1--3 of these sources in each of our \chandra\ images,
but even without optical detections we should be able to identify
these as galaxies in the near-infrared, as all but one of the sources
were detected at $K_s$ band (where we have our deepest observations)
and the X-ray spectra are harder than those of most neutron stars
(also see \citealt{gcfj03}).  In
addition, these galaxies appear to just be the tail end of the
X-ray-to-optical flux ratio distribution, with most neutron stars
being significantly higher.

\subsubsection{Execution}
Therefore, after the detection of the X-ray sources we obtained
progressively deeper optical and IR data of the SNR fields.  Starting
with small telescopes (Palomar 60-inch, Las Campanas 40-inch) we moved
to larger telescopes (Palomar 200-inch, NTT, Magellan, and finally
Keck) as we identified counterparts to more and more of the X-ray
sources (see below).  The details of the optical/IR observations are
in Tables~\ref{tab:opt}--\ref{tab:optd}.

\begin{figure*}
%\centering
%\includegraphics[width=0.5\textwidth]{f16a.ps}\includegraphics[width=0.5\textwidth]{f16b.ps}
%\includegraphics[width=0.5\textwidth]{f16c.ps}\includegraphics[width=0.5\textwidth]{f16d.ps}
%\includegraphics[width=0.5\textwidth]{f16e.ps}\includegraphics[width=0.5\textwidth]{f16f.ps}
\includegraphics[width=\hsize]{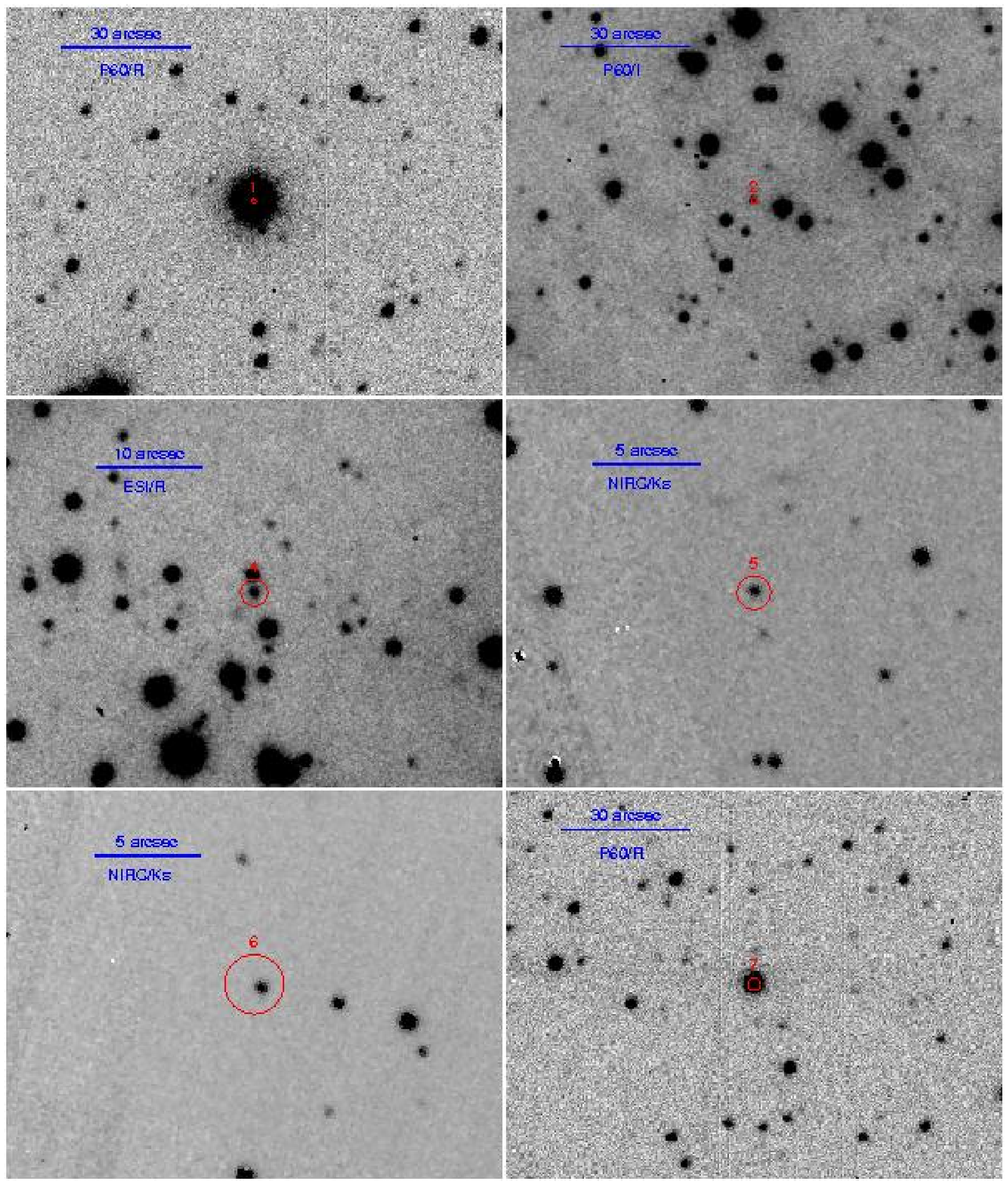}
\caption{Optical/IR images of counterparts to X-ray sources in \snr.  North is up, East is to the left, and a scale-bar is in
  the upper-left corner.}
\label{fig:opta1}
\end{figure*}

\begin{figure*}
%\centering
%\includegraphics[width=0.5\textwidth]{f17a.ps}\includegraphics[width=0.5\textwidth]{f17b.ps}
%\includegraphics[width=0.5\textwidth]{f17c.ps}\includegraphics[width=0.5\textwidth]{f17d.ps}
%\includegraphics[width=0.5\textwidth]{f17e.ps}\includegraphics[width=0.5\textwidth]{f17f.ps}
\includegraphics[width=\hsize]{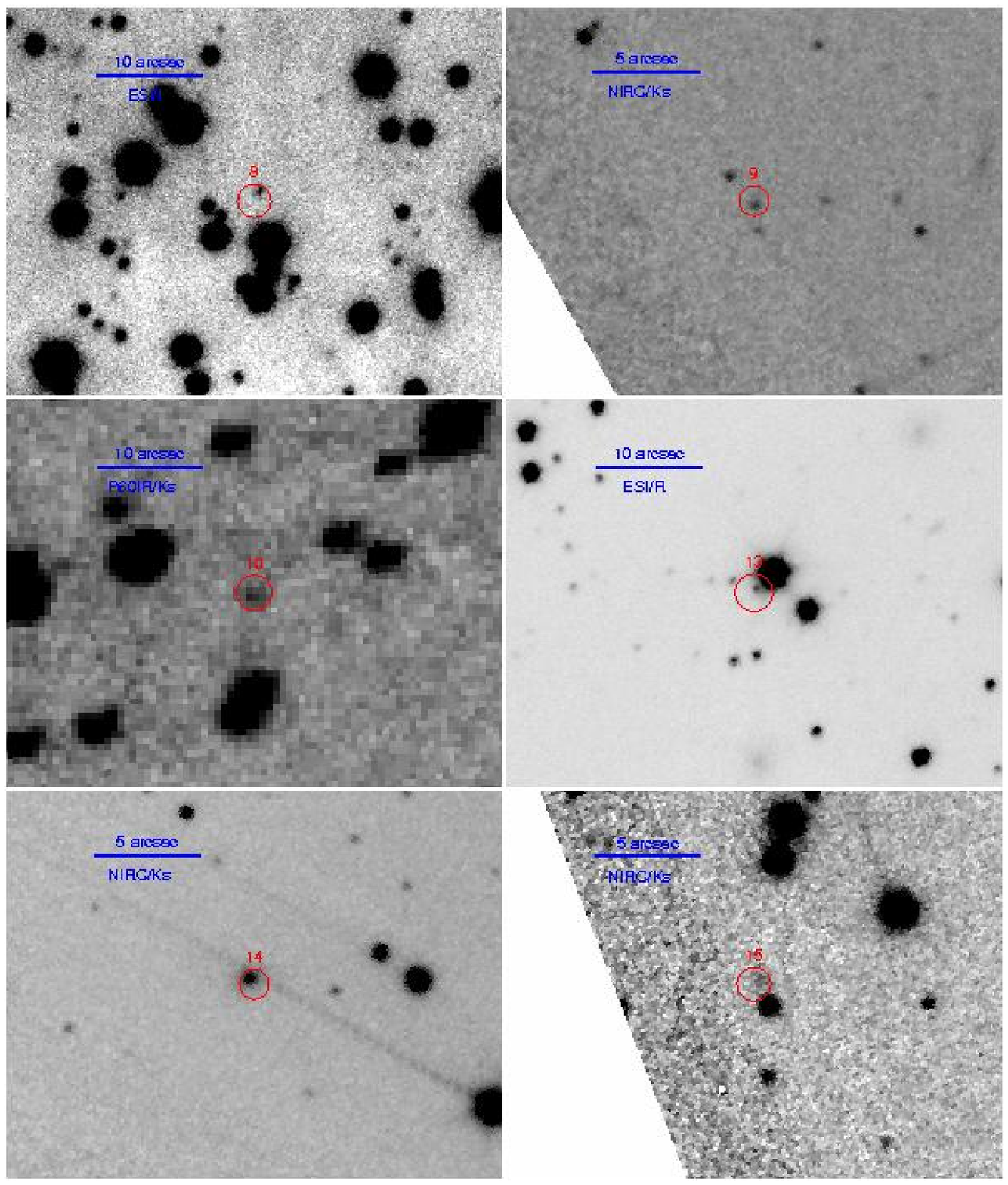}
\caption{Optical/IR images of counterparts to X-ray sources in \snr\ (cont.).  North is up, East is to the left, and a scale-bar is in
  the upper-left corner.}
\label{fig:opta2}
\end{figure*}

\begin{figure*}
%\centering
%\includegraphics[width=0.5\textwidth]{f18a.ps}\includegraphics[width=0.5\textwidth]{f18b.ps}
%\includegraphics[width=0.5\textwidth]{f18c.ps}\includegraphics[width=0.5\textwidth]{f18d.ps}
%\includegraphics[width=0.5\textwidth]{f18e.ps}\includegraphics[width=0.5\textwidth]{f18f.ps}
\includegraphics[width=\hsize]{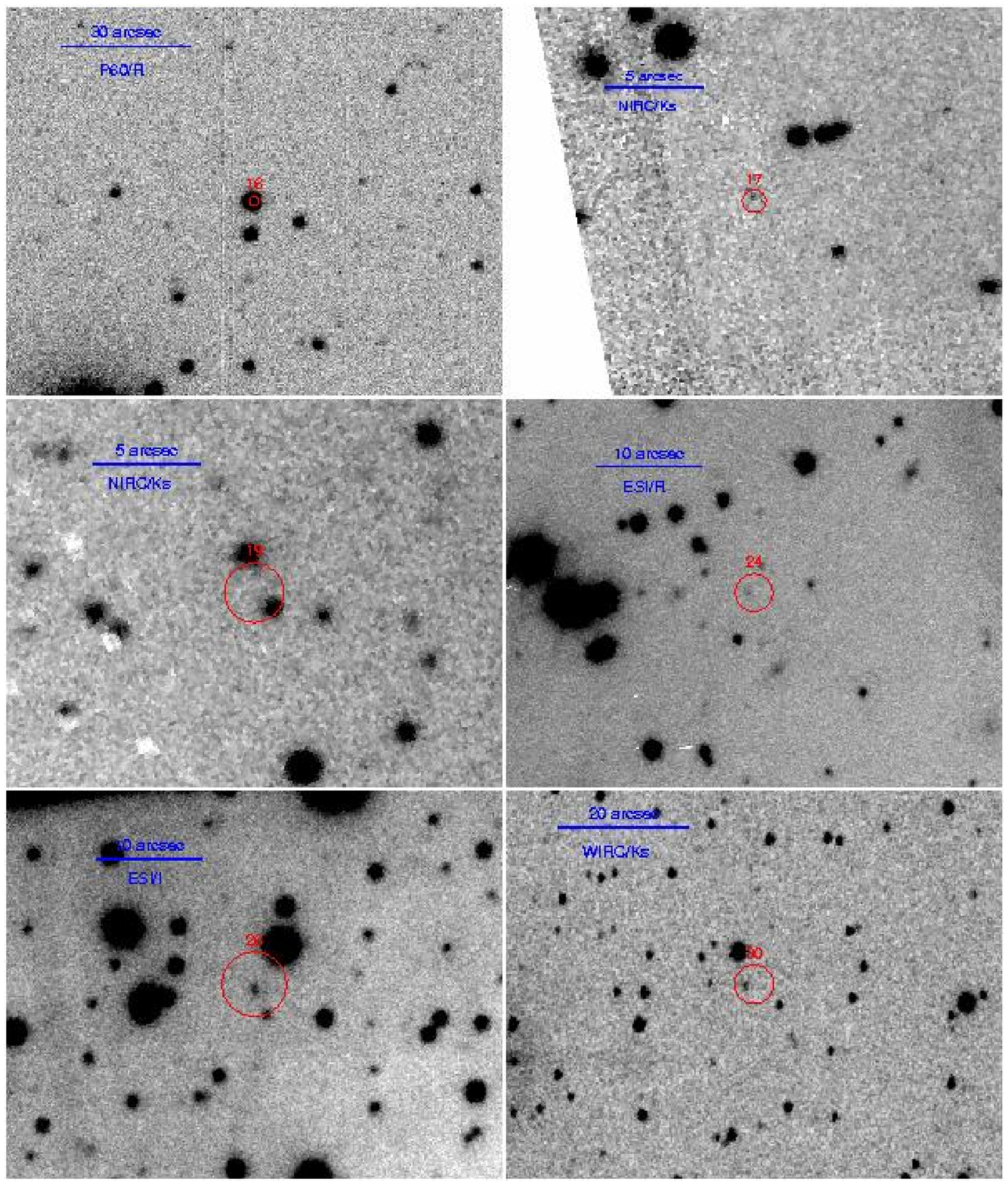}
\caption{Optical/IR images of counterparts to X-ray sources in \snr\ (cont.). North is up, East is to the left, and a scale-bar is in
  the upper-left corner.}
\label{fig:opta3}
\end{figure*}

\begin{figure*}
%\centering
%\includegraphics[width=0.5\textwidth]{f19a.ps}\includegraphics[width=0.5\textwidth]{f19b.ps}
%\includegraphics[width=0.5\textwidth]{f19c.ps}\includegraphics[width=0.5\textwidth]{f19d.ps}
%\includegraphics[width=0.5\textwidth]{f19e.ps}\includegraphics[width=0.5\textwidth]{f19f.ps}
\includegraphics[width=\hsize]{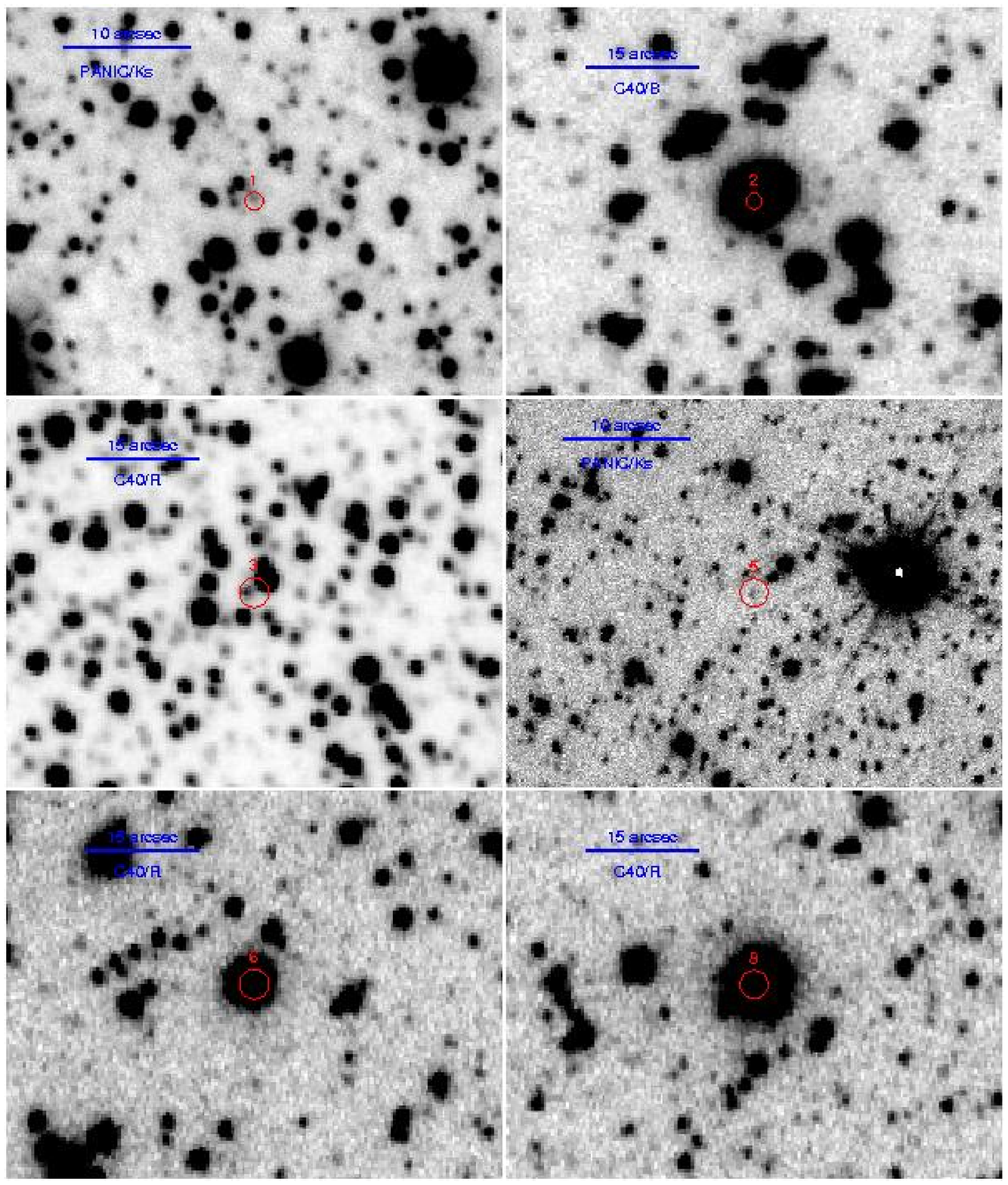}
\caption{Optical/IR images of counterparts to X-ray sources in \snrb.
  North is up, East is to the left, and a scale-bar is in the
  upper-left corner.}
\label{fig:optb1}
\end{figure*}

\begin{figure*}
%\centering
%\includegraphics[width=0.5\textwidth]{f20a.ps}\includegraphics[width=0.5\textwidth]{f20b.ps}
%\includegraphics[width=0.5\textwidth]{f20c.ps}\includegraphics[width=0.5\textwidth]{f20d.ps}
%\includegraphics[width=0.5\textwidth]{f20e.ps}\includegraphics[width=0.5\textwidth]{f20f.ps}
\includegraphics[width=\hsize]{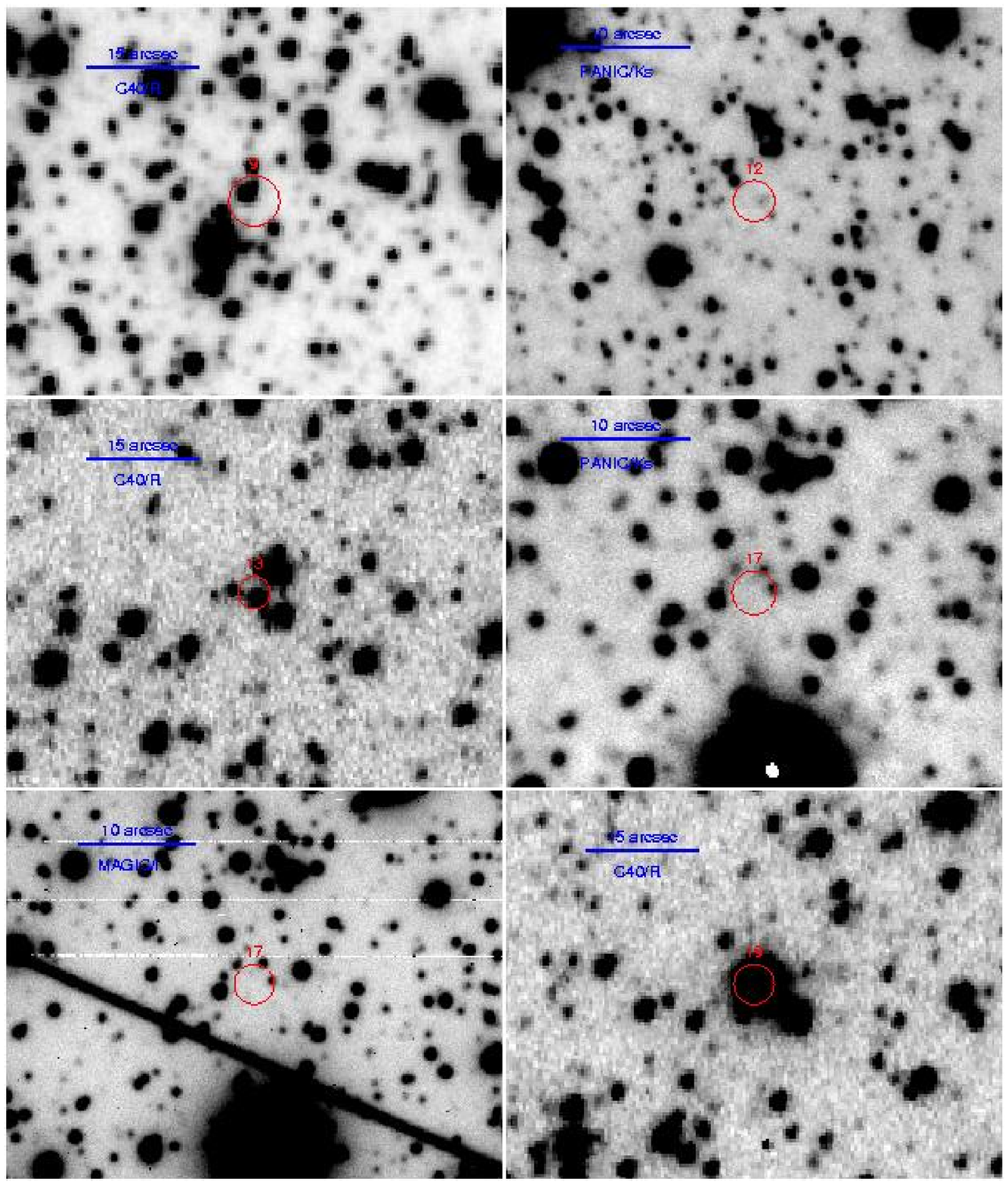}
\caption{Optical/IR images of counterparts to X-ray sources in
  \snrb\ (cont.). North is up, East is to the left, and a scale-bar is in the
  upper-left corner.  }
\label{fig:optb2}
\end{figure*}

\begin{figure*}
%\centering
%\includegraphics[width=0.5\textwidth]{f21a.ps}\includegraphics[width=0.5\textwidth]{f21b.ps}
%\includegraphics[width=0.5\textwidth]{f21c.ps}\includegraphics[width=0.5\textwidth]{f21d.ps}
%\includegraphics[width=0.5\textwidth]{f21e.ps}
\includegraphics[width=\hsize]{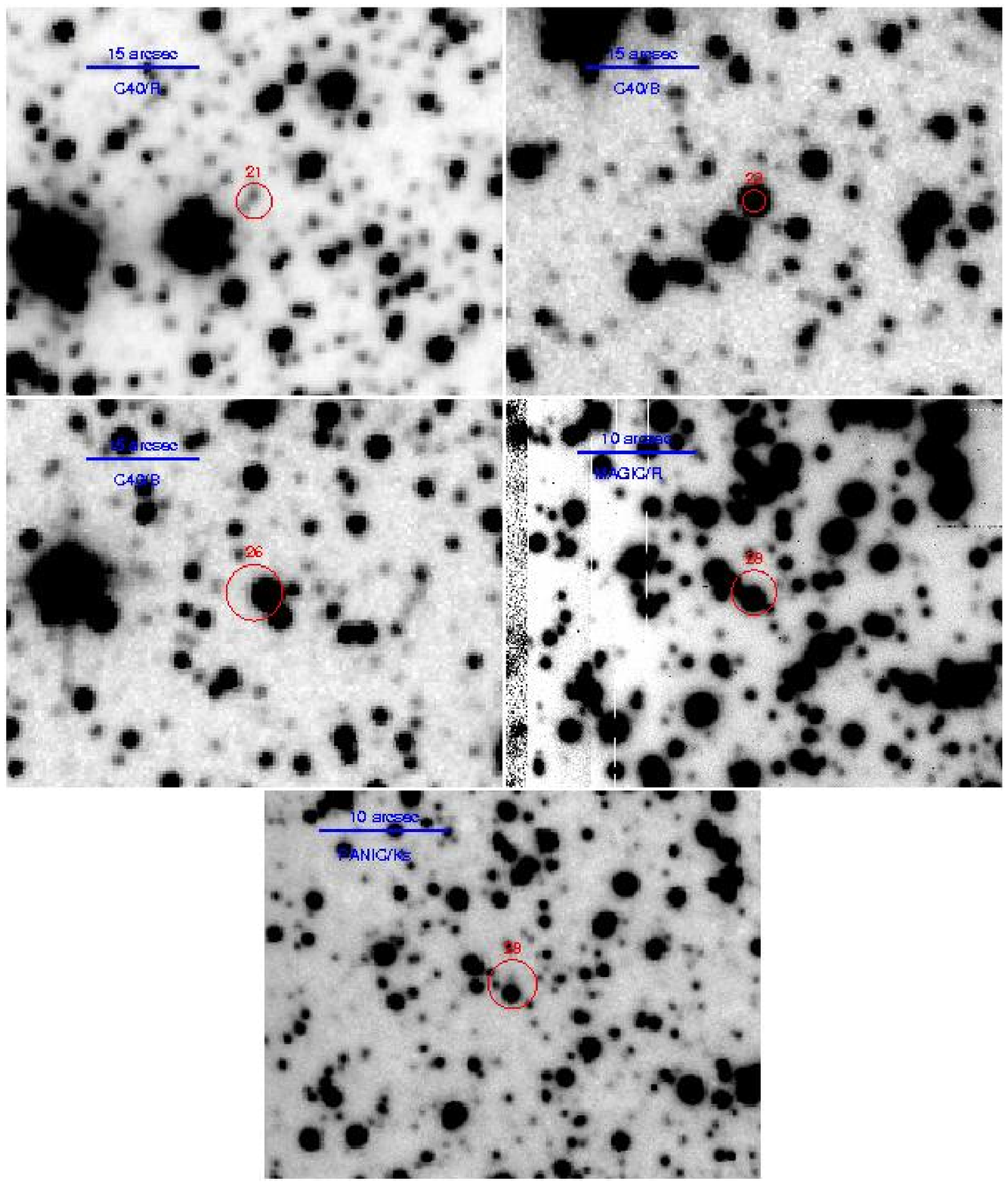}
\caption{Optical/IR images of counterparts to X-ray sources in \snrb\ (cont.).
North is up, East is to the left, and a scale-bar is in
  the upper-left corner.}
\label{fig:optb3}
\end{figure*}

\begin{figure*}
%\centering
%\includegraphics[width=0.5\textwidth]{f22a.ps}\includegraphics[width=0.5\textwidth]{f22b.ps}
%\includegraphics[width=0.5\textwidth]{f22c.ps}\includegraphics[width=0.5\textwidth]{f22d.ps}
%\includegraphics[width=0.5\textwidth]{f22e.ps}\includegraphics[width=0.5\textwidth]{f22f.ps}
\includegraphics[width=\hsize]{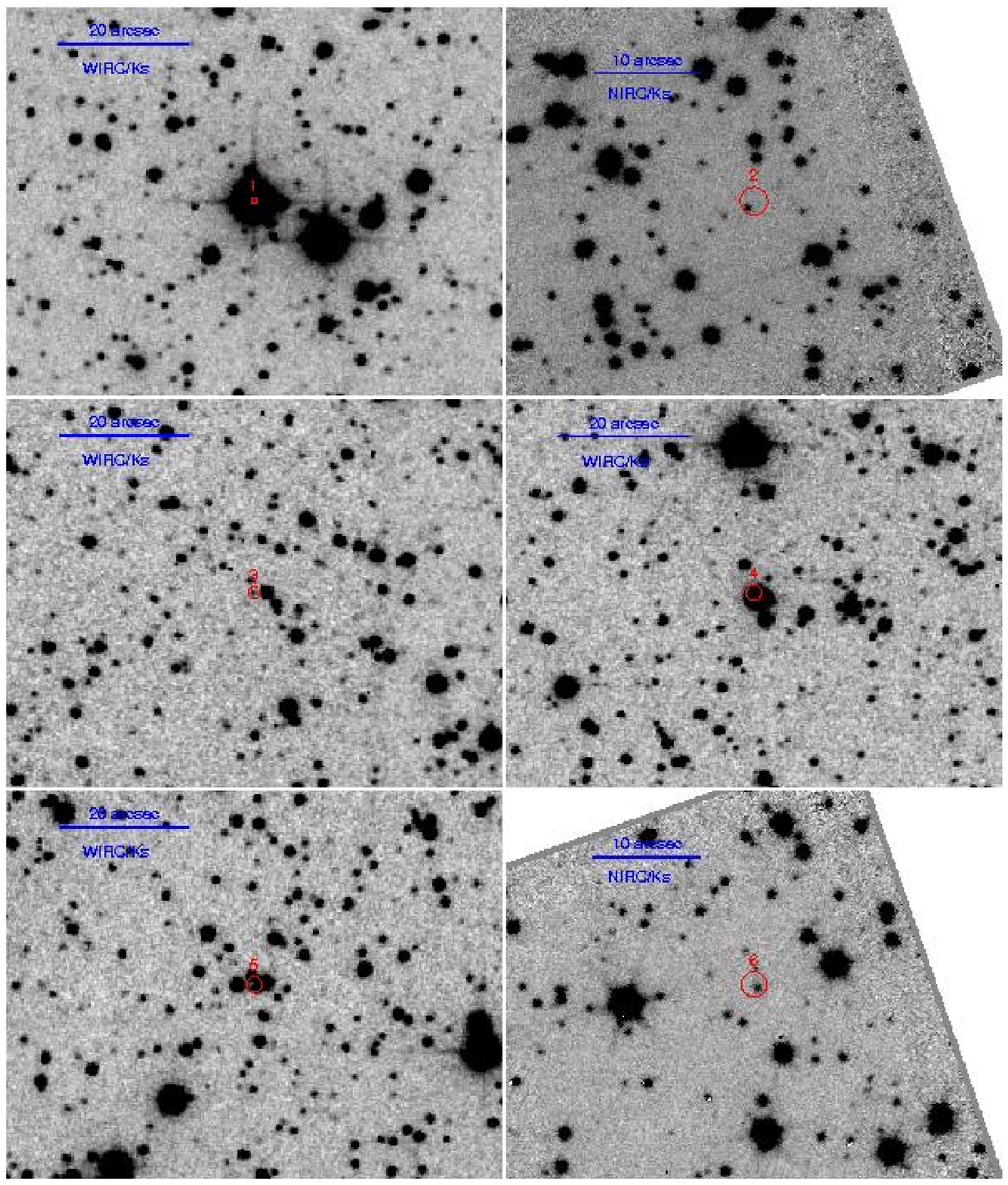}
\caption{Optical/IR images of counterparts to X-ray sources in \snrc.  North is up, East is to the left, and a scale-bar is in
  the upper-left corner.}
\label{fig:optc1}
\end{figure*}

\begin{figure*}
%\centering
%\includegraphics[width=0.5\textwidth]{f23a.ps}\includegraphics[width=0.5\textwidth]{f23b.ps}
%\includegraphics[width=0.5\textwidth]{f23c.ps}\includegraphics[width=0.5\textwidth]{f23d.ps}
%\includegraphics[width=0.5\textwidth]{f23e.ps}\includegraphics[width=0.5\textwidth]{f23f.ps}
\includegraphics[width=\hsize]{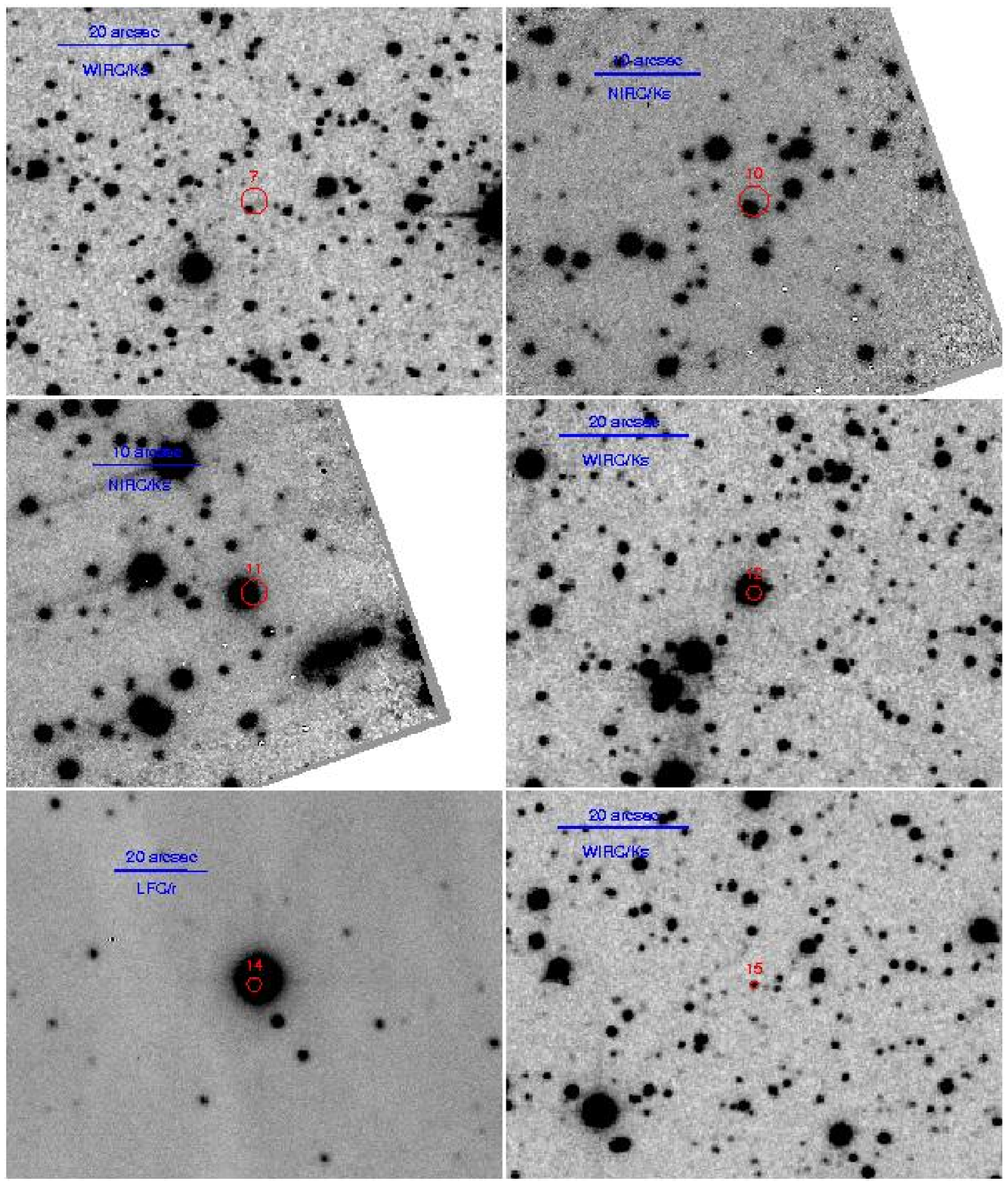}
\caption{Optical/IR images of counterparts to X-ray sources in \snrc\ (cont.).  North is up, East is to the left, and a scale-bar is in
  the upper-left corner.}
\label{fig:optc2}
\end{figure*}

\begin{figure*}
%\centering
%\includegraphics[width=0.5\textwidth]{f24a.ps}\includegraphics[width=0.5\textwidth]{f24b.ps}
%\includegraphics[width=0.5\textwidth]{f24c.ps}\includegraphics[width=0.5\textwidth]{f24d.ps}
\includegraphics[width=\hsize]{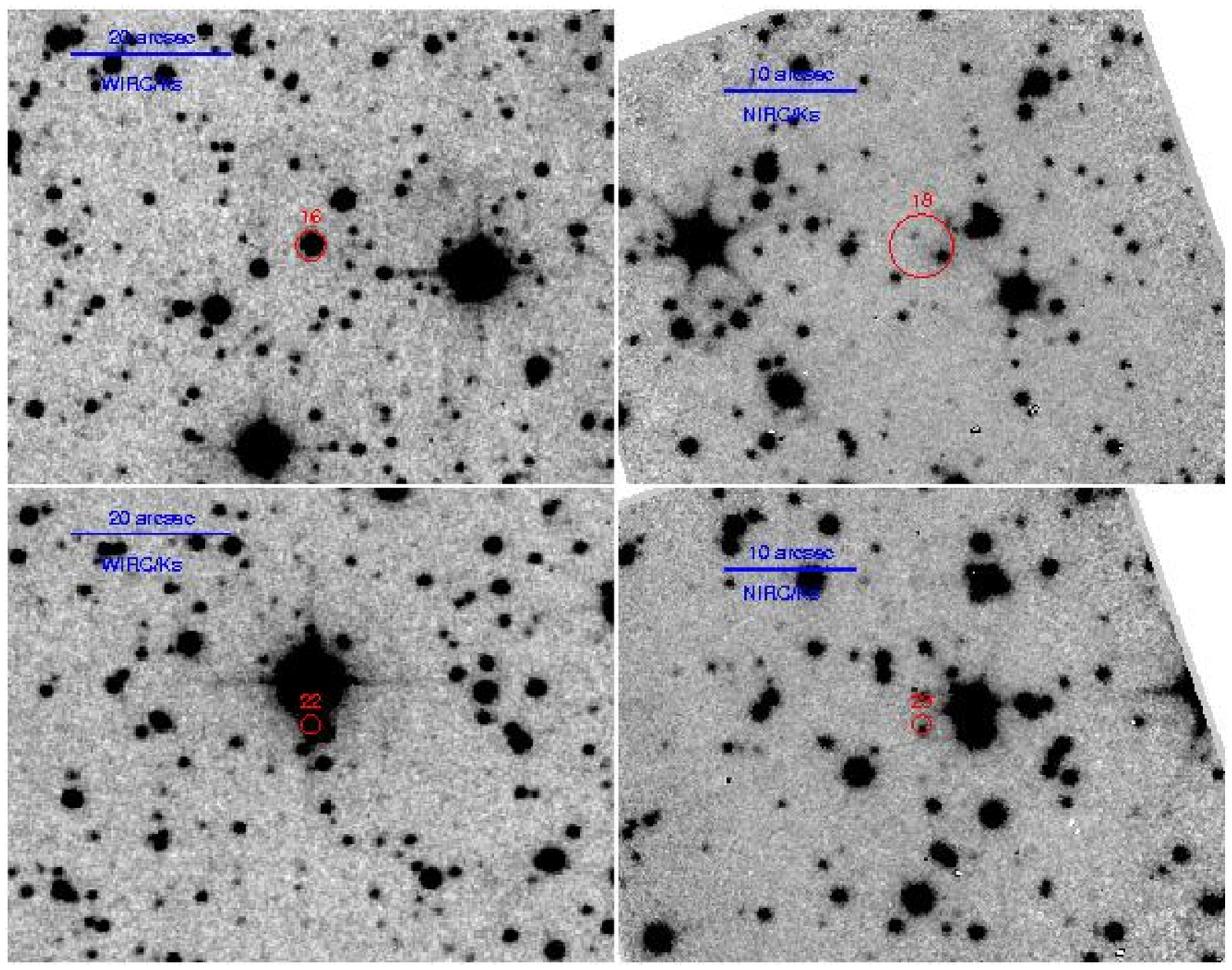}
\caption{Optical/IR images of counterparts to X-ray sources in \snrc\ (cont.).  North is up, East is to the left, and a scale-bar is in
  the upper-left corner.}
\label{fig:optc3}
\end{figure*}

\begin{figure*}
%\centering
%\includegraphics[width=0.5\textwidth]{f25a.ps}\includegraphics[width=0.5\textwidth]{f25b.ps}
%\includegraphics[width=0.5\textwidth]{f25c.ps}\includegraphics[width=0.5\textwidth]{f25d.ps}
%\includegraphics[width=0.5\textwidth]{f25e.ps}\includegraphics[width=0.5\textwidth]{f25f.ps}
\includegraphics[width=\hsize]{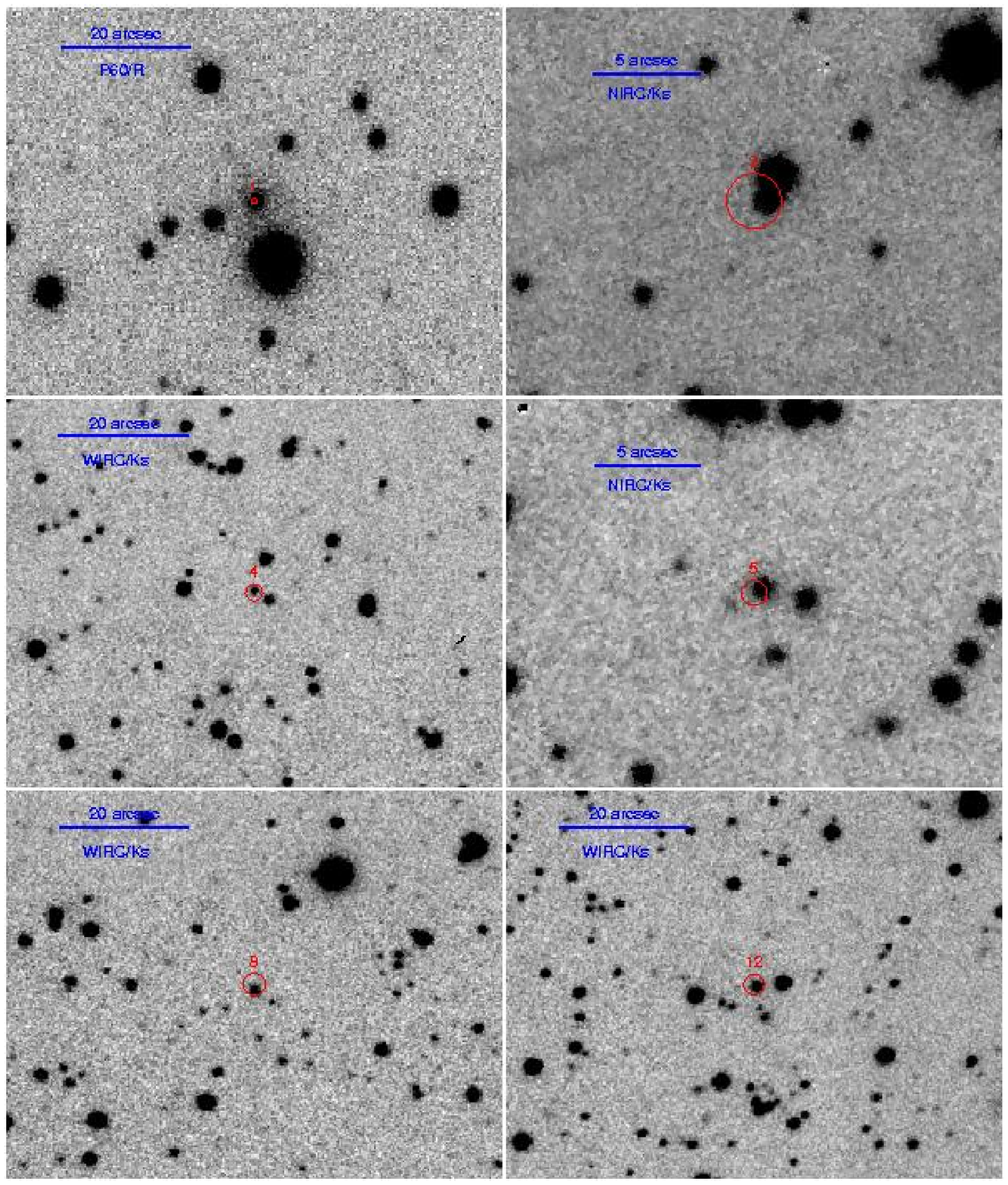}
\caption{Optical/IR images of counterparts to X-ray sources in \snrd.  North is up, East is to the left, and a scale-bar is in
  the upper-left corner.}
\label{fig:optd1}
\end{figure*}

\begin{figure*}
%\centering
%\includegraphics[width=0.5\textwidth]{f26a.ps}\includegraphics[width=0.5\textwidth]{f26b.ps}
%\includegraphics[width=0.5\textwidth]{f26c.ps}\includegraphics[width=0.5\textwidth]{f26d.ps}
%\includegraphics[width=0.5\textwidth]{f26e.ps}\includegraphics[width=0.5\textwidth]{f26f.ps}
\includegraphics[width=\hsize]{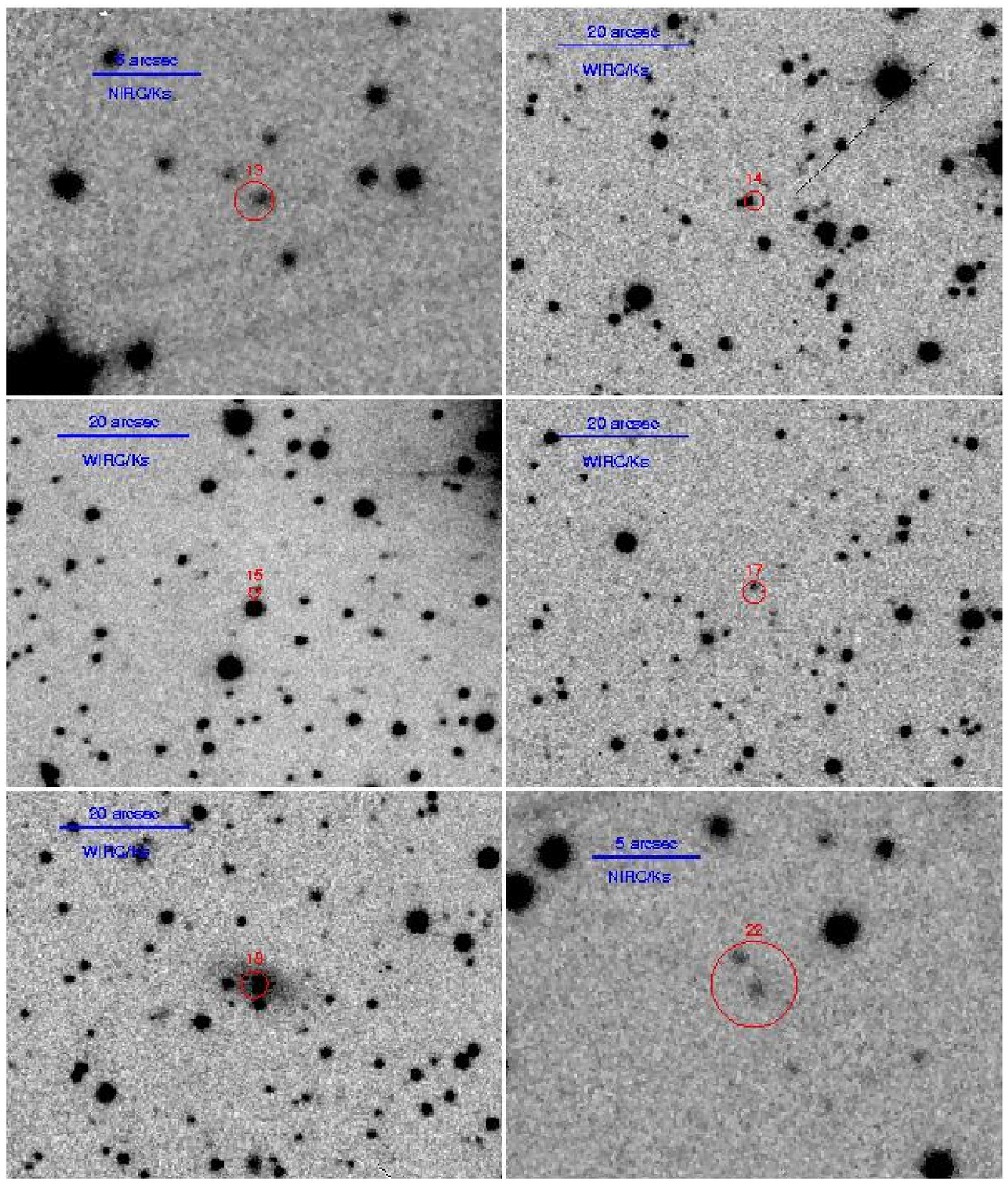}
\caption{Optical/IR images of counterparts to X-ray sources in \snrd\ (cont.).  North is up, East is to the left, and a scale-bar is in
  the upper-left corner.}
\label{fig:optd2}
\end{figure*}

We began by registering all of the data for a SNR (we also included 2MASS $J$-
and $K_{s}$-band images) to the same frame, and then searched for
optical/IR counterparts to the X-ray sources using \texttt{sextractor}
\citep{ba96}.  We considered a source as a potential counterpart if it
was within the 90\% confidence radius of the X-ray source combined
with a $0\farcs2$ uncertainty for the optical astrometry.  We also
inspected all X-ray sources to see if there was an optical source at
the edge of the error circle, if there were multiple sources, etc.  We
then determined the positions and magnitudes of all of the detected
sources, which we list in Table~\ref{tab:match}--\ref{tab:matchd}.
The positions in Tables~\ref{tab:match}--\ref{tab:matchd} are the
averages of the positions for all images where the source was
detected, except for sources that were saturated in several bands but
which had unambiguous 2MASS detections, where we used the 2MASS
position alone.  For sources that had multiple detections in the same
band but by different instruments (e.g.\ C40 and EMMI $R$-band), we
used the detection that had the highest S/N without being saturated.
We show postage-stamp images of the optical/IR counterparts to the
X-ray sources in Figures~\ref{fig:opta1}--\ref{fig:opta3} (for \snr),
Figures~\ref{fig:optb1}--\ref{fig:optb3} (for \snrb),
Figures~\ref{fig:optc1}--\ref{fig:optc3} (for \snrc), and
Figures~\ref{fig:optd1}--\ref{fig:optd2} (for \snrd).  The
instrument(s)/band(s) chosen for each source are those that best
illustrate the detection.

\begin{deluxetable}{l l l l c r l}
\tablecaption{Optical/IR Observations of \snr\label{tab:opt}}
\tablewidth{0pt}
\tabletypesize{\small}
\tablehead{
\colhead{Telescope} & \colhead{Instrument} & \colhead{Band(s)} &
\colhead{Date} & \colhead{Seeing} & \colhead{Exposure} & \colhead{Sources\tablenotemark{a}}\\
&  & & \colhead{(UT)} & \colhead{(arcsec)} & \colhead{(sec)} & \\
}
\startdata
P60 & P60CCD & $BVRI$ & 2001-Jul-23 & 1.9--2.5 & 300 &$-4$\\
P200 & LFC & $g\arcmin r\arcmin$ & 2002-Jan-18 & 1.9 & 600 & $-13$ \\
Keck I & LRIS & $g$ & 2002-Jun-15 & 0.9 & 2300 & 5,14,15,17\\
 & & $R$&  &0.9 &  2000 & 5,14,15,17\\
 & & $g$ &  &0.9 & 1380 &  2,5,7,8,9,16,19,26\\
 & & $R$ &  &1.0 & 1200 &  2,5,7,9,16,19,26\\
Keck I & NIRC & $K_s$ & 2002-Jun-02 &0.5& 250 & 5,6,9,14,15,17\\
P60 & P60IR & $K_s$ & 2002-Jul-26 & 2.0 & 900 & 4,10,13,19\\
Keck II & ESI & $R$ & 2002-Sep-03 &0.8& 540 & 4 \\
 & & $I$ & &0.8& 900 & 4 \\
 & & $I$ & &0.8& 2520 & 8,26\\
 & & $R$ & &0.7& 660 & 10,24\\
 & & $I$ & &0.6& 600 & 10\\
 & & $R$& &0.7& 960 & 13 \\
Keck I & NIRC & $K_s$ & 2002-Nov-16 & 0.5 & 2950 & 19\\
P200 & WIRC & $K_s$ & 2003-Jul-24 & 0.8 & 1080 &
$-6$,$-10$,$-16$ \\
\enddata
\tablenotetext{a}{Indicates which sources from Table~\ref{tab:srcs}
  were on which images.  Negative numbers indicate that all sources
  but the negated one(s) were on the image.}
\tablecomments{The telescopes/instruments used were P60CCD: the CCD
  camera on the Palomar 60-inch; P60IR: the Infrared camera on the
  Palomar 60-inch \citep{mpp+95}; LFC: the Large Format Camera on the
  Palomar 200-inch; LRIS: the Low-Resolution Imaging Spectrometer on
  the 10-m Keck~I telescope \citep{o+95}; NIRC: the Near-Infrared Camera on
  the 10-m Keck~I telescope \citep{ms94}; ESI: the Echellette Spectrograph
  and Imager on the10-m  Keck~II telescope \citep{sbe+02}; WIRC: the
  Wide-field Infrared Camera on the Palomar 200-inch.
}
\end{deluxetable}

\begin{deluxetable}{l l l l c r l}
\tablecaption{Optical/IR Observations of \snrb\label{tab:optb}}
\tablewidth{0pt}
\tabletypesize{\small}
\tablehead{
\colhead{Telescope} & \colhead{Instrument} & \colhead{Band(s)} &
\colhead{Date} & \colhead{Seeing} & \colhead{Exposure} & \colhead{Sources\tablenotemark{a}}\\
&  & & \colhead{(UT)} & \colhead{(arcsec)} & \colhead{(sec)} & \\
}
\startdata
C40 & C40CCD & $BVRI$ & 2002-Apr-18 & 1.3 & 3600 & $-$3 \\
 & & $BR$ & 2002-Apr-20 & 1.5 & 3600 & $-$21 \\
NTT & EMMI & $VRI$ & 2002-Jun-14 & 0.8 & 1130 & 1,5,17,23\\
Magellan II & MagIC & $BR$ & 2003-Apr-03 & 0.8 &1800 & 1\\
 & & $R$ & 2003-Apr-03 & 0.8 &3000 & 12,28 \\
 & & $R$ & 2003-Apr-03 & 0.8 &3000 & 6,17\\
 & & $I$ & 2003-Apr-04 & 0.7 &1500 & 1 \\
 & & $I$ & 2003-Apr-04 & 0.6 &900 & 6,17\\
Magellan II & PANIC & $K_s$ & 2003-Apr-18 & 0.8 & 2160 & 1 \\
 & & $K_s$ & 2003-Apr-18 &0.9  & 2160 & 12,28 \\
 & & $K_s$ & 2003-Apr-18 &1.0 & 2160 & 6,17 \\
 & & $K_s$ & 2003-Apr-20 & 0.6 & 2160 & 1 \\
 & & $K_s$ & 2003-Apr-20 & 0.5 & 2160 & 12,28 \\
 & & $K_s$ & 2003-Jun-07 & 0.4 & 900 & 5\tablenotemark{b} \\
\enddata
\tablenotetext{a}{Indicates which sources from Table~\ref{tab:srcsb}
  were on which images.  Negative numbers indicate that all sources
  but the negated one(s) were on the image.}
\tablenotetext{b}{Observed by M.~van Kerkwijk.}
\tablecomments{The telescopes/instruments used were C40CCD: the
  direct-imaging CCD camera on the Las Campanas 40-inch; EMMI: the ESO
  Multi-Mode Instrument (red imaging arm only) on the 3.5-m New
  Technology Telescope (NTT) at La Silla; MagIC: Raymond and Beverly
  Sackler Magellan Instant Camera on the 6.5-m Clay (Magellan II)
  telescope; PANIC: Persson's Auxiliary Nasmyth Infrared Camera on
  the 6.5-m Clay (Magellan II) telescope.}
\end{deluxetable}

\begin{deluxetable}{l l l l c r l}
\tablecaption{Optical/IR Observations of \snrc\label{tab:optc}}
\tablewidth{0pt}
\tabletypesize{\small}
\tablehead{
\colhead{Telescope} & \colhead{Instrument} & \colhead{Band(s)} &
\colhead{Date} & \colhead{Seeing} & \colhead{Exposure} & \colhead{Sources\tablenotemark{a}}\\
&  & & \colhead{(UT)} & \colhead{(arcsec)} & \colhead{(sec)} & \\
}
\startdata
P200 & LFC & $r^{\prime}$ & 2002-Oct-07 & 1.5 & 2730 & all \\
P200 & WIRC & $J$ & 2002-Nov-01 & 1.2 & 4500 & all \\
   & & $K_s$ & 2002-Nov-01 & 0.9 & 2640 & all \\
Keck~I & NIRC & $K_s$ & 2003-Aug-11 & 0.5 & 450 & 2,6,10,11,14,18,23 \\
% & & $K_s$ & 2003-Sep-04 & & 1750 & 18 \\
\enddata
\tablenotetext{a}{Indicates which sources from Table~\ref{tab:srcsc}
  were on which images.  Negative numbers indicate that all sources
  but the negated one(s) were on the image.}
\tablecomments{The telescopes/instruments used follow from Table~\ref{tab:opt}.
}
\end{deluxetable}

\begin{deluxetable}{l l l l c r l}
\tablecaption{Optical/IR Observations of \snrd\label{tab:optd}}
\tablewidth{0pt}
\tabletypesize{\small}
\tablehead{
\colhead{Telescope} & \colhead{Instrument} & \colhead{Band(s)} &
\colhead{Date} & \colhead{Seeing} & \colhead{Exposure} & \colhead{Sources\tablenotemark{a}}\\
&  & & \colhead{(UT)} & \colhead{(arcsec)} & \colhead{(sec)} & \\
}
\startdata
P60 & P60CCD & $R$ & 2001-Jul-23 & 1.9 & 3900 & $-22$\\
 & & $I$ & 2001-Jul-23 & 1.5 & 300 & $-22$ \\
P200 & LFC & $r\arcmin$ & 2002-Oct-07 & 1.9 & 2700 & all \\
P200 & WIRC & $J$ & 2002-Oct-28 & 0.9 & 4050 & all\\
 &  & $K_s$ & 2002-Oct-28 & 0.8 & 2700 & all \\
Keck~I & NIRC & $K_s$ & 2003-Aug-11 & 0.6 & 900 & 3,5,13,22 \\
% & & & 2003-Sep-04 & 0.5 & 3150 & 57 \\
\enddata
\tablenotetext{a}{Indicates which sources from Table~\ref{tab:srcsd}
  were on which images.  Negative numbers indicate that all sources
  but the negated one(s) were on the image.}
\tablecomments{The telescopes/instruments used follow from Table~\ref{tab:opt}.
}
\end{deluxetable}

To determine limiting magnitudes, we simulated Gaussian stars with the
same FWHM as the average seeing for the image.  We placed the stars
randomly in regions that were not too crowded, similar to the regions
where the X-ray sources actually were, and determined the $3\,\sigma$
limiting flux by performing photometry (again using
\texttt{sextractor}) on the stars.

We define optical/IR fluxes using the zero-point calibrations of
\citet*{bcp98}, where $F(m)=F_{0} 10^{-m/2.5}$ and
$F_{0}=(\expnt{2.0}{-5},\expnt{1.4}{-5},\expnt{9.0}{-6},\expnt{8.7}{-7})\mbox{
  ergs s}^{-1}\mbox{ cm}^{-2}$ for $(V,R,I,K_{s})$, respectively.  We
also use the reddening coefficients of \citet{bcp98}, such that
$A_{\lambda}/A_{V}=(1.0,0.82,0.62,0.11)$ again for $(V,R,I,K_{s})$,
respectively.

\subsubsection{Counterpart Evaluation}
It is certainly possible that there will be an unrelated optical/IR
source in the error circle of an X-ray source (e.g., \citealt*{kkm01}),
especially if the X-ray sources are off-axis and/or have few counts,
and therefore have large position uncertainties.  There is no entirely accurate
way to prevent this from happening and leading to the false
association of what is actually a neutron star with another source,
which would cause us to reject the neutron star from our sample.

\begin{figure*}
% plot_srcs2.m
%\centering
%\includegraphics[width=0.5\textwidth]{f27a.eps}\includegraphics[width=0.5\textwidth]{f27b.eps}
%\includegraphics[width=0.5\textwidth]{f27c.eps}\includegraphics[width=0.5\textwidth]{f27d.eps}
\includegraphics[width=\textwidth]{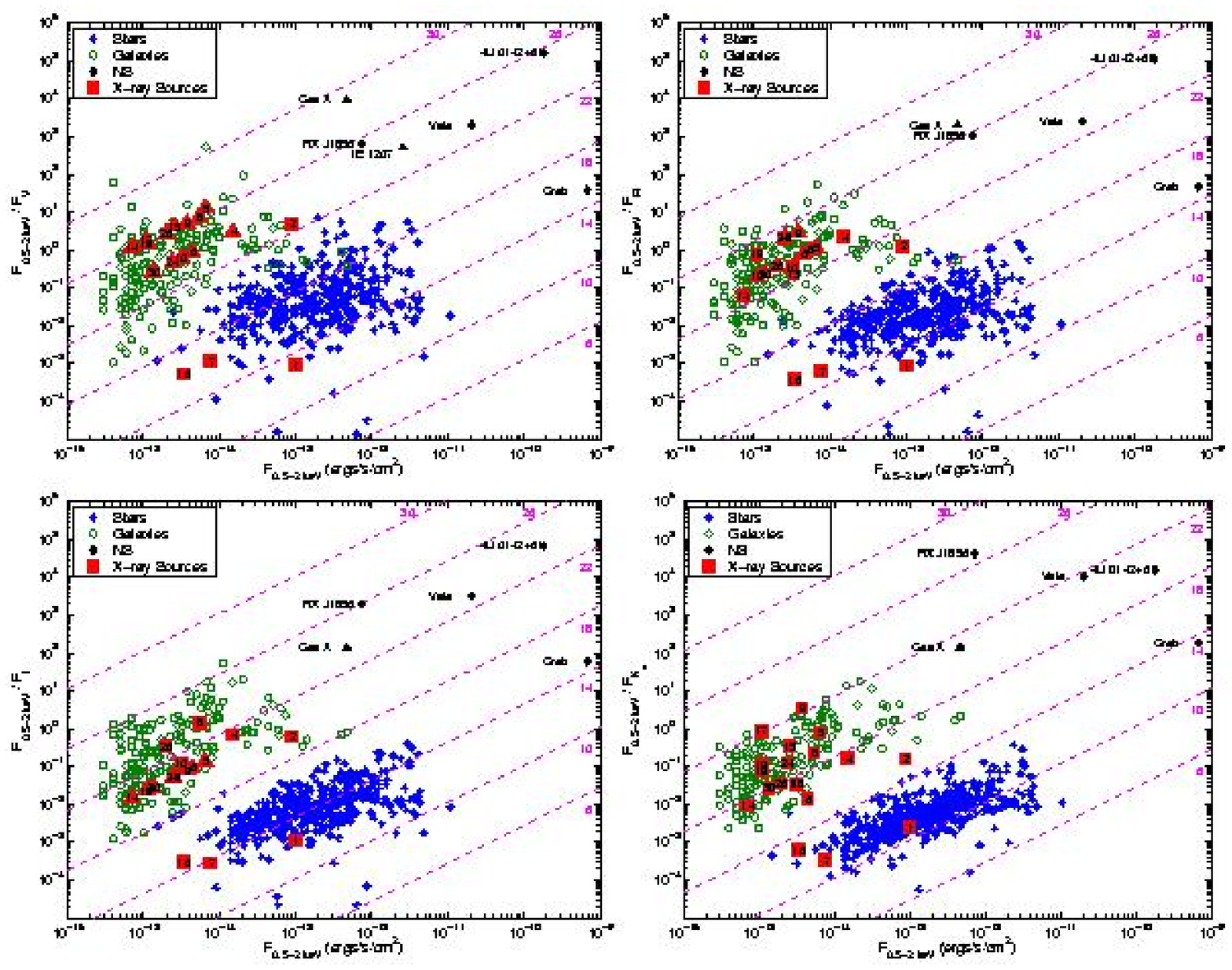}
\caption{X-ray-to-optical/IR flux ratio vs.\ X-ray flux for sources in
  \snr\ (Tabs.~\ref{tab:srcs} and \ref{tab:match}) with sources from
  the CDF/Orion studies and selected neutron stars. Upper left:
  $V$-band; upper right: $R$-band; lower left: $I$-band; lower right:
  $K_s$-band. Stars from CDF/Orion are blue asterisks, galaxies are
  green circles.  Selected neutron stars are black diamonds/limits,
  and are labeled.  The unidentified X-ray sources in \snr\ are the
  red squares/limits and are also labeled (in the case of multiple
  possible counterparts, the source is plotted multiple times).  The
  diagonal lines represent constant magnitude, and are labeled by
  that magnitude.  For the sources in \snr\ the counts were converted
  to a flux by $F_{0.5-2.0\mbox{ \scriptsize keV}}={\rm
  counts}_{0.5-2.0\mbox{ \scriptsize keV}} \times
  \expnt{3.4}{-16}\mbox{ ergs s}^{-1}\mbox{ cm}^{-1}$, appropriate for
  a blackbody with $kT_\infty=0.25$~keV and $\nh=\expnt{2}{21}\mbox{
  cm}^{-2}$.  As seen in Table~\ref{tab:limits}, the X-ray fluxes
  change can be a factor of $\approx 2$ higher for a power-law spectrum.}
\label{fig:xopt}
\end{figure*}

\begin{figure*}
% plot_srcs2.m
\centering
\includegraphics[width=\textwidth]{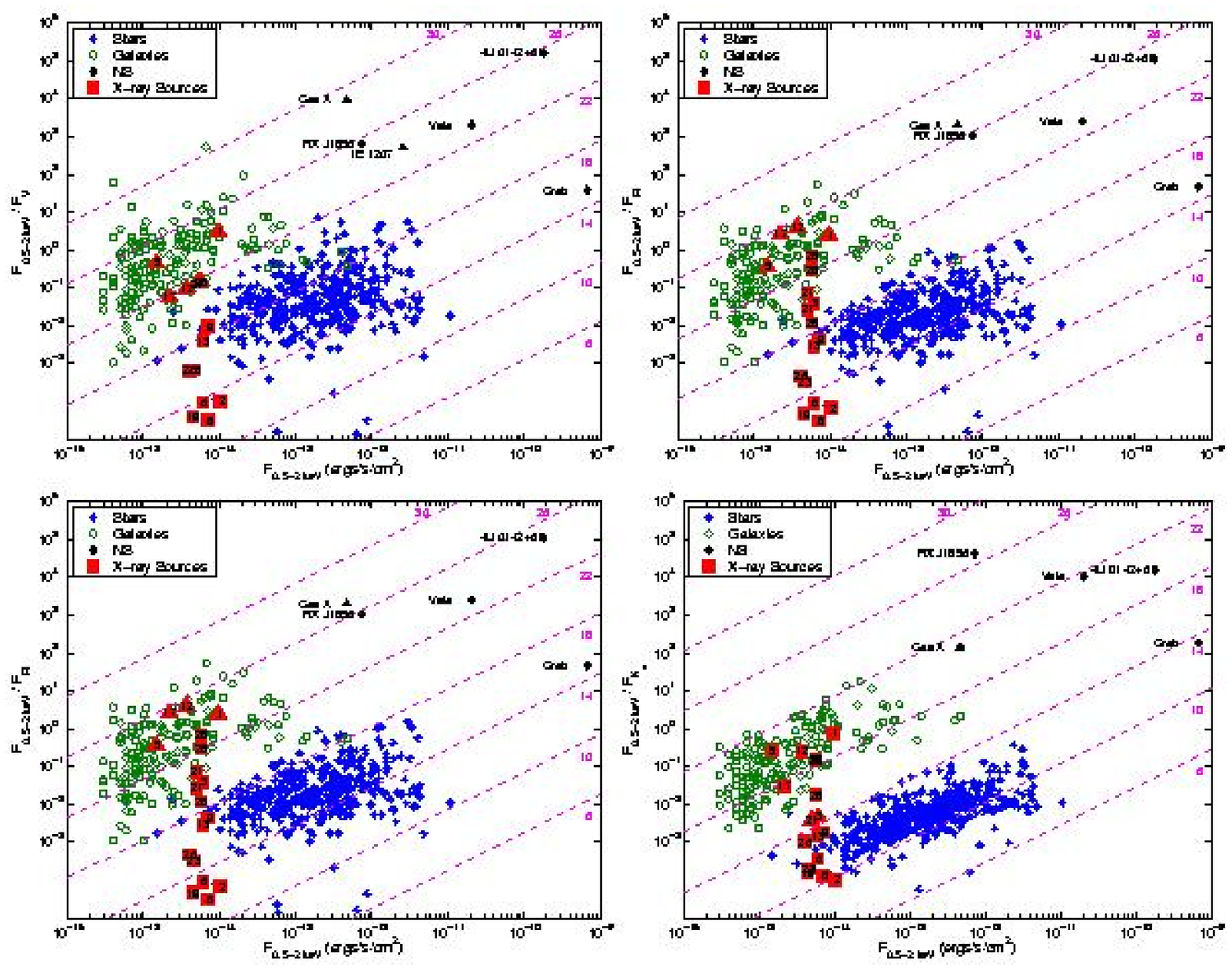}
\caption{X-ray-to-optical/IR flux ratio vs.\ X-ray flux for sources in
  \snrb\ (Tabs.~\ref{tab:srcsb} and \ref{tab:matchb}) with sources
  from the CDF/Orion studies and selected neutron stars, following
  Figure~\ref{fig:xopt}. Upper left: $V$-band; upper right: $R$-band;
  lower left: $I$-band; lower right: $K_s$-band. For the sources in
  \snrb\ the counts were converted to a flux by $F_{0.5-2.0\mbox{
  \scriptsize keV}}={\rm counts}_{0.5-2.0\mbox{ \scriptsize keV}}
  \times \expnt{5.1}{-16}\mbox{ ergs s}^{-1}\mbox{ cm}^{-1}$,
  appropriate for a blackbody with $kT_\infty=0.25$~keV and
  $\nh=\expnt{2}{21}\mbox{ cm}^{-2}$.  }
\label{fig:xoptb}
\end{figure*}

\begin{figure*}
% plot_srcs2.m
%\includegraphics[width=0.5\textwidth]{f29a.eps}\includegraphics[width=0.5\textwidth]{f29b.eps}
\includegraphics[width=\textwidth]{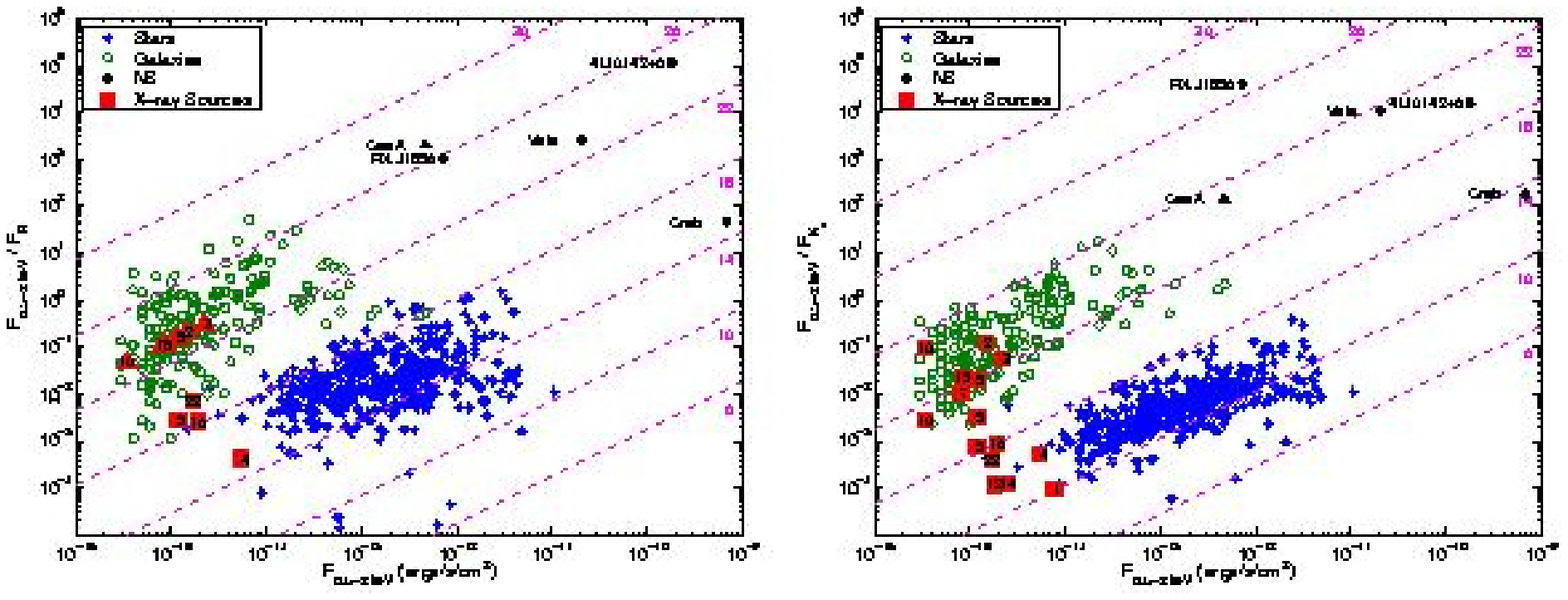}
\caption{X-ray-to-optical/IR flux ratio vs.\ X-ray flux for sources in
  \snrc\ (Tabs.~\ref{tab:srcsc} and \ref{tab:matchc}) with sources from
  the CDF/Orion studies and selected neutron stars, following
  Figure~\ref{fig:xopt}. Left:
  $R$-band; right:
  $K_s$-band. For the
  sources in \snr\ the counts were converted to a flux by
  $F_{0.5-2.0\mbox{ \scriptsize keV}}={\rm counts}_{0.5-2.0\mbox{
  \scriptsize keV}} \times \expnt{2.0}{-16}\mbox{ ergs s}^{-1}\mbox{
  cm}^{-1}$, appropriate for a blackbody with $kT_\infty=0.25$~keV and
  $\nh=\expnt{2}{21}\mbox{ cm}^{-2}$.  }
\label{fig:xoptc}
\end{figure*}

\begin{figure*}
% plot_srcs2.m
%\includegraphics[width=0.5\textwidth]{f30a.eps}\includegraphics[width=0.5\textwidth]{f30b.eps}
\includegraphics[width=\textwidth]{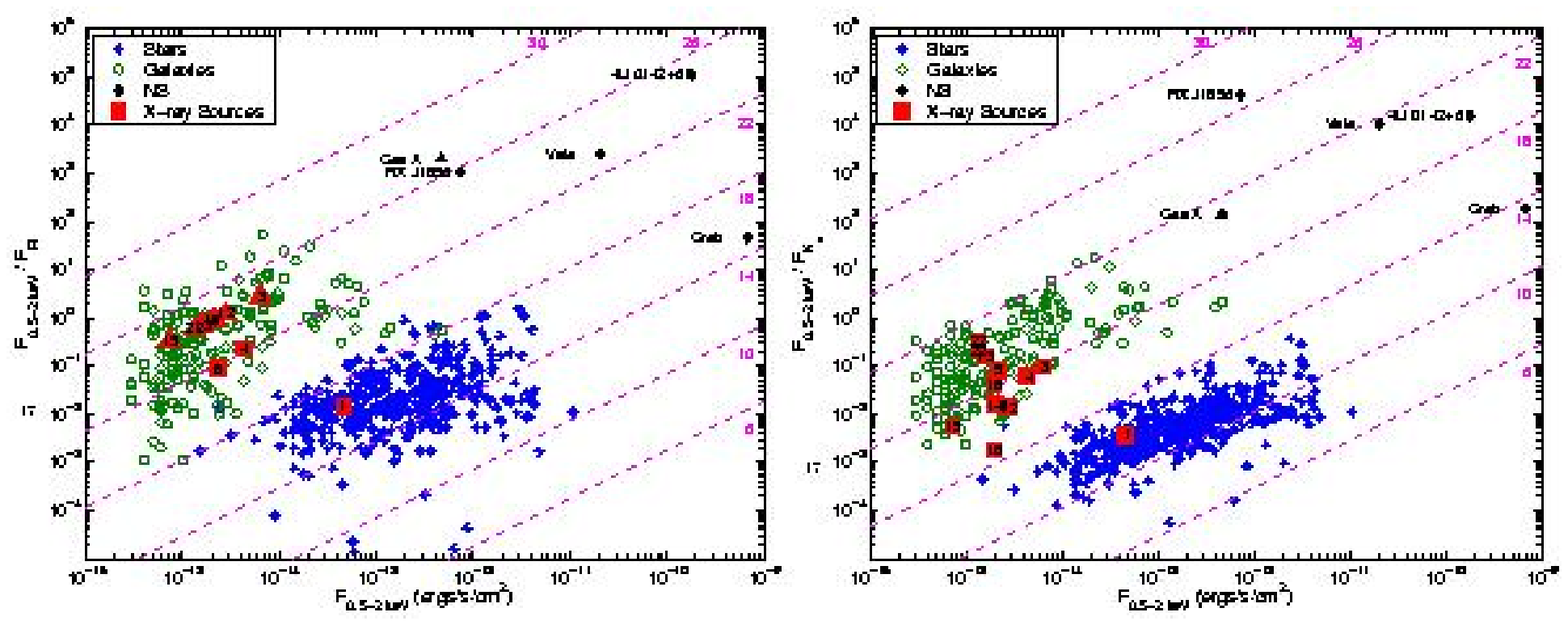}
\caption{X-ray-to-optical/IR flux ratio vs.\ X-ray flux for sources in
  \snrd\ (Tabs.~\ref{tab:srcsd} and \ref{tab:matchd}) with sources from
  the CDF/Orion studies and selected neutron stars, following
  Figure~\ref{fig:xopt}. Left:
  $R$-band; right:
  $K_s$-band. For the
  sources in \snrd\ the counts were converted to a flux by
  $F_{0.5-2.0\mbox{ \scriptsize keV}}={\rm counts}_{0.5-2.0\mbox{
  \scriptsize keV}} \times \expnt{2.6}{-16}\mbox{ ergs s}^{-1}\mbox{
  cm}^{-1}$, appropriate for a blackbody with $kT_\infty=0.25$~keV and
  $\nh=\expnt{2}{21}\mbox{ cm}^{-2}$.  }
\label{fig:xoptd}
\end{figure*}

However, there are ways that we can guard against this and incorporate
our uncertainties into our limits.  If enough is known about the
optical/IR source (e.g.\ colors), it is possible to determine what the
source is \citep{kkm01} and thereby assign some likelihood to its
association with an X-ray source.  This is not always possible for the
sources in our sample as the information is often limited to
optical/IR detections in one or two bands (Tabs.~\ref{tab:match}--
\ref{tab:matchd}).  Similarly, for the few sources with enough X-ray
counts a determination can be made based on the X-ray spectrum (or
just hardness) and lightcurve (Fig.~\ref{fig:lc}).  For the majority
of the sources, though, we must examine them relative to the sources
in other samples, specifically the CDF and Orion samples (many of the
stars in the Orion sample are younger than the general Galactic
population, but they do cover a wide range of stellar masses).  For
this reason we plot the optical/IR-to-X-ray flux ratio against X-ray
flux in Figures~\ref{fig:xopt}--\ref{fig:xoptd} (we compute the X-ray
flux from the observed count rate by using conversion factors
determined from W3PIMMS for a blackbody with $kT_\infty=0.25$~keV and
$\nh$ appropriate for each SNR; a change to a power-law with photon
index of 1.5 raises the implied fluxes by a factor of $\approx 2$).
Sources that fall in the loci defined by the other samples are likely
to be of similar type.  This is of course not a definitive assignment,
but it should work most of the time.  For instance, if we had
associated Star~A from \citet{kkm01} with the X-ray flux from
CXO~J232327.8+584842 (the central source in Cas A), it would appear in
Figure~\ref{fig:xopt} with the same X-ray flux as the compact source
in Cas~A and $<1$ magnitude brighter than the limit we have plotted.
In other words, it would be far outside the Orion/CDF loci.  This
argument, that Star~A could not be the source of the X-rays from
CXO~J232327.8+584842 just on the basis of its X-ray-to-optical flux
ratio, was made by \citet{kkm01}, who then followed it up by a
detailed analysis of multi-band photometry of Star~A (see also
\citealt{fcht02}).

%\begin{landscape}
\begin{deluxetable}{r c c c c c c c c c c c r}
\setlength{\tabcolsep}{0.05in}
%\rotate
\tablecaption{Optical/IR Matches to X-ray Sources in \snr\label{tab:match}}
\tablewidth{0pt}
\tabletypesize{\scriptsize}
\tablehead{ \colhead{ID\tablenotemark{a}}  & \colhead{$\alpha$\tablenotemark{b}} & \colhead{$\delta$\tablenotemark{b}} & \colhead{$\Delta r$\tablenotemark{c}} & \colhead{$B$} & \colhead{$V$} & \colhead{$R$} & \colhead{$I$} &
 \colhead{$g/g^{\prime}$\tablenotemark{d}} & \colhead{$r^{\prime}$} & \colhead{$J$} & \colhead{$K_s$}& \colhead{$N(<K_s)$\tablenotemark{e}} \\
 & \multicolumn{2}{c}{(J2000)} & \colhead{(arcsec)} & \multicolumn{8}{c}{(mag)} \\
}
\startdata
1 & 20:52:22.90 & +55:23:43.7 & 0.1 & 13.34(7) & 13.16(3) &
12.76(3) & 12.535(10) & sat & sat & 11.172(9) &
10.859(11) & $<0.01$\\
2 & 20:52:31.01 & +55:14:37.6 & 0.3 & $>$23.5 & $>$22.5 & 20.74(5)
& 19.52(4) & 22.61(10) & 21.05(6) & $>$15.8 & 15.52(1) & $<0.01$\\
4 & 20:52:25.83 & +55:27:41.7 & 0.1 & \nodata & \nodata & 23.30(11)
& 21.60(10) & $>$24.0 & $>$23.5 & $>$15.8 & 17.46(17) & $<0.01$\\
5 & 20:52:22.01 & +55:15:16.9 & 0.1 & $>$23.5 & $>$22.5 & $>$25.0 &
$>$20.7 & $>$26.5 & $>$23.5 & $>$15.8 & 20.1(4) & $<0.01$ \\
6 & 20:51:31.33 & +55:20:31.4 & 0.4 & $>$23.5 & $>$22.5 & $>$22.3 &
$>$20.7 & $>$24.0 & $>$23.5 & $>$15.8 & 16.1(4) & $<0.01$\\[0.1in]
7 & 20:52:50.22 & +55:16:06.7 & 0.2 & 17.67(7) & 16.23(3) &
15.18(3) & 13.857(10) & sat & sat & 12.363(16) &
11.475(16) & $<0.01$\\
8 & 20:52:42.13 & +55:14:10.7 & 1.1 & $>$23.5 & $>$22.5 & $>$22.3 & 23.43(13) & $>$26.1 & $>$23.5 & $>$15.8 & 18.9(1)& 0.06 \\
9 & 20:52:36.48 & +55:17:45.8 & 0.2 & $>$23.5 & $>$22.5 & $>$25.1 & $>$20.7 & $>$26.1 & $>$25.1 & $>$15.8 & 22.3(4) & $<0.01$\\
10 & 20:51:39.85 & +55:25:53.1 & 0.3 & $>$23.5 & $>$22.5 & 22.99(11) & 21.36(10) & $>$24.0 & $>$23.5 & $>$15.8 & 17.45(11) & $<0.01$\\
13 & 20:52:42.23 & +55:26:07.4 & 0.4 & \nodata & \nodata & 22.5(5) & \nodata & \nodata & \nodata & $>$15.8 &17.44(6)& $<0.01$ \\[0.1in]
14 & 20:52:05.98 & +55:17:58.6 & 0.2 & $>$23.5 & $>$22.5 & 22.72(6) & $>$20.7 & $>$26.3 & $>$23.5 & $>$15.8 & 17.04(5) & $<0.01$\\
15 & 20:52:09.83 & +55:18:21.4 & 0.4 & $>$23.5 & $>$22.5 & $>$22.3 &
$>$20.7 & $>$26.3 & $>$25.2 & $>$15.8 & 20.2(4) & 0.01\\
16 & 20:52:50.16 & +55:20:25.3 & 0.7 & 17.63(7) & 16.22(3) & 15.49(3) & 14.732(10) & sat & sat & 13.61(3) & 13.02(3) & $<0.01$\\
17 & 20:52:12.66 & +55:18:54.6 & 0.2 & $>$23.5 & $>$22.5 & $>$25.2 &
$>$20.7 & $>$26.3 & $>$23.5 & $>$15.8 & 21.1(4) & $<0.01$ \\
19 & 20:52:36.79 & +55:15:28.1 & 1.1 & $>$23.5 & $>$22.5 & 24.99(14)
& $>$20.7 & $>$26.1 & $>$23.5 & $>$15.8 & 19.9(4) & 0.09\\
" & 20:52:36.92 & +55:15:30.5 & 1.8 & $>$23.5 & $>$22.5 & $>$22.3 &
$>$20.7 & $>$26.1 & $>$23.5 & $>$15.8 & 19.5(4) & 0.22\\[0.1in]
24 & 20:51:52.31 & +55:26:02.8 & 0.7 & $>$23.5 & $>$22.5 & 25.2(2) & $>$20.7 & $>$24.0 & $>$23.5 & $>$15.8 & 19.1(1)& 0.03 \\
26 & 20:52:48.31 & +55:14:21.8 & 0.5 & $>$23.5 & $>$22.5 & $>$22.3 & 23.00(11) & $>$26.1 & $>$23.5 & $>$15.8 & 17.98(7)& 0.01 \\
30 & 20:51:29.88 & +55:15:48.2 & 1.4 & $>$23.5 & $>$22.5 & $>$22.3 & $>$20.7 & $>$24.0 & $>$23.5 & $>$15.8 & 18.1(1)& 0.07 \\
\enddata
\tablenotetext{a}{Source number from Table~\ref{tab:srcs}.}
\tablenotetext{b}{Position of optical/IR source, averaged over all the
  bands in which there were detections.}
\tablenotetext{c}{Position difference between the X-ray and optical/IR source.}
\tablenotetext{d}{For brevity, we give both $g$-band (from LRIS) and
  $g^\prime$-band 
  (from LFC) data in the same column.  The limits of
  24.0-mag are $g$-band, while the limits of $26.1/26.3$-mag are
  $g^\prime$-band.}  
\tablenotetext{e}{The chance of finding a star within $\Delta r$ of the
  X-ray source given the $K_s$ magnitude,
  using the star-count model of \citet{nim+00}.}
\end{deluxetable}
%\end{landscape}

%\begin{landscape}
\begin{deluxetable}{r c c c c c c c c c r}
\setlength{\tabcolsep}{0.05in}

%\rotate
\tablecaption{Optical/IR Matches to X-ray Sources in \snrb\label{tab:matchb}}
\tablewidth{0pt}
\tabletypesize{\scriptsize}
\tablehead{ \colhead{ID\tablenotemark{a}}  & \colhead{$\alpha$\tablenotemark{b}} & \colhead{$\delta$\tablenotemark{b}} & \colhead{$\Delta r$\tablenotemark{c}} & \colhead{$B$} & \colhead{$V$} & \colhead{$R$} & \colhead{$I$} &
  \colhead{$J$} & \colhead{$K_s$} & \colhead{$N(<K_s)$\tablenotemark{d}}\\
 & \multicolumn{2}{c}{(J2000)} & \colhead{(arcsec)} & \multicolumn{6}{c}{(mag)} \\
}
\startdata
1 & 14:43:19.33 & $-$62:28:04.1 & 0.2 & $>22.3$ & $>24.5$ & $>23.9$ &
$>22.5$ & $>15.8$ & 19.60(8) & 0.04\\
2 & 14:43:33.73 & $-$62:29:27.7 & 0.5 &14.52(2) & 13.23(4)  &
12.41(8) & 11.5(1) & 10.62(2) & 9.84(2)  & $<0.01$\\
% 2MASS 14433372-6229277
3 & 14:41:51.73 & $-$62:28:33.4 & 1.0 & $>22.8$ & \nodata &
19.88(6) & \nodata & $>15.8$ & $>14.6$ & \nodata\\
5  & 14:42:19.56 & $-$62:28:34.6 & 0.2 & $>22.3$ & $>24.5$ & $>23.9$ &
$>22.5$ & $>15.8$ & 20.5(4) & 0.05 \\
6 & 14:43:20.72 & $-$62:33:08.4 & 0.7 & 14.49(2) & 13.73(4) &
13.27(8) & 12.63(10) & 12.23(2) & 11.77(3) & $<0.01$\\[0.1in]
% 2MASS 14432072-6233083
8 & 14:43:40.99 & $-$62:31:38.3 & 0.4 & sat & 12.39(4) &
11.92(8) & 11.35(10) & 10.88(2) & 10.50(2) & $<0.01$\\
% 2MASS 14434099-6231383 
9 & 14:43:46.51 & $-$62:24:12.1 & 1.4 & 20.11(3) & 18.58(3) &
17.35(4) & 16.12(7) & 14.50(6) & 13.40(5) & 0.02\\
%    2MASS 14434652-6224121
12 & 14:43:15.26 & $-$62:21:28.7 & 0.7 & $>22.3$ & $>21.8$ & $>25.5$ &
$>19.7$ & $>15.8$ & 19.44(10) & 0.38\\
13 & 14:43:50.38 & $-$62:30:40.8 & 0.9 & 19.71(3) & 17.79(5) &
17.04(8) & 15.55(10) &14.33(5) & 13.31(5) & 0.01 \\
% 2MASS 14435037-6230408
17  & 14:43:21.90 & $-$62:32:19.0 & 1.5 & $>22.3$ & $>21.8$ & $>25.6$
& 20.73(5) & $>15.8$ & 17.71(6) & 0.43\\[0.1in]
19 & 14:42:10.79 & $-$62:22:02.9 & 1.1 & 13.63(2) & 13.10(2) &
12.92(4) & 12.29(7) & 11.52(2) & 11.19(3) & $<0.01$\\
% 2MASS 14421079-6222028 
21 & 14:42:52.36 & $-$62:21:08.1 & 1.7 & $>22.3$ & $>21.8$ &
20.76(2) & $>19.7$ &$>15.8$ & $>14.6$ & \nodata\\
" & 14:42:52.19 & $-$62:21:06.5 & 0.7 & $>22.3$ & $>21.8$ &
19.69(8) & $>19.7$ &$>15.8$ & $>14.6$ & \nodata\\
23 & 14:42:13.11 & $-$62:32:20.6 & 0.2 & 17.83(2) & 16.18(4) &
15.00(8) & 13.76(10) & 12.46(3) & 11.42(3) & $<0.01$\\
% 2MASS 14421311-6232206
26 & 14:42:06.91 & $-$62:22:04.7 & 1.8 & 17.18(2) & 16.26(2) &
15.53(4) & 14.87(7) & 14.02(4) & 13.37(6) & 0.03 \\[0.1in]
% 2MASS 14420691-6222046?
28 & 14:43:20.25 & $-$62:21:38.7 & 1.3 & $>22.3$ & $>21.8$ &
23.14(6) & $>19.7$ & $>15.8$ &  18.44(6) & 0.62\\
" & 14:43:19.88 & $-$62:21:40.2 & 2.0 & $>22.3$ & $>21.8$ &
22.23(6) & $>19.7$ & $>15.8$ & 18.38(6) & 1.41\\
" & 14:43:20.07 & $-$62:21:39.4 & 0.7 & $>22.3$ & $>21.8$ &
18.66(5) & $>19.7$ & $>15.8$ & 16.12(5) & 0.02\\
\enddata
\tablenotetext{a}{Source number from Table~\ref{tab:srcsb}.}
\tablenotetext{b}{Position of optical/IR source, averaged over all the
  bands in which there were detections.}
\tablenotetext{c}{Position difference between the X-ray and optical/IR source.}
\tablenotetext{d}{The chance of finding a star within  $\Delta r$ of the
  X-ray source given the $K_s$ magnitude,
  using the star-count model of \citet{nim+00}.}
\end{deluxetable}
%\end{landscape}

%\begin{landscape}
\begin{deluxetable}{r c c c c c c r}
\setlength{\tabcolsep}{0.05in}
\tablecaption{Optical/IR Matches to X-ray Sources in \snrc\label{tab:matchc}}
\tablewidth{0pt}
\tabletypesize{\scriptsize}
\tablehead{ \colhead{ID\tablenotemark{a}}  & \colhead{$\alpha$\tablenotemark{b}} & \colhead{$\delta$\tablenotemark{b}} &
  \colhead{$\Delta r$\tablenotemark{c}} &\colhead{$r^{\prime}$} & \colhead{$J$} &
  \colhead{$K$} & \colhead{$N(<K_s)$\tablenotemark{d}}\\
 & \multicolumn{2}{c}{(J2000)} & \colhead{(arcsec)} & \multicolumn{3}{c}{(mag)} \\
}
\startdata
% 2MASS 20532897+4326587, LFC sat
1 & 20:53:28.97 & +43:26:58.7 & 0.4 &  sat & 10.36(2) & 10.11(2) & $<0.01$ \\
% NIRC detection only
2 & 20:53:57.97 & +43:24:58.6 & 0.9 &  $>$23.2 & $>$20.8 & 19.68(6) & 0.41\\
% WIRC Ks detection
3 & 20:53:49.20 & +43:25:51.1 & 0.2 & $>$23.2 & $>$20.8 & 18.32(7) & $<0.01$\\
% 2MASS 20525051+4330293, LFC not sat
4 & 20:52:50.51 & +43:30:29.3 & 0.8 &  15.08(7) & 12.99(2) & 12.25(3) & $<0.01$\\
% complex: multiple sources in LFC, J, Ks
5 & 20:53:52.06 & +43:28:10.8 & 1.2 &  $>23.2$ & 17.44(4) & 15.97(2) &0.05\\[0.1in]
'' & 20:53:51.87 & +43:28:09.9 & 1.7 &  18.83(7) & 15.45(2)& 14.39(1)&0.03\\
'' & 20:53:51.99 & +43:28:08.9 & 0.8 &  $>$23.2 & $>$20.8 & 17.95(5) &0.10\\
% NIRC Ks, but near bleed trail
6 & 20:53:06.86 & +43:32:59.3 & 0.5 &  $>$23.2 & $>$20.8 & 19.45(6) &0.11\\
% faint WIRC J, WIRC Ks, plus diffuse?
7 & 20:52:48.32 & +43:32:12.7 & 1.5  & $>$23.2 & 20.67(13) & 17.64(3)&0.27\\
% faint WIRC J, WIRC/NIRC Ks, plus second faint source in NIRC Ks?
10 & 20:53:51.79 & +43:25:37.1 & 0.9  & $>$23.2 & 20.69(15) & 17.10(4)&0.06\\[0.1in]
'' & 20:53:51.65 & +43:25:37.2 & 1.3  & $>$23.2 & $>$20.8 & 21.0(2)&1.53\\
% WIRC J,Ks off-center
11 & 20:53:17.76 & +43:22:06.6 & 1.1  & $>$23.2 & 18.43(4) & 15.332(10)&0.03\\
% 2MASS 20535317+4327520, LFC sat/in gap 
12 & 20:53:53.17 &  +43:27:52.0 & 0.7  & sat & 13.39(3) & 11.88(2) & $<0.01$\\
% 2MASS 20530275+4332085, LFC sat; may be multiple IR sources
14 & 20:53:02.75 & +43:32:08.5 & 1.5  & sat &  11.97(2) & 11.61(2) & $<0.01$ \\
% WIRC Ks
15 & 20:53:06.49 & +43:28:21.9 & 0.3  & $>$23.2 & $>$20.8 & 18.37(5)&0.02\\[0.1in]
% 2MASS 20533561+4334275, LFC not sat
16 & 20:53:35.61 & +43:34:27.5 & 0.6  & 18.22(7) & 14.84(4) & 14.01(7) & $<0.01$\\
% >= 2 WIRC/NIRC Ks sources, in bleed trail
18 & 20:53:46.08 & +43:33:46.4 & 0.9  & $>$23.2 & $>$20.8 & 20.91(14) &0.71 \\
'' & 20:53:45.89 & +43:33:44.8 & 1.8  & $>$23.2 & $>$20.8 & 18.08(3) &0.56 \\
% 2MASS 20524255+4324505, LFC not sat
22 & 20:52:42.55 & +43:24:50.5 & 0.8  & 19.40(7) & 14.73(7) & 13.26(5) & $<0.01$\\
% WIRC/NIRC Ks, in bleed trail
23 & 20:53:32.25 & +43:23:55.2 & 0.5  & $>$23.2 & $>$20.8 & 18.23(4) &0.05\\
\enddata
\tablenotetext{a}{Source number from Table~\ref{tab:srcsc}.}
\tablenotetext{b}{Position of optical/IR source, averaged over all the
  bands in which there were detections.}
\tablenotetext{c}{Position difference between the X-ray and optical/IR source.}
\tablenotetext{d}{The chance of finding a star within  $\Delta r$ of the
  X-ray source given the $K_s$ magnitude,
  using the star-count model of \citet{nim+00}.}
\end{deluxetable}
%\end{landscape}

\clearpage

%\begin{landscape}
\begin{deluxetable}{r c c c c c c c c r}
\setlength{\tabcolsep}{0.05in}
\tablecaption{Optical/IR Matches to X-ray Sources in \snrd\label{tab:matchd}}
\tablewidth{0pt}
\tabletypesize{\scriptsize}
\tablehead{ \colhead{ID\tablenotemark{a}}  &
  \colhead{$\alpha$\tablenotemark{b}} &
  \colhead{$\delta$\tablenotemark{b}} & \colhead{$\Delta
    r$\tablenotemark{c}}  &
  \colhead{$R$} & \colhead{$I$} & \colhead{$r^{\prime}$} &
  \colhead{$J$} & \colhead{$K_s$} & \colhead{$N(<K_s)$\tablenotemark{d}}\\ 
 & \multicolumn{2}{c}{(J2000)} & \colhead{(arcsec)} & \multicolumn{5}{c}{(mag)} \\
}
\startdata
1 & 01:28:30.58 & +63:06:29.7 & 0.4 & 16.59(4) & sat &17.76(5) & 13.96(12) & 12.09(8) &$<0.01$\\
3 & 01:27:36.51 & +63:03:45.5 & 0.6 & $>$23.0 & $>$21.3 &  $>$24.5 &blended & 17.82(3)&0.03\\
4 & 01:28:07.70 & +63:01:50.6 & 0.2 & 22.18(9) & 20.69(10) & 22.64(7) & 19.45(6) & 17.74(4)&$<0.01$\\
5 & 01:28:42.25 & +63:08:25.8 & 0.4 & $>$23.0 & $>$21.3  & $>$24.5 & $>$21.1 & 18.80(4)&0.02\\
8 & 01:29:25.42 & +63:05:34.2 & 1.1 & 21.80(8) & 20.48(8)  & 22.43(6) & 18.39(2) & 16.86(3)&0.04\\[0.1in]
12 & 01:29:10.80 & +63:10:14.4 & 0.6 & $>$23.0 & $>$21.3 & $>$24.5 & 18.91(5) & 16.60(3)&0.01\\
13 & 01:28:13.86 & +63:06:21.2 & 0.3 & $>$23.0 & $>$21.3 & $>$24.5 & $>$21.1 & 19.85(8)&0.02\\
14 & 01:27:55.35 & +63:01:41.5 & 0.8 & $>$23.0 & $>$21.3 & $>$24.5 & 20.32(7) & 17.08(3)&0.03\\
15 & 01:28:37.77 & +63:06:03.3 & 0.7 & $>$23.0 & $>$21.3 & $>$24.5 &blended & 18.53(8)&0.06\\
17 & 01:27:35.74 & +63:02:42.4 & 0.8 & $>$23.0 & $>$21.3 & $>$24.5 & $>$21.1 & 18.17(8)&0.06\\[0.1in]
18 & 01:29:17.27 & +63:02:42.3 & 1.0 & $>$23.0 & $>$21.3 & $>$24.5 & 16.717(14) & 14.774(15)&0.01\\
22 &01:28:01.81 & +63:12:44.2 & 0.3 & $>$23.0 & $>$21.3  & $>$24.5 & $>$21.1 & 20.34(13)&0.03\\
" &01:28:01.93 & +63:12:45.7 & 1.4 & $>$23.0 & $>$21.3 & $>$24.5 & $>$21.1 & 20.81(15)&0.75\\
%57 & \nodata & \nodata & \nodata &  $>$23.0 & $>$21.3 & $>$24.5 & $>$21.1 & $>$22.1\\
\enddata
\tablenotetext{a}{Source number from Table~\ref{tab:srcsd}.}
\tablenotetext{b}{Position of optical/IR source, averaged over all the
  bands in which there were detections.}
\tablenotetext{c}{Position difference between the X-ray and optical/IR source.}
\tablenotetext{d}{The chance of finding a star within $\Delta r$ of the
  X-ray source given the $K_s$ magnitude,
  using the star-count model of \citet{nim+00}.}
\end{deluxetable}
%\end{landscape}

\clearpage

This will not work so well for the sources here, as the X-ray fluxes
are significantly less than that of CXO~J232327.8+584842 and
consequently the location in Figure~\ref{fig:xopt} for a random
optical/IR counterpart would be closer to the Orion/CDF loci.

\begin{figure}
% uses starcounts.m
% which needs get_starcounts.pl to figure out the data
\plotone{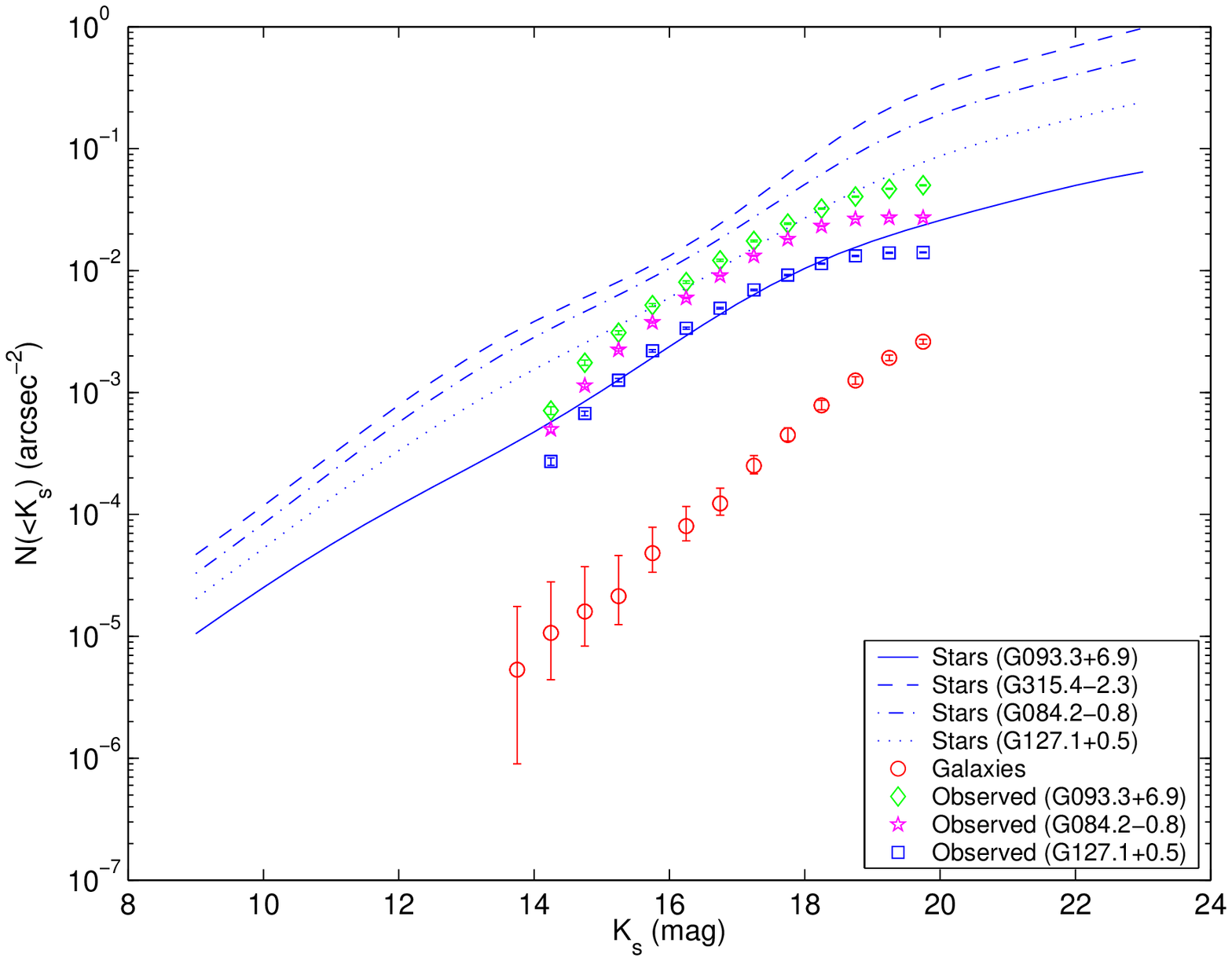}
\caption{Number density of IR sources.  The number 
  per square arcsecond brighter than a given $K_s$ magnitude is
  plotted against $K_s$ magnitude.  The data for stars are from
  the model of \citet{nim+00}, and are
  plotted for  \snr\ (solid line),  \snrb\ (dashed line), \snrc\
  (dash-dotted line), and \snrd\ (dotted line).  
For galaxies (circles) the counts are from the K-20 project
  \citep{cmd+02}.  The observed data for \snr, \snrc, and \snrd\  are from the WIRC
  observations, and are likely not complete for $K_s \gsim 19$.  The
  predicted star counts are within a factor of 3 of the observed counts.
 }
\label{fig:stars}
\end{figure}

\begin{figure}
% plot_rmk.m
\plottwo{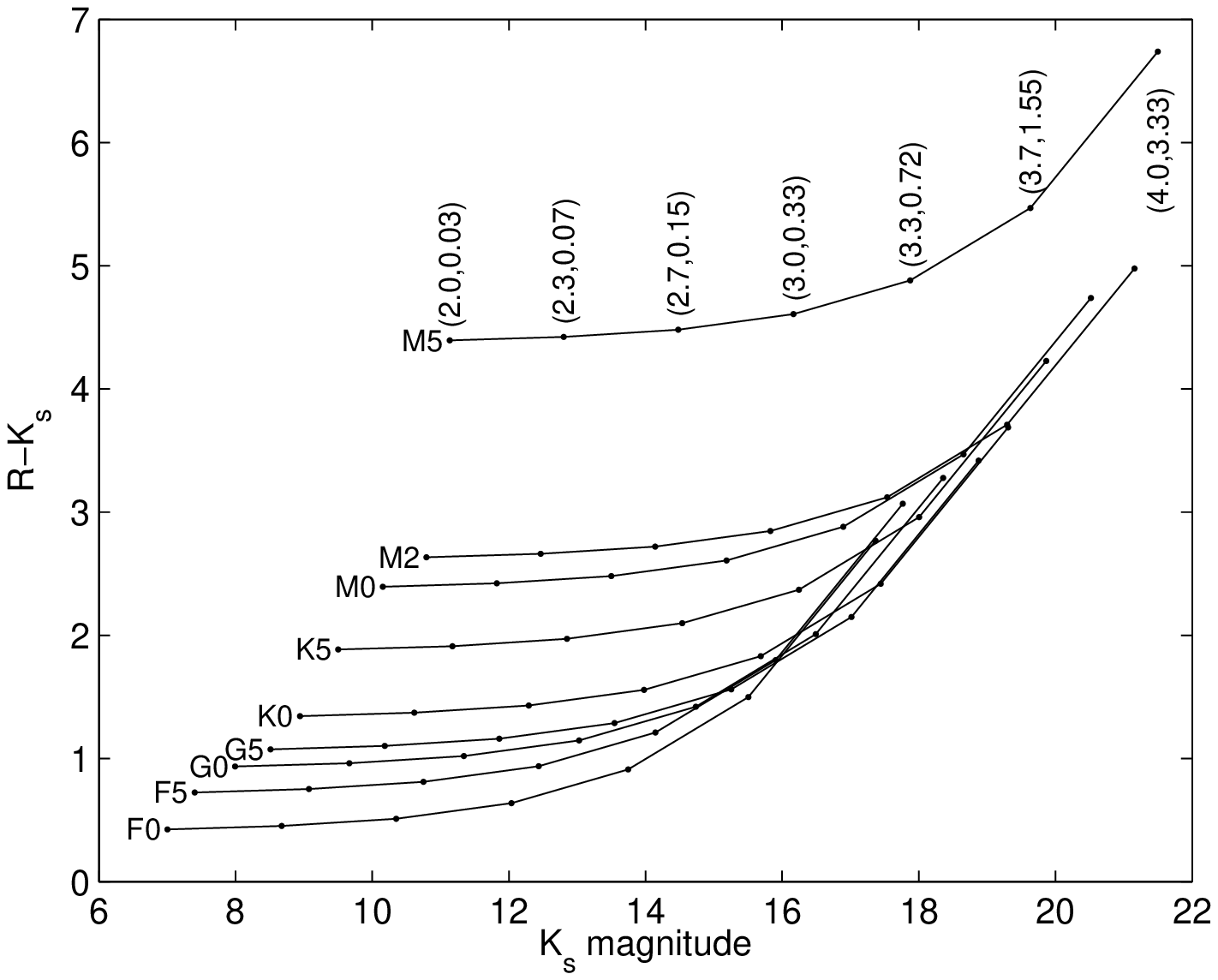}{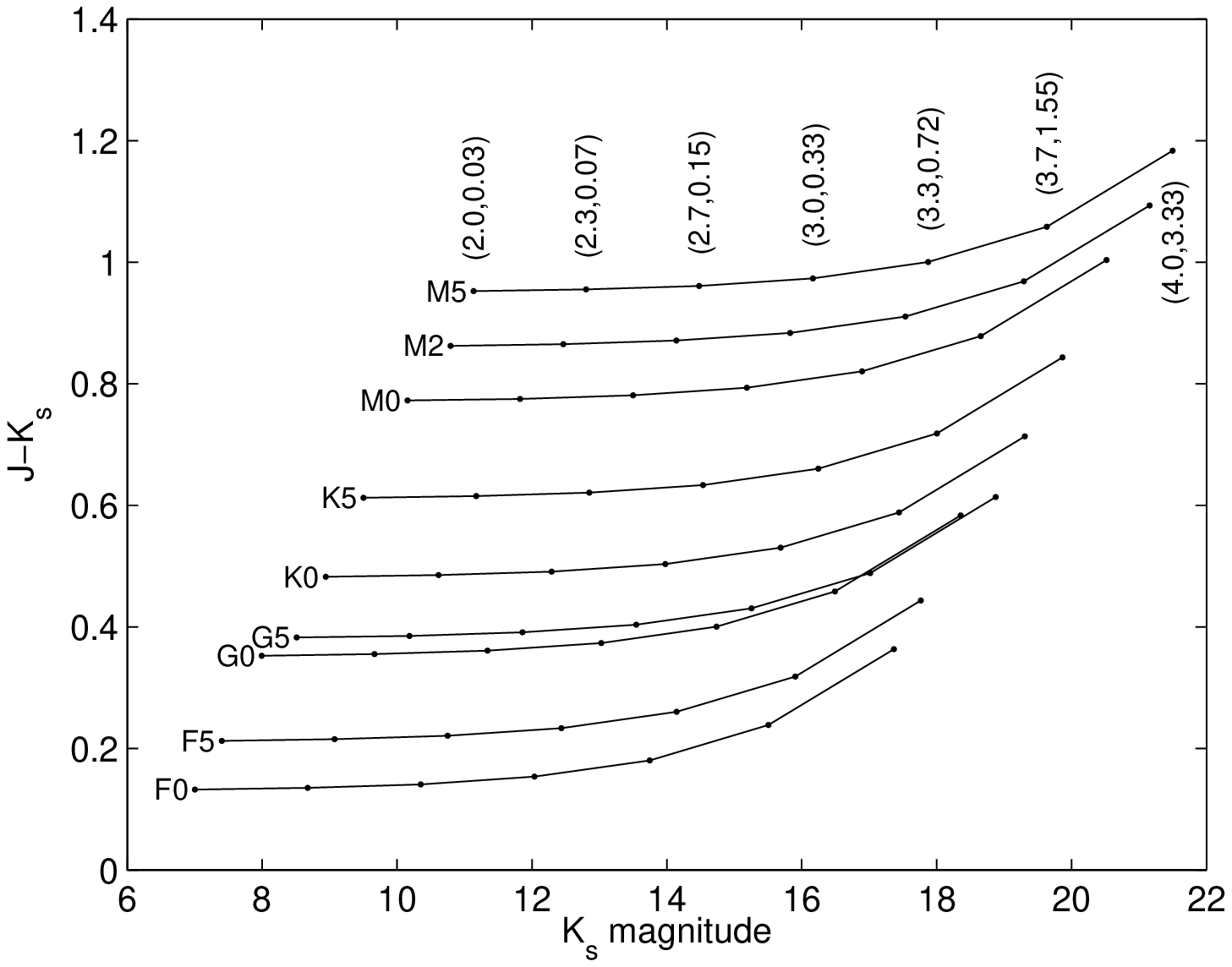}
\caption{Stellar color vs.\ magnitude for main-sequence stars, from
  \citet[][pp.\ 151 \& 388]{allen}: we have assumed that
  $A_{V} \propto d$, normalized to $A_{V}=1$~mag at $d=3$~kpc (roughly
  appropriate for \snr\ and \snrb).  Left: $R-K_s$ color vs.\ $K_s$
  magnitude. Right: $J-K_s$ color vs.\ $K_s$
  magnitude.  Shown are tracks for luminosity class V stellar
  types F0, F5, G0, G5, K0, K5, M0, M2, and M5 (as labeled).  Points along the
  tracks are labeled by $(\log_{10} (d/{\rm pc}),A_{V})$, and the
  tracks progress from (2.0,0.03) to (4.0,3.3).    The maximum values of $A_V$
  expected along the lines of sight to \snr\ and \snrb\ are 2~mag and
  5~mag, respectively \citep{ps95}.  The majority of the stars from
  \citet{nim+00} are of spectral type M0--M3.
}
\label{fig:color}
\end{figure}

\begin{figure}
% plot_rmk.m
\plottwo{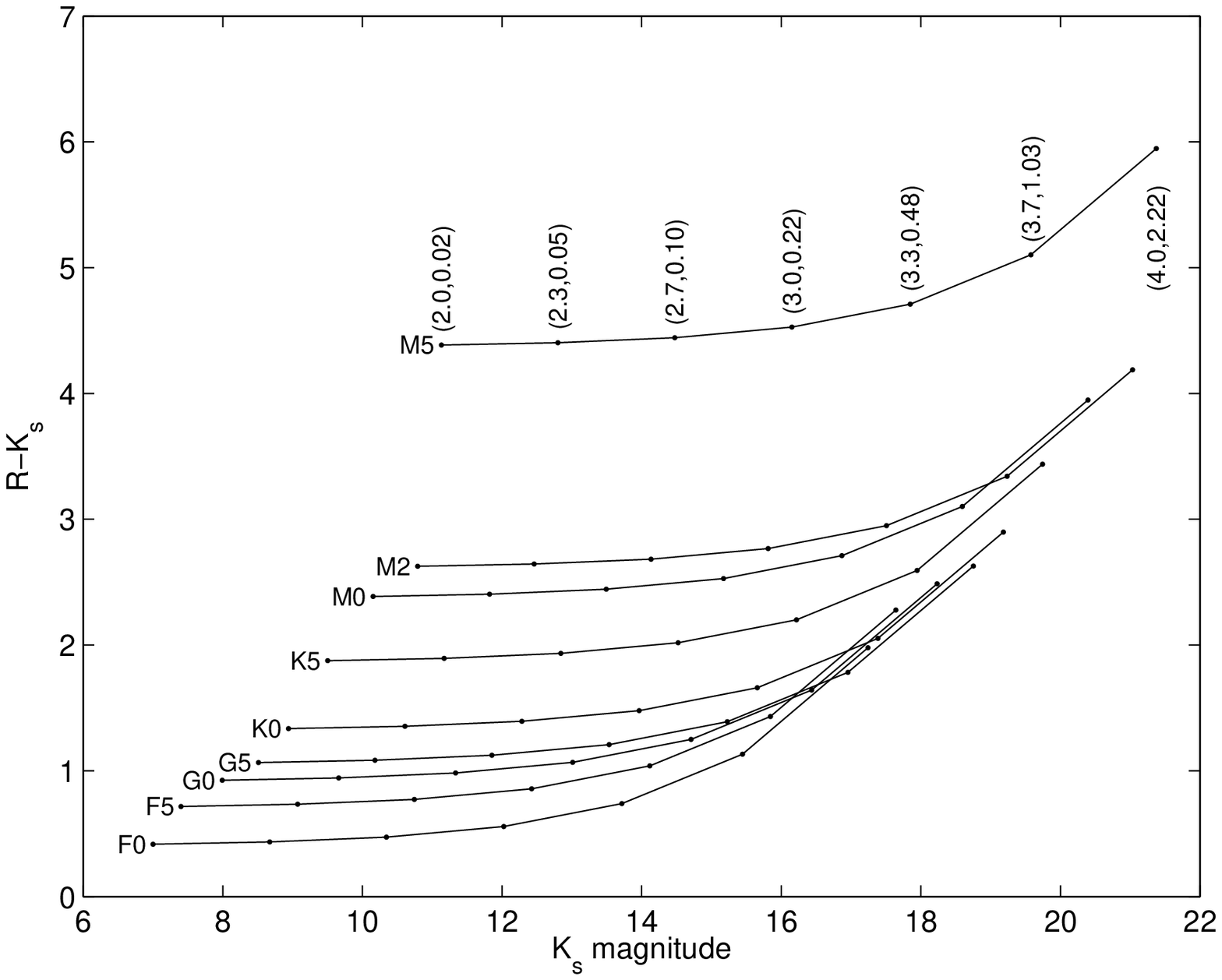}{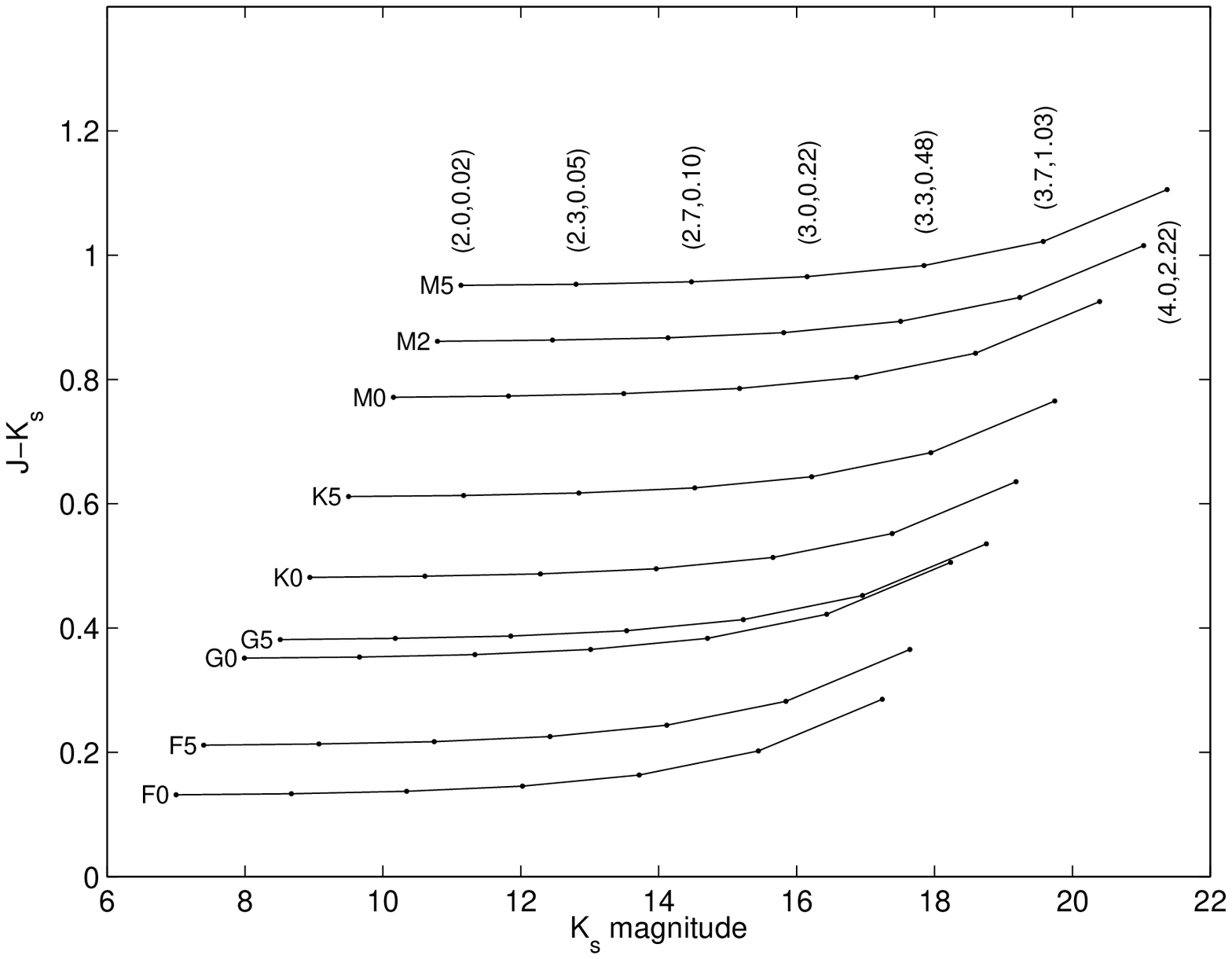}
\caption{Stellar color vs.\ magnitude for main-sequence stars, from
  \citet[][pp.\ 151 \& 388]{allen}: we have assumed that $A_{V}
  \propto d$, normalized to $A_{V}=1$~mag at $d=4.5$~kpc (roughly
  appropriate for \snrc). Otherwise the figures are the same as
  Figure~\ref{fig:color}.  The maximum value of $A_V$ expected along
  the line of sight to \snrc\ is 5~mag \citep{ps95}.   }
\label{fig:colorc}
\end{figure}

\begin{figure}
% plot_rmk.m
\plottwo{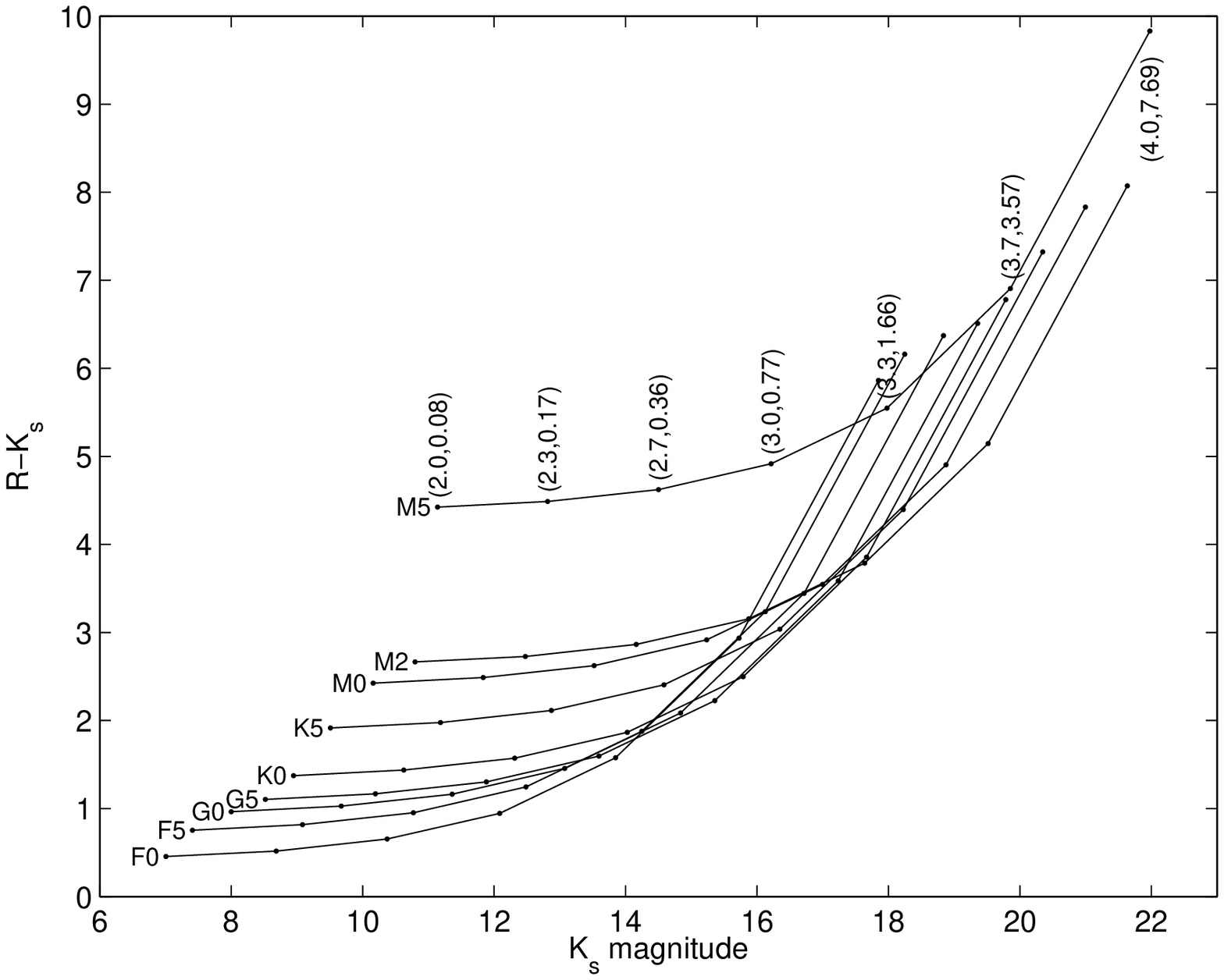}{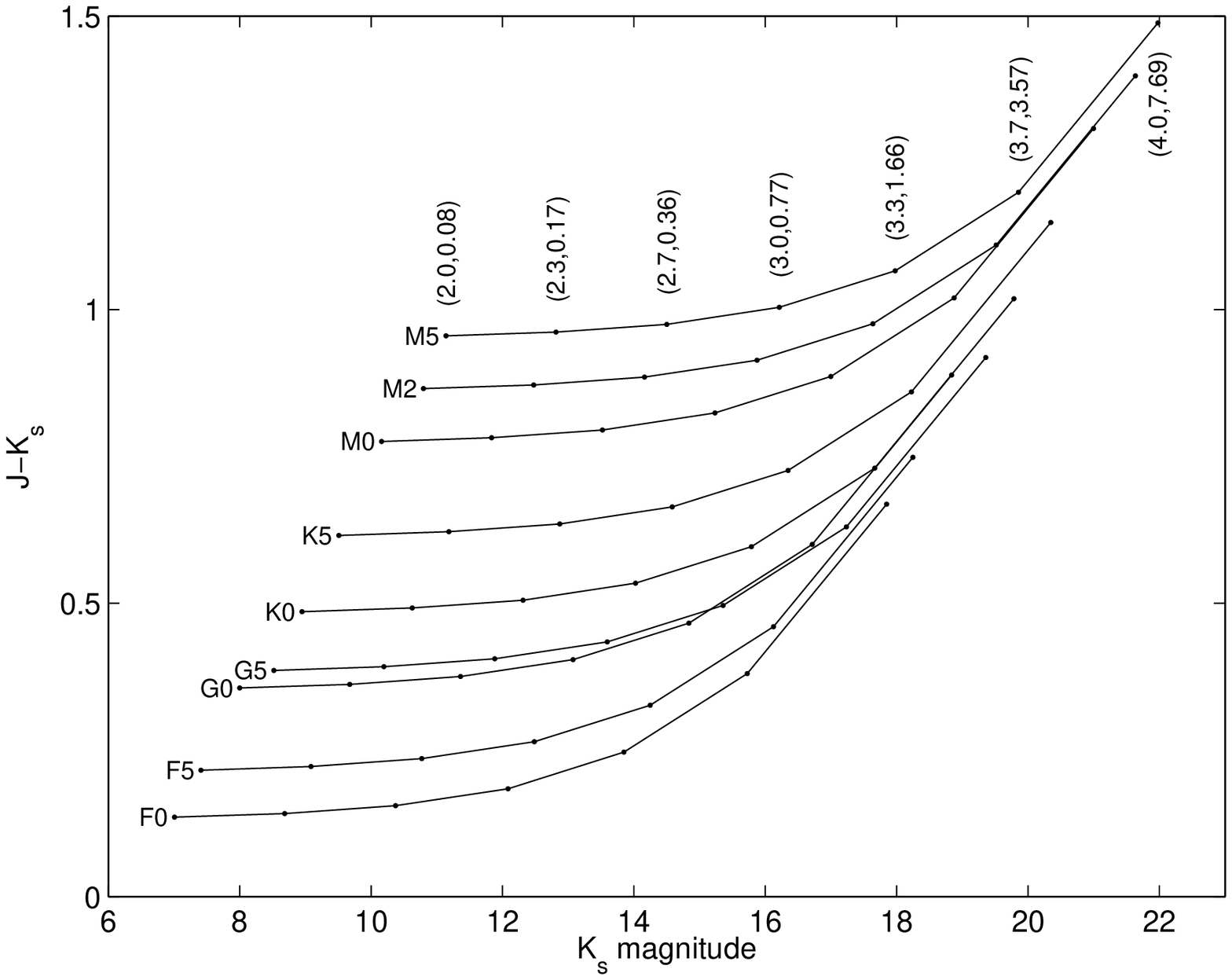}
\caption{Stellar color vs.\ magnitude for main-sequence stars, from
  \citet[][pp.\ 151 \& 388]{allen}: we have assumed that
  $A_{V} \propto d$, normalized to $A_{V}=1$~mag at $d=1.3$~kpc (roughly
  appropriate for \snrd).  Otherwise the figures are the same as
  Figure~\ref{fig:color}.  The maximum value of $A_V$
  expected along the line of sight to \snrd\ is
  5~mag \citep{ps95}.  
}
\label{fig:colord}
\end{figure}

To determine the probabilities of random false associations, we have
used the IR star-counts as modeled by \citet{nim+00} and galaxy-counts
observed by \citet{cmd+02}.  We plot in Figure~\ref{fig:stars} the
number of sources brighter than a given $K_s$ magnitude per square
arcsecond for all four SNRs (due to their lower Galactic latitudes,
the numbers for \snrb\ and \snrc\ are a factor of 5--15 higher than
those for \snr, and \snrd\ is in between).  We chose to plot the
counts in the $K_s$ band as these are least affected by extinction.
In Tables~\ref{tab:match}--\ref{tab:matchd} we also give, for sources
with $K_s$ detections, the expected number of random stars\footnote{As
seen in Figure~\ref{fig:stars}, the chance of random associations with
a galaxy is quite small compared to the chance of association with a
star, and is usually negligible.}  brighter than the detected source
within $\Delta r$ (the distance between the $K_s$ source and the X-ray
source).  For most cases, especially the bright 2MASS stars, these
numbers are very low ($<0.01$), but for a few of the fainter sources
the numbers can become significant ($\gsim 0.5$).  For those cases
with roughly even chances of random associations we also examine (in
\S\S~\ref{sec:indva}, \ref{sec:indvb}, \ref{sec:indvc}, and
\ref{sec:indvd}) other factors like the source spectrum and optical/IR
colors, where available.  In Figures~\ref{fig:color}--\ref{fig:colord}
we show plots of expected stellar color vs.\ magnitude for different
distances and extinctions.  Using these plots can help us determine
the approximate type of the stellar counterparts in
Tables~\ref{tab:match}--\ref{tab:matchd}, as demonstrated in
Sections~\ref{sec:indva}, \ref{sec:indvb}, \ref{sec:indvc}, and
\ref{sec:indvd}.  These plots are not exact, though: they assume
$A_{V} \propto D$ and they are only for main-sequence stars.  When
possible, the $J-K_s$ color should be used instead of other
combinations as it is least susceptible to reddening effects
($E(J-K_s)=0.18$, compared to $E(R-K_s)=0.71$), but it is less
sensitive to stellar type than other combinations.

\begin{figure}
% plot_offets.m
\plottwo{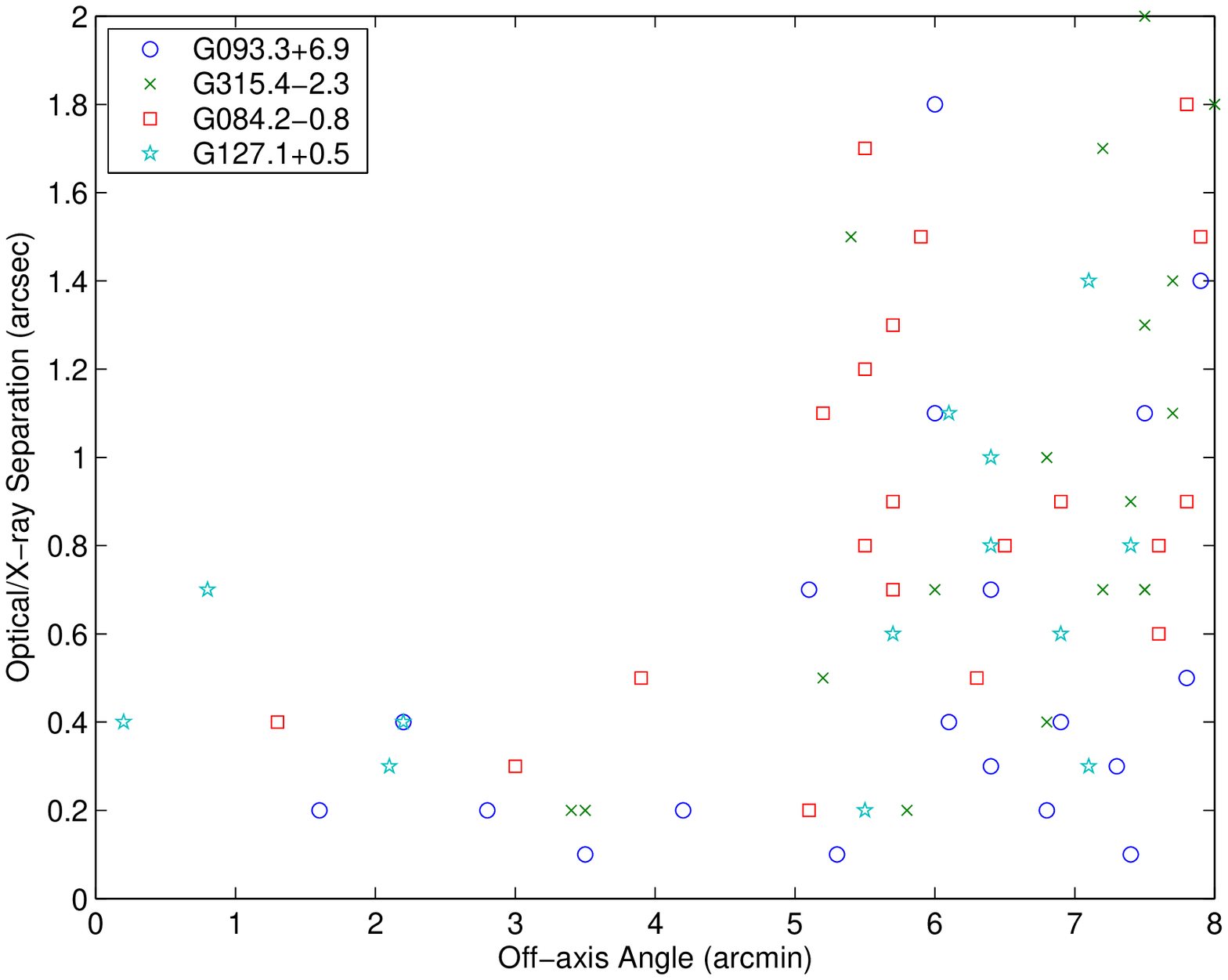}{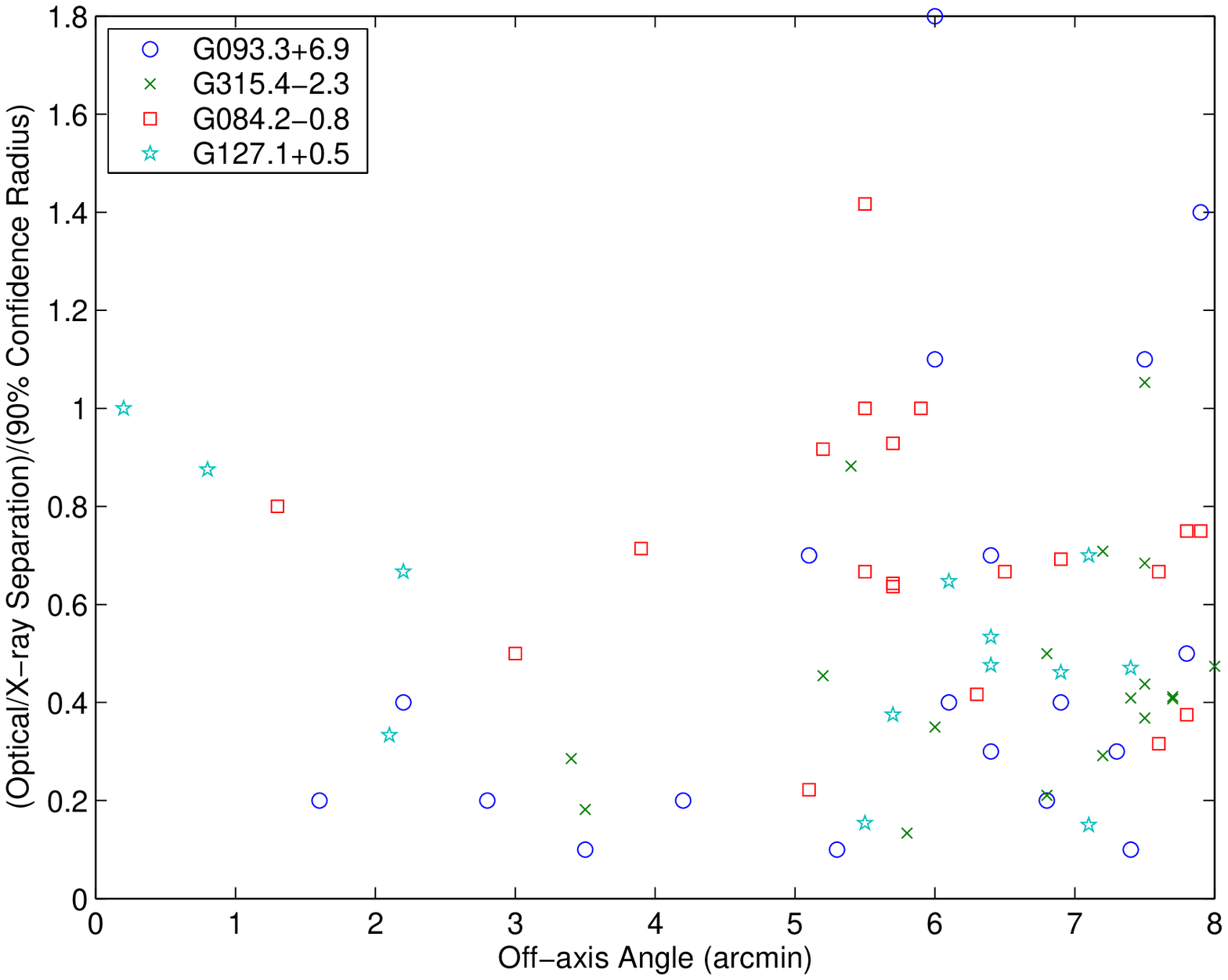}
\caption{Distribution of offset between X-ray sources and their
  proposed optical/IR counterparts as a function of off-axis angle.
  Left: offset (in arcsec) vs.\ off-axis angle (in arcmin).  Right:
  offset normalized to the 90\% radius (from
  Tabs.~\ref{tab:srcs}--\ref{tab:srcsd}) vs.\ off-axis angle (in
  arcmin).  Sources from \snr\ are circles, those from \snrb\ are x's,
   those from \snrc\ are squares, and those from \snrd\ are stars.  }
\label{fig:offset}
\end{figure}

\begin{figure}
% plot_offets.m
\plotone{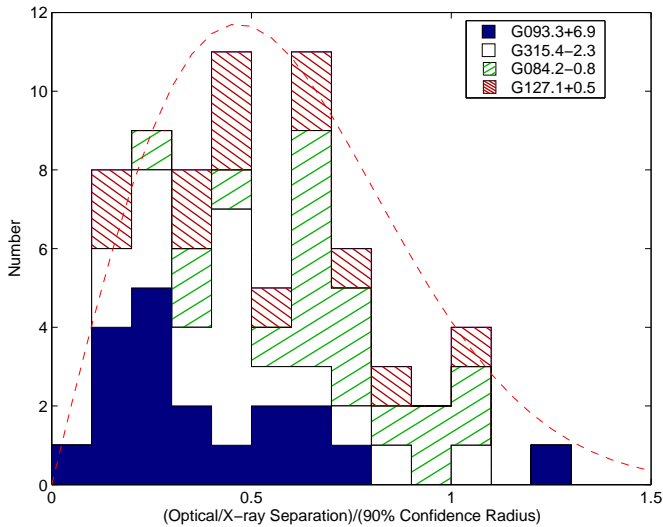}
\caption{Histogram of the distribution of offset between X-ray sources
  and their proposed optical/IR counterparts normalized to the 90\%
  radius (from Tabs.~\ref{tab:srcs}-- \ref{tab:srcsd}).  Sources from
  \snr\ are dark blue, those from \snrb\ are in white, those from
  \snrc\ are the green hatched region, and those from \snrd\ are the
  maroon hatched region.  The red dashed line is the expected
  distribution ($f(r)\propto r \exp(-r^2)$).  }
\label{fig:offsethst}
\end{figure}

In Figures~\ref{fig:offset} and \ref{fig:offsethst} we examine the
X-ray-to-optical offset.  We see that even for sources with large
offsets, the values are reasonably consistent with the expected
distribution.  Compare to Figure~1 of \citet{bcc+03}.

We can see some general trends from the discussions for the different
SNRs (\SS~\ref{sec:indva}, \ref{sec:indvb}, \ref{sec:indvc}, and
\ref{sec:indvd}) and from Figures~\ref{fig:xopt}--\ref{fig:xoptd}.
Namely, \snr\ and \snrd\ have the highest fractions of probable
galaxies, with \snrc\ intermediate, and \snrb\ the lowest (summarized
in Tab.~\ref{tab:class}).  This is largely due to the differing
Galactic latitudes and longitudes of the four SNRs, and is also seen
somewhat in Figure~\ref{fig:stars}.  \snrb, at a relatively low $|b|$
and well in quadrant IV, is along the line of sight of many stars and
the short exposure time of the observation helps to keep the number of
background galaxies down.  \snrc\ is at a lower $|b|$, but it's
position near $l=90\degr$ and at the edge of a spiral arm \citep{fg93}
lowers the number of stars, while the longer exposure identified more
galaxies (many of whom are heavily reddened).  The total \nh\ along
this like of sight helps to delineate galaxies and stars even without
optical counterparts, as can be seen from Figure~\ref{fig:hardc}.
Finally, \snr\ is at a relatively high $|b|$ and also does not look
toward the inner Galaxy, and \snrd\ is decidedly toward the outer
Galaxy, so the X-ray sources are predominantly background galaxies.

\begin{deluxetable}{c c c c}
\tablecaption{Classifications of X-ray Sources in  SNRs \Ga, \Gb, \Gc, and \Gd\label{tab:class}}
\tablewidth{0pt}
\tablehead{\colhead{SNR} & \colhead{Stars} & \colhead{Galaxies} &
  \colhead{Uncertain\tablenotemark{a}} \\}
\startdata
G093.3+6.9 & 3 & 15 & 0\\
G315.4$-$2.3 & 8 & 4 & 3: 17, 21, 28\\
G084.2$-$0.8 & 6 & 8 & 2: 5, 10\\
G127.1+0.5 & 0 & 9 & 3: 12, 18, 22\\
\tableline
Total & 17 & 36 & 8\\
\enddata
\tablecomments{Classifications follow the discussion in
  \S\S~\ref{sec:indva}, \ref{sec:indvb},
\ref{sec:indvc}, and \ref{sec:indvd}.}
\tablenotetext{a}{Uncertain sources are those for which the
  classification as a star or galaxy was unclear.  This encompasses
  both sources that definitely have counterparts but where the type is
  uncertain  and 
  sources where the counterpart itself is uncertain.  The uncertain
  sources for each SNR are listed.}
\end{deluxetable}

\section{Results}
\label{sec:results}
Here we discuss the results of applying the techniques in
\S~\ref{sec:obs} to the first four SNRs from our \chandra\ sample:
SNRs~\Ga, \Gb, \Gc, and \Gd.
For each SNR, we first discuss its general properties, then details of
the X-ray analysis, optical observations and analysis, and finally
identification of counterparts.

\subsection{\snr}
\label{sec:snra}
The radio source \snr\, (also known as DA~530) was first identified as
a SNR by \citet{rc76}, whose radio observations showed a shell
$27\arcmin$ in diameter with bright rims and high polarization
\citep*{hsp80} --- see Figure~\ref{fig:radio}.  The kinematic distance
based on \ion{H}{1} emission gives a distance of $2.5\pm 0.4$~kpc, but
allowing for non-circular motion relaxes the distance limits and
constrains the distance to be 1.0--3.5~kpc.

While infrared and optical emission have been only marginally
detected, \snr\ has been detected by \rosat\ \citep{lrr+99}.  Based on
fits to the X-ray spectrum, \citet{lrr+99} prefer a distance of
3.5~kpc.  In X-rays, it appears superficially similar to SN~1006,
which has been suggested to be the remnant of a Type~Ia~SN
(\citealt{fwlh88}, \citealt*{apg01}).  Taken together with the high Galactic latitude of
$7\degr$, some authors believe that \snr\ is also the young ($\approx
5000$~yr) remnant from a Type~Ia~SN \citep[e.g.,][]{lrr+99}, but there
are some problems with this, as none of the X-ray spectral models give
fully consistent results and they imply that the SNR may have occurred
in a especially low-density bubble such as might exist around a massive star
(i.e.\ the progenitor of a Type~II SN).  

Most of the X-ray fits in \citet{lrr+99} assumed a hydrogen column
density of $\nh=\expnt{2.1}{21}\mbox{ cm}^{-2}$, which is consistent
with the value inferred from the \ion{H}{1} data.  This is the nominal
value that we adopt for this source below.  However, some of the fits
required larger values of $\nh$, up to $\expnt{6}{21}\mbox{ cm}^{-2}$,
while the total Galactic column density\footnote{Determined using
  Colden, 
\url{http://asc.harvard.edu/toolkit/colden.jsp}.} in this direction is
only $\expnt{4}{21}\mbox{ cm}^{-2}$.

\citet*{llc98} searched \snr\ unsuccessfully for a radio pulsar down to
a limit of 0.8~mJy at 606~MHz, implying a luminosity limit of $Sd^2 <
9.8\mbox{ mJy kpc}^2$.  Extrapolating to a frequency of 1400~MHz using
an average spectral index of $-1.8$ \citep{mkkw00}, this is still a
factor of almost 10 brighter than the luminosities of the emerging
class of faint radio pulsars such as PSR~J0205+6449 in 3C~58
\citep{csl+02}.  \citet{lrr+99} identified six point sources in the
\rosat\ PSPC data, three of which are inside the remnant (see
\S~\ref{sec:xraya}), but they conclude that none of these is likely
the compact remnant of the explosion as they are all relatively hard
(however, neutron stars such as the Crab can also have hard spectra).

\subsubsection{X-ray Observations}
\label{sec:xraya}
We observed \snr\ with the \textit{Chandra X-ray Observatory}
\citep{wtvso00} on 2001 December 17.  Based on examination of the
\rosat/PSPC data, we determined the center of the remnant\footnote{The
geometric centers of SNRs were identified by eye, with the
understanding that the actual site of the SNR is not always at the
geometric center \citep[e.g.,][]{gvar02}.} to be at J2000 position
$20^{\rm h} 52^{\rm m} 14^{\rm s}$, $+55\degr 20\arcmin 30\arcsec$,
about $1.5\arcmin$ away from the nominal position from
\citet{green01}.  The aim-point was on the ACIS-I imaging array, and
the final exposure time was 16.4~ksec.  See Figure~\ref{fig:radio} for
the placement of the ACIS-I detector relative to the SNR.  A smoothed
of the data is shown in Figure~\ref{fig:acis} with the X-ray sources
that we identified (Tab.~\ref{tab:srcs}) labeled.

The analysis of the \chandra\ data proceeded according to
Section~\ref{sec:proc}.  In Figure~\ref{fig:hard} we plot the H-band
counts vs.\ the L-band counts for the sources in Table~\ref{tab:srcs}.
We also plot the lines for sources with power-laws having $\Gamma=0,\,
1.0,\, 1.5,\,2.0$ and blackbody models having
$kT_\infty=0.2,\,0.4,\,1.0$~keV, all with $\nh=\expnt{2}{21}\mbox{
cm}^{-2}$, and the median and 25-/75-percentiles of count-ratios
($C_{\rm H}/C_{\rm L}$) for the sources from the CDF and Orion.  Most
of the sources are consistent with power-laws having indices from
0.0--2.0, such as one would expect for energetic pulsars like the Crab
or for AGN.  A few sources (\snr:1, \snr:7, and \snr:16) are softer,
with implied temperatures (for blackbody models) of $kT_\infty\approx
0.4$~keV.

Three of the sources were also detected by \citet{lrr+99} in \rosat\
data.  Source \snr:1 is source 6 from \citet{lrr+99},  \snr:2 is source 5
from \citet{lrr+99}, and \snr:7 is source 4 from \citet{lrr+99}.
While it is difficult to compare directly because of the small numbers
of counts involved, the count-rates and hardness ratios are roughly
comparable between those that we observed here and those predicted by
converting the \rosat\ count-rates to \chandra\  using
\texttt{W3PIMMS}\footnote{See \url{http://heasarc.gsfc.nasa.gov/Tools/w3pimms.html}.}

\subsubsection{Optical/IR Observations}
\label{sec:opta}
We observed the field of \snr\ a number of times with a number of
instruments, as described in Table~\ref{tab:opt}.  The aim of the
observations was to identify counterparts to the X-ray sources in
Table~\ref{tab:srcs} with progressively deeper exposures
(\S~\ref{sec:cpt}).  Data reduction for the optical data used standard
\texttt{IRAF} routines to subtract the bias, flat-field, and then
combine separate exposures.  For the LFC and LRIS-B data, where
significant focal-plane distortion prevented simple addition of data
and where there are multiple CCDs, we used the \texttt{IRAF}
\texttt{MSCRED} package to flatten each image with custom distortion
maps prior to addition.  For the infrared data, we subtracted dark
frames, then produced a sky frame for subtraction by taking a sliding
box-car window of 4 exposures on either side of a reference exposure.
We then added the exposures together, identified all the stars, and
produced masks for the stars that were used to improve the sky frames
in a second round of sky subtraction.

We determined $BVRI$ photometric zero-points for the P60CCD data using
observations of the Stetson
fields\footnote{\url{http://cadcwww.dao.nrc.ca/cadcbin/wdbi.cgi/astrocat/stetson/query}}
PG1657 and NGC 6823 \citep{s00}.  We then determined zero-points for the other
optical data referenced to the P60CCD observations, employing
appropriate transformations \citep{jor94,stk+02}.

We determined a $K_s$ zero-point for the P60IR
data using observations of the standard stars SJ 9101, SJ 9166, SJ
9177, and  SJ 9188 \citep{pmk+98}.  For the NIRC observations, we
determined zero-points using 1--4 2MASS \citep{2mass} stars in each
field (the $K_s$ magnitudes include a 0.3-mag
systematic uncertainty arising from zero-point calibration).  For the
WIRC observations, we determined zero-points using 400 unsaturated
2MASS stars.

We performed absolute astrometry on the R-band P60CCD data, the LFC
data, the LRIS data, and the WIRC data.  After applying distortion
corrections to the LFC and LRIS data (we did not use any distortion
for the P60CCD data or WIRC), we solved for plate-scale, rotation, and
central position relative to stars from version 2.2 of the Guide Star
Catalog
(GSC-2.2\footnote{\url{http://www-gsss.stsci.edu/support/data\_access.htm}})
for all but WIRC and relative to 2MASS stars for WIRC, getting
residuals in each coordinate of $0\farcs22$ (427 stars), $0\farcs08$
(135 stars), $0\farcs13$ (613 stars), and $0\farcs10$ (530 stars) for
P60CCD, LRIS, LFC, and WIRC respectively.  For the remaining data
(ESI, P60IR, and NIRC) we used non-saturated stars from the other
observations to boot-strap the astrometry, getting typical residuals
of $<0\farcs05$ in each coordinate.

\subsubsection{Notes on Individual Sources}
\label{sec:indva}
\begin{description}
\item[\snr:1]  This source is almost certainly an active star:  the
  infrared colors are indicative of class G0 or so, the source falls
  on the locus of stars in Figure~\ref{fig:xopt}, and the variability
  (Fig.~\ref{fig:lc}) is typical of that seen for stars
  \citep*{mmp00}.  The X-ray luminosity expected of such a star,
  $10^{29-31}\mbox{ ergs s}^{-1}$ \citep{kc00} is implies that it
  is $\sim 500$~pc away, consistent with its $K_s$ magnitude (Fig.~\ref{fig:color}).
\item[\snr:2] This is likely a galaxy, given its position in
  Figure~\ref{fig:xopt} and its constant X-ray flux (Fig.~\ref{fig:lc}).
\item[\snr:4]  This is likely a galaxy, given its position in
  Figure~\ref{fig:xopt}.
\item[\snr:5]  This is likely a galaxy, given its position in
  Figure~\ref{fig:xopt}.
\item[\snr:6] The IR counterpart places it midway between the stars
  and galaxies in Figure~\ref{fig:xopt}.  It could be a spurious
  counterpart, but there are other sources in its vicinity in
  Figure~\ref{fig:xopt}.  Based on its hard spectrum and extremely red
  $R-K_s$ colors, it is likely a galaxy.
\item[\snr:7] While redder than \snr:1, it is otherwise similar,
  suggesting that this is very likely also an
  active star of type M2 or so.
\item[\snr:8] The optical counterpart is near the edge of the error
  circle, but it is far off-axis and therefore this is not unexpected.
  This source is otherwise consistent with being a galaxy.
\item[\snr:9] This source has the most extreme X-ray-to-IR flux ratio
  of the sources in Table~\ref{tab:match}, but it most likely is not
  the associated neutron star.  The optical counterpart has a very
  small ($0\farcs2$) offset from the X-ray source, suggesting that
  while it may be 2~mag fainter than other counterparts it is still
  likely to be a real association.  It is still reasonably consistent
  with the CDF locus and it is somewhat red ($R-K_s>3$), contrary to
  known neutron stars \citep{kvkm+03} but similar to that of other
  galaxy candidates like \snr:10.  Also, the X-ray spectrum is
  moderately hard, consistent with the power-law expected from an AGN.
  Without an optical counterpart, it might be a candidate for a
  low-luminosity Crab-like pulsar ($L_X$ would be $\approx
  \expnt{2}{31}\mbox{ ergs s}^{-1}$), but with the counterpart it is very
  likely an AGN.
\item[\snr:10] This is likely a galaxy, given its position in
  Figure~\ref{fig:xopt}.
\item[\snr:13] This is likely a galaxy, given its position in
  Figure~\ref{fig:xopt}.
\item[\snr:14] This is likely a galaxy, given its position in
  Figure~\ref{fig:xopt} and hard spectrum.
\item[\snr:15]  This is likely a galaxy, given its position in
  Figure~\ref{fig:xopt}.  It may be extended, indicating a low
  redshift (Fig.~\ref{fig:opta2}).
\item[\snr:16] While redder than \snr:1, it is otherwise similar,
  suggesting that this is very likely also an
  active star of type K5.
\item[\snr:17] This is likely a galaxy, given its hard spectrum and
  position in Figure~\ref{fig:xopt}.  The IR source may also be
  extended (Fig.~\ref{fig:opta3}).
\item[\snr:19] There are multiple IR sources listed in
  Table~\ref{tab:match}, one just in the error circle and one outside
  (see Fig.~\ref{fig:opta3}).  The true counterpart is likely the one
  in the circle (the first source in Tab.~\ref{tab:match}), which is
  likely a galaxy, given its position in Figure~\ref{fig:xopt} and its
  colors.
\item[\snr:24] This is likely a galaxy, given its position in
  Figure~\ref{fig:xopt}.
\item[\snr:26] This is likely a galaxy, given its position in
  Figure~\ref{fig:xopt}.
\item[\snr:30] This is likely a galaxy, given its hard spectrum and
  position in Figure~\ref{fig:xopt}, similar to  \snr:19.
  The X-ray hardness ratio is, while harder than most galaxies, not
  unheard of in that context \citep{bva+02}.  
\end{description}

\subsection{\snrb}
\label{sec:snrb}
\snrb\ (RCW~86, or MSH~14$-$63) is a large ($45\arcmin$), non-thermal
Galactic radio shell (Fig.~\ref{fig:radiob}) identified as a SNR by
\citet{hill67} and considered as the remnant of the historic supernova
explosion SN~185 \citep{cs77}.

The identification of \snrb\ with the historic supernova explosion of
AD~185 is a matter of some controversy\footnote{While some recent
reviews of the Chinese record even question the association of SN~185
with a supernova explosion event \citep{ch94,schae95,smith97}, this is
not universally accepted \citep{sg02}.}.  Geometric considerations
would require a distance to the remnant of less than 1 kpc
\citep{strom94}, while the kinematic distance of the remnant is found
to be $2.8\pm0.4$~kpc \citep{raclcm96}. The latter distance suggests a
physical connection with an OB association \citep{west69}, an
interpretation which is supported by the light elemental abundance
which favors a Type~II supernova and by the measured interstellar
column density.  However, some recent modeling of X-ray data suggest a
distance as close as 1.2~kpc and abundances more typical of a Type~Ia
explosion \citep{bvf+00}, more in line with that expected if \snrb\
were the remnant of SN~185, although these interpretations are by no
means secure \citep{rdbr02}.  In what follows, we assume the 2.8~kpc
distance; if the SNR were closer (and younger), then our luminosity
limits would be lower and even more constraining if \snrb\ is the
result of a core-collapse event.

In addition to radio emission, \snrb\ shows thin Balmer-dominated
filaments \citep{lb90,smith97} and X-ray emission \citep*{phk84,vkb97}
that have the same general morphology.  There is X-ray spectral
variation over the remnant, but the hydrogen column density is likely
$\approx \expnt{2-3}{21}\mbox{ cm}^{-2}$ \citep{vkb97} (the total
Galactic column density from Colden in this direction is
$\expnt{9}{21}\mbox{ cm}^{-2}$).  The age of \snrb\ is somewhat
uncertain, as it has features of both young (a few thousand years) and
old ($>10^4$~yrs) remnants \citep*{dsm01}.  On average, probably the
best estimate for its age is 4000~yrs, although ages up to $10^4$~yrs
are not impossible \citep{raclcm96,petruk99,bvf+00,brrd01}.

\citet{kmj+96} searched \snrb\ unsuccessfully for a radio pulsar down
to a limit of 1.3~mJy at 436~MHz and 0.2~mJy at 1520~MHz, both
implying a 1400-MHz luminosity limit of $Sd^2 < 1.5\mbox{ mJy kpc}^2$.
This is quite faint --- a factor of three fainter than  PSR~J0205+6449.

Based on \rosat\ and \textit{Einstein} data, \citet{vbdk00} identified
an X-ray point source inside \snrb\ that they conclude is likely an
active star.  Unfortunately, this source is outside our X-ray
observations so we cannot confirm or deny their conclusion.
Similarly, \citet{gv03} used \chandra\ data of the bright south-west
region to search for point sources, motivated by their hypothesis that
\snrb\ was a significantly off-center cavity SN. They find one source
at the edge of the remnant (Fig.~\ref{fig:radiob}) without an optical
counterpart, although the only limit they cite is from the Digital Sky
Survey ($m_{\rm lim}\sim 21$), and conclude based on this and the
source's X-ray spectrum that it may be a neutron star (we have
examined 2MASS and see no clear counterpart to this source there,
implying limits of $J>15.8$ and $K_s > 14.6$).  However, these
optical/IR limits are far from constraining (\S~\ref{sec:optb}) and it
is quite possible that, while not an active star like the other
source they identify, this source is an active galaxy.  In any case,
all of these sources are quite far out from the nominal center of the
remnant.  If the explosion occurred near what we define as the center
of the SNR, none of the sources in \citet{vbdk00} or \citet{gv03}
could be the compact remnant without transverse velocities in excess
of $1500\mbox{ km s}^{-1}$ --- not an unheard of velocity
(\S~\ref{sec:vel}), but certainly large.

\subsubsection{X-ray Observations}
We observed \snrb\ with \chandra\ on 2002-December-02.  We determined
the center of the remnant to be at (J2000) $14^{\rm h}42^{\rm
m}50^{\rm s}$, $-62\degr28\arcmin20\arcsec$.  The aim-point was on the
ACIS-I imaging array, and the final exposure time was 10.0~ksec.  See
Figure~\ref{fig:radiob} for the placement of the ACIS-I detector
relative to the SNR.  A smoothed image of the data is shown in
Figure~\ref{fig:acisb} with the X-ray sources that we
identified (Tab.~\ref{tab:srcsb}) labeled.

The analysis of the \chandra\ data proceeded according to
Section~\ref{sec:proc}.  
In Figure~\ref{fig:hardb} we plot the H-band counts vs.\ the L-band
counts for the sources in Table~\ref{tab:srcsb}.  Most of the sources
are consistent with power-laws having indices from 0.0--2.0, but the
sources are slightly softer than those in \snr.  A few sources
(\snrb:2, \snrb:6, \snrb:8, \snrb:9, \snrb:26, and \snrb:28) are soft,
with implied temperatures (for blackbody models) of $kT_\infty\leq 0.4$~keV.

\subsubsection{Optical/IR Observations}
\label{sec:optb}
For \snrb\ we also observed the field with a variety of instruments,
listed in Table~\ref{tab:optb}.  The reduction proceeded as in
\S~\ref{sec:opta}.   

We determined $BVRI$ zero-points for the C40 data using
observations of the Stetson
fields\footnote{\url{http://cadcwww.dao.nrc.ca/cadcbin/wdbi.cgi/astrocat/stetson/query}}
E4 and L107 \citep{s00}.  We then determined zero-points for the other
optical data referenced to the C40 observations, employing
appropriate transformations \citep{jor94,stk+02}.  
We determined a $K_s$ zero-point for the PANIC
data with $\approx 20$ 2MASS stars in each field.

Astrometry was performed relative to 2MASS\footnote{The 2MASS data for
this field were released before those for \snr, \snrc, and \snrd,
where we used the GSC-2.2 for some of the astrometry.  We have
compared the results of 2MASS and GSC-2.2 astrometry and found them
indistinguishable.}.  For the C40 and EMMI data, where there is some
optical distortion, we computed solutions (plate-scale, rotation, and
central position) locally around each X-ray source, limiting the
fields to $\pm 1\arcmin$.  These solutions typically used 45 stars and
had residuals of $0\farcs09$ in each coordinate.  For the MagIC and
PANIC images we computed solutions for the entire image, using
$\approx 80$ stars and giving residuals of $0\farcs05$ in each
coordinate.

\subsubsection{Notes on Individual Sources}
\label{sec:indvb}
\begin{description}
\item[\snrb:1] This is likely a galaxy, given its position in
  Figure~\ref{fig:xoptb} and its hard spectrum
  (Fig.~\ref{fig:hardb}).  However, we only detect the counterpart in
  a single band and it is quite faint, so it is possible that the
  counterpart is a coincidence.
\item[\snrb:2] This is likely a star, given its position in
  Figure~\ref{fig:xoptb}  and its soft spectrum
  (Fig.~\ref{fig:hardb}).  Based on its IR colors and magnitude, this
  may be a K2III giant star.
\item[\snrb:3] The optical counterpart places it midway between the stars
  and galaxies in Figure~\ref{fig:xoptb}.  It could be a spurious
  counterpart, but there are other sources in its vicinity in
  Figure~\ref{fig:xoptb}.  Based on its hard spectrum, it is likely a galaxy.
\item[\snrb:5] This is likely a galaxy, given its position in
  Figure~\ref{fig:xoptb} and its hard spectrum (Fig.~\ref{fig:hardb}).
\item[\snrb:6] This is likely a star, given its position in
  Figure~\ref{fig:xoptb}  and its very soft spectrum
  (Fig.~\ref{fig:hardb}).  Based on its IR colors, it is likely type
  K0 or so.
\item[\snrb:8]  This is likely a star, given its position in
  Figure~\ref{fig:xoptb}  and its soft spectrum
  (Fig.~\ref{fig:hardb}).  Based on its IR colors, it is likely type
  G5 or so.
\item[\snrb:9]  This is likely a star, given its position in
  Figure~\ref{fig:xoptb}  and its soft spectrum
  (Fig.~\ref{fig:hardb}).  Based on its IR colors, it is likely type
  M7 or so.
\item[\snrb:12] This is likely a galaxy, given its position in
  Figure~\ref{fig:xoptb} and its hard spectrum
  (Fig.~\ref{fig:hardb}).  While the chance of a random star in the
  error circle is non
  negligible, only the reddest stars have $R-K_s >6$
  (Fig.~\ref{fig:color}) and these stars make up only a small fraction
  of those in this line of sight \citep{nim+00}.
\item[\snrb:13] This is likely a star, given its position in
  Figure~\ref{fig:xoptb}.  Based on its IR colors, it is likely type
  M6 or so. 
\item[\snrb:17] The optical/IR source is on the edge of the X-ray error
  circle in Figure~\ref{fig:optb2} (to the west of center).  There is also a faint smudge that
  is below the detection threshold and may be an extension of the
  source just to the north of the error circle.  If the source to the
  west is the counterpart, the offset is  large but not entirely
  unreasonable (Fig.~\ref{fig:offsethst}; see also \citealt{bcc+03}).  The source would then be a
  galaxy, given its position in
  Figure~\ref{fig:xoptb},  hard spectrum (Fig.~\ref{fig:hardb}), and
  red colors ($R-K_s > 8$; see \citealt{tbf+99}).  It
  is also possible that the northern source is the correct
  counterpart, in which case the source would also likely be a
  galaxy.  A third possibility is that the X-ray and optical/IR
  emission are not entirely spatially coincident, such as would be the
  case for a nearby interacting binary star or a low-redshift galaxy.
  This might explain the X-ray-to-optical offset.  It is unlikely that
  this source is a neutron star, as its spectrum is quite hard --- harder
  than that expected of a Crab-like pulsar.
\item[\snrb:19]  This is likely a star, given its position in
  Figure~\ref{fig:xoptb}.  Based on its IR colors, it is likely type
  F7 or so.
\item[\snrb:21] There are two optical sources in the X-ray error
  circle (Fig.~\ref{fig:optb3}), or perhaps one extended source.
  Separately, each source is near the star/galaxy boundary in
  Figure~\ref{fig:xoptb}, and the X-ray spectrum is intermediate, so
  no quick classification is possible. This may be an interacting
  binary (CV, X-ray binary, etc.), sources that would not
  have been in the Orion sample but are present in the general
  Galactic/extragalactic populations \citep{hg88,gzh+03}.
\item[\snrb:23] This is likely a star, given its position in
  Figure~\ref{fig:xoptb}.  Based on its IR colors, it is likely type
  M5 or so.
\item[\snrb:26] This is likely a star, given its position in
  Figure~\ref{fig:xoptb}.  Based on its IR colors, it is likely type
  K5 or so. 
\item[\snrb:28] There are three possible counterparts in
  Figure~\ref{fig:optb3}.  The brightest is the closest to the center
  of the error circle and would be consistent with being a star, and
  it may be extended to the north, possibly indicating a binary
  companion.  The source to the east of the circle's center is more
  consistent with a galaxy.  Finally, the source to the south-west of
  the circle's center, on the edge of the circle, is also more
  consistent with a galaxy.  Given the soft X-ray spectrum of this
  source, the bright stellar companion is likely the correct one.  It
  is also possible that there is no companion (the random probability
  for a star is somewhat high, even for the closest source), in which
  case it would be a candidate neutron star, but a definitive answer
  will have to await additional X-ray observations (assessing the
  spectrum and variability and improving the position).

\end{description}

\subsection{\snrc}
\label{sec:snrc}
\snrc\ was identified as a Galactic SNR by \citet{mbw+77}, who
observed a well-defined non-thermal radio shell
(Fig.~\ref{fig:radioc}) with diameters of about $20\arcmin \times
14\arcmin$.  \citet{fg93} identified CO and \ion{H}{1} emission
interacting with \snrc, giving the remnant a kinematic distance of
4.5~kpc and a size of $28\mbox{ pc} \times 22\mbox{ pc}$.  It has not
been detected in X-rays so there are no spectral fits to determine its
age or temperature.  The hydrogen column density to \snrc\ is
$\nh\approx\expnt{2}{21}$ (again using Colden) and integrating the
appropriate velocity range), while the total hydrogen through the
Galaxy is $\nh\approx\expnt{1}{22}$.  Assuming a Sedov-phase remnant,
we find (see \S~\ref{sec:Chandra-survey}) $t_4\approx 0.6
T_{7}^{-1/2}$, and since $T_7\approx 1$ holds for most SNRs, we can
say that $t_{4}\approx 0.3$--1.0 (with $T_7=0.3$--3).

\citet{llc98} searched \snrc\ unsuccessfully for a radio pulsar down
to a limit of 1.1~mJy at 606~MHz, implying a luminosity limit of $Sd^2
< 44\mbox{ mJy kpc}^2$.  Extrapolating to a frequency of 1400~MHz
using an average spectral index of $-1.8$, this is a factor of almost
$>50$ brighter than the luminosity of PSR~J0205+6449.

\subsubsection{X-ray Observations}
We observed \snrc\ with \chandra\ on 2002-November-24.  We determined
the center of the remnant to be at (J2000) $20^{\rm h}53^{\rm
m}21^{\rm s}$, $+43\degr26\arcmin55\arcsec$.  The aim-point was on the
ACIS-I imaging array, and the final exposure time was 24.6~ksec.  See
Figure~\ref{fig:radioc} for the placement of the ACIS-I detector
relative to the SNR.  A smoothed image of the data is shown in
Figure~\ref{fig:acisc} with the X-ray sources that we identified
(Tab.~\ref{tab:srcsc}) labeled.

The analysis of the \chandra\ data proceeded according to
Section~\ref{sec:proc}.  In Figure~\ref{fig:hardc} we plot the H-band
counts vs.\ the L-band counts for the sources in
Table~\ref{tab:srcsc}.  About half of the sources are consistent with
power-laws having indices from 0.0--2.0, but a number of the sources
are significantly harder.  These are likely distant AGN that have had
their soft photons heavily absorbed by Galactic gas --- a source with
$\Gamma=0.5$ and $\nh=\expnt{2}{22}\mbox{ cm}^{-2}$ would have ${\rm
HR}_{\rm L,H}=0.8$, similar to the hardest sources.  A few sources
(\snrc:1, \snrc:4, \snrc:14, \snrc:16) are soft, with implied
temperatures (for blackbody models) of $kT_\infty\leq 0.4$~keV.

\subsubsection{Optical/IR Observations}
\label{sec:optc}
We observed the field of \snrc\ a number of times with a number of
instruments, as described in Table~\ref{tab:optc}.  
The reduction proceeded as in
\S~\ref{sec:opta}.   

We determined the photometric zero-points for the LFC data by
bootstrapping from $VRI$ Palomar 60-inch observations of the Stetson
fields\footnote{\url{http://cadcwww.dao.nrc.ca/cadcbin/wdbi.cgi/astrocat/stetson/query}}
L98, NGC~7654, and PG~0231 \citep{s00} and employing appropriate
transformations \citep{jor94,stk+02} to LFC observations of these
fields.  For the WIRC observations, we determined zero-points using
1700 unsaturated 2MASS \citep{2mass} stars.  For the NIRC
observations, we determined zero-points using about 60 stars from the
WIRC image.

We performed absolute astrometry on the LFC
data and the WIRC data.  After applying distortion
corrections to the LFC data (we did not use any distortion correction
for WIRC), we solved for plate-scale, rotation, and
central position relative to stars from version 2.2 of the Guide Star
Catalog
(GSC-2.2\footnote{\url{http://www-gsss.stsci.edu/support/data\_access.htm}})
for all but WIRC and relative to 2MASS stars for WIRC, getting
residuals in each coordinate of $0\farcs13$ (600 stars),  and $0\farcs19$ (2500 stars) for
LFC and WIRC respectively.  For the NIRC data
we used non-saturated stars from the WIRC to boot-strap the astrometry, getting typical residuals
of $<0\farcs06$ in each coordinate with about 60 stars.

\subsubsection{Notes on Individual Sources}
\label{sec:indvc}
\begin{description}
\item[\snrc:1]  This source is almost certainly an active star:  the
  infrared colors are indicative of class F6 or so, the X-ray spectrum is
  quite soft, and the source falls
  on the locus of stars in Figure~\ref{fig:xoptc}.
\item[\snrc:2] This is likely a galaxy, given its position in
  Figure~\ref{fig:xoptc} and its hard X-ray spectrum.
\item[\snrc:3]  This is likely a galaxy, given its position in
  Figure~\ref{fig:xoptc} and its hard X-ray spectrum.
\item[\snrc:4]  This source is probably an active star:  the
  infrared colors are indicative of class M0 or so, the X-ray spectrum is
  quite soft, and the source falls
  on the locus of stars in Figure~\ref{fig:xoptc}.  There is an offset
  between the X-ray and IR positions, but this is consistent with the
  X-ray positional uncertainty and the number of chance stars of this
  brightness is quite low.
\item[\snrc:5] This is probably a galaxy, given its position in
  Figure~\ref{fig:xoptc} and its hard X-ray spectrum.  However, the
  multiple possible IR counterparts make a clear identification impossible.
\item[\snrc:6]  This is likely a galaxy, given its  very hard X-ray
  spectrum.  It does not appear in Figure~\ref{fig:xoptc} as there are
  no L-band counts, but its $K_s$ magnitude is similar to that of
  other galaxies such as \snrc:2.
\item[\snrc:7] This is likely a galaxy, given its position in
  Figure~\ref{fig:xoptc} and its hard X-ray spectrum.
\item[\snrc:10] This is likely a galaxy, given its position in
  Figure~\ref{fig:xoptc} and its hard X-ray spectrum.  However, the
  multiple possible IR counterparts may a clear identification impossible.
\item[\snrc:11] This source may be an active star, given how bright
  its IR counterpart is (it is possible, but not likely, that the
  counterpart is due to chance).  However, the X-ray spectrum is quite
  hard and the IR colors are far redder than those typical for stars
  (an M7 star would have $J-K_s\approx 1.3$, which would require
  $A_V\approx 10$~mag to get to the observed value of $J-K_s$).  
  Therefore, this source is likely an active galaxy, where intrinsic
  $J-K_s\gsim 2$ is not that unusual \citep[e.g.,][]{flr+03} and
  a larger foreground \nh\ is expected.
\item[\snrc:12] This source is probably an active star:  the
  infrared colors are redder than typical for a main-sequence star of
  its magnitude, but it could be a distant M5 giant.  In addition, the X-ray spectrum is
  quite soft, and the source falls
  on the locus of stars in Figure~\ref{fig:xoptc}.  
\item[\snrc:14]  This source is almost certainly an active star:  the
  infrared colors are indicative of class G0, the X-ray spectrum is
  quite soft, and the source falls
  on the locus of stars in Figure~\ref{fig:xoptc}.
\item[\snrc:15] This is likely a galaxy, given its position in
  Figure~\ref{fig:xoptc} and its hard X-ray spectrum.
\item[\snrc:16] This source is probably an active star:  the
  infrared colors are indicative of class M1 (or possible a K giant), the X-ray spectrum is
  quite soft, and the source falls
  on the locus of stars in Figure~\ref{fig:xoptc}.
\item[\snrc:18]  This may be a galaxy, given its  hard X-ray
  spectrum and the faint IR counterparts.  However, the error circle
  is large, and the IR counterparts could be spurious.  
Even so, it would still probably be extragalactic in origin 
 as the X-ray source is
  too hard to be Galactic.  For a power-law source with $\Gamma=0.7$
  (very hard for a neutron star, but plausible for an AGN) we would
  need $\nh \gsim \expnt{5}{22}\mbox{ cm}^{-2}$ to give the observed
  hardness ratio.  This value is significantly higher than the column
  density expected to \snrc, and is in fact even somewhat higher than
  the Galactic column density in this direction (although the AGN
  spectrum could have $\Gamma < 0.7$).
\item[\snrc:22]  This source is probably an active star:  the
  infrared colors are redder than typical for a main-sequence star of
  its magnitude, but it could be a distant M5 giant.  In addition, the X-ray spectrum is
  quite soft, and the source falls
  on the locus of stars in Figure~\ref{fig:xoptc}. 
\item[\snrc:23]  This is probably a galaxy, given its  hard X-ray
  spectrum and the faint IR counterparts.
\end{description}

\subsection{\snrd}
\label{sec:snrd}
\snrd\ (R5) is a $45\arcmin$-diameter radio shell
(Fig.~\ref{fig:radiod}), also known as R5, which was identified as an
SNR by \citet{caswell77}.  It is remarkable in that it has a bright
(396~mJy at 1.4~GHz), flat-spectrum radio point source (G127.11+0.54)
near the center (\citealt{caswell77}, \citealt*{jrd89}), but optical observations
(\citealt{kc78}, \citealt*{ssh79}) and \ion{H}{1} absorption measurements
\citep{pvgg+82,gvg84} instead favor an identification with a massive
elliptical galaxy at $\approx 100$~Mpc distance, similar to M87.

The distance is likely 1.2--1.3~kpc, if it is in fact associated with
the open cluster NGC~559 \citep{pauls77}.  It has not been detected in
X-rays so there are no spectral fits to determine its age or
temperature.  The hydrogen column density to \snrd\ is
$\nh\approx\expnt{2}{21}$ (again using Colden) and integrating the
appropriate velocity range), while the total hydrogen through the
Galaxy is $\nh\approx\expnt{1}{22}$.  Again assuming a Sedov-phase
remnant, we find (see \S~\ref{sec:Chandra-survey}) $t_4\approx 0.6
T_{7}^{-1/2}$, or $t_{4}\approx 0.2$--0.8 (with $T_7=0.3$--3).

\citet{llc98} searched \snrd\ unsuccessfully for a radio pulsar down
to a limit of 0.8~mJy at 606~MHz, implying a luminosity limit of $Sd^2
< 1.4\mbox{ mJy kpc}^2$.  Extrapolating to a frequency of 1400~MHz
using an average spectral index of $-1.5$, this is a factor of almost
$>500$ brighter than the luminosity of PSR~J0205+6449.

\subsubsection{X-ray Observations}
We observed \snrd\ with \chandra\ on 2002-September-14.  We determined
the center of the remnant to be at (J2000) $01^{\rm h}28^{\rm
m}32^{\rm s}$, $+63\degr06\arcmin34\arcsec$.  The aim-point was on the
ACIS-I imaging array, and the final exposure time was 19.5~ksec.  See
Figure~\ref{fig:radiod} for the placement of the ACIS-I detector
relative to the SNR.  
 A smoothed image of the data is shown in
Figure~\ref{fig:acisd} with the X-ray sources that we
identified (Tab.~\ref{tab:srcsd}) labeled.

The analysis of the \chandra\ data proceeded according to
Section~\ref{sec:proc}.  
In Figure~\ref{fig:hardd} we plot the H-band counts vs.\ the L-band
counts for the sources in Table~\ref{tab:srcsd}.  Virtually all of the
sources are consistent with power-laws having indices from 0.0--2.0.
It may be that we have underestimated the \nh\ to \snrd, which would
put the distribution of hardness ratios closer to what is seen in the
other SNRs, but this does not affect our analysis of the sources.

\subsubsection{Optical/IR Observations}
\label{sec:optd}
We observed the field of \snrd\ a number of times with a number of
instruments, as described in Table~\ref{tab:optd}.  
The reduction proceeded as in
\S~\ref{sec:opta}.   

We determined the photometric zero-points for the LFC data from VRI
Palomar 60-inch observations of the Stetson
fields\footnote{\url{http://cadcwww.dao.nrc.ca/cadcbin/wdbi.cgi/astrocat/stetson/query}}
L98, NGC~7654, and PG~0231 \citep{s00} and employing appropriate
transformations \citep{jor94,stk+02} to LFC observations of the SNR.
For the WIRC observations, we determined zero-points using 900
unsaturated 2MASS \citep{2mass} stars.  For the NIRC observations, we
determined zero-points using about 20--30 stars from the WIRC image.

We performed absolute astrometry on the LFC data and the WIRC data.
After applying distortion corrections to the LFC data, we solved for
plate-scale, rotation, and central position relative to stars from
version 2.2 of the Guide Star Catalog
(GSC-2.2\footnote{\url{http://www-gsss.stsci.edu/support/data\_access.htm}})
for LFC and relative to 2MASS stars for WIRC, getting residuals in
each coordinate of $0\farcs12$ (1200 stars), and $0\farcs22$ (1100
stars) for LFC and WIRC respectively.  For the NIRC data we used
non-saturated stars from the WIRC to boot-strap the astrometry,
getting typical residuals of $<0\farcs07$ in each coordinate with
about 20 stars.

\subsubsection{Notes on Individual Sources}
\label{sec:indvd}
\begin{description}
\item[\snrd:1] This source is coincident with the non-thermal radio
  source G127.11+0.54 and with its optical counterpart
  \citep{kc78,ssh79}.  The X-ray spectrum is quite hard, fitting with
  the identification as an active galaxy.  Interestingly, this is the
  only source for \snrd\ to be in the stellar locus in
  Figure~\ref{fig:xoptd}, when we know it to be extragalactic.  The
  optical counterpart is somewhat brighter than those of most galaxies
  relative to the X-ray flux (similar to \snrc:11).  This could be due
  to the extended stellar emission seen from this source, or due to
  the orientation of the optical jets.
\item[\snrd:3] This is probably a galaxy, given its  hard X-ray
  spectrum and the faint IR counterpart.
\item[\snrd:4] This is probably a galaxy, given its  hard X-ray
  spectrum and the faint IR counterpart.
\item[\snrd:5] This is probably a galaxy, given its  hard X-ray
  spectrum and the faint IR counterpart.
\item[\snrd:8] This is probably a galaxy, given its  hard X-ray
  spectrum and the faint IR counterpart.
\item[\snrd:12] This source is among the softest of the sources in
  \snrd\ and its position in Figure~\ref{fig:xoptd} is intermediate
  between the stars and galaxies, so it could be an active star.
  However, the spectrum is not actually all that soft, and it is more
  likely to be a galaxy, given its faint, red IR counterpart.
\item[\snrd:13] This is probably a galaxy, given its  hard X-ray
  spectrum and the faint IR counterpart.
\item[\snrd:14] This is probably a galaxy, given its  hard X-ray
  spectrum and the faint IR counterpart.
\item[\snrd:15] This is probably a galaxy, given its  hard X-ray
  spectrum and the faint IR counterpart.
\item[\snrd:17] This is probably a galaxy, given its  hard X-ray
  spectrum and the faint IR counterpart.
\item[\snrd:18] This source is among the softest of the sources in
  \snrd\ and its position in Figure~\ref{fig:xoptd} is intermediate
  between the stars and galaxies, so it could be an active star.  Like
  \snrd:12, though, the spectrum is still reasonably hard.  The source
  could be a galaxy, given its X-ray spectrum and the faint, red IR
  counterpart.  If it were a star, it would have to be a distant M
  giant in order to produce the observed $J-K_s$ value.  It could also
  be a chance coincidence, given how crowded the region is in
  Figure~\ref{fig:optd2}.
\item[\snrd:22]  This could be a galaxy, given its  hard X-ray
  spectrum and the faint IR counterparts.  The second counterpart in
  Table~\ref{tab:srcsd} is almost certainly a star, as suggested by
  the last column of Table~\ref{tab:srcsd}.  The first source may also
  be spurious, but the chances of a real association are reasonable.
\end{description}

\section{Limits on Central Sources}
\label{sec:limits}
\begin{figure*}
% plot_l.m
\includegraphics[width=\hsize]{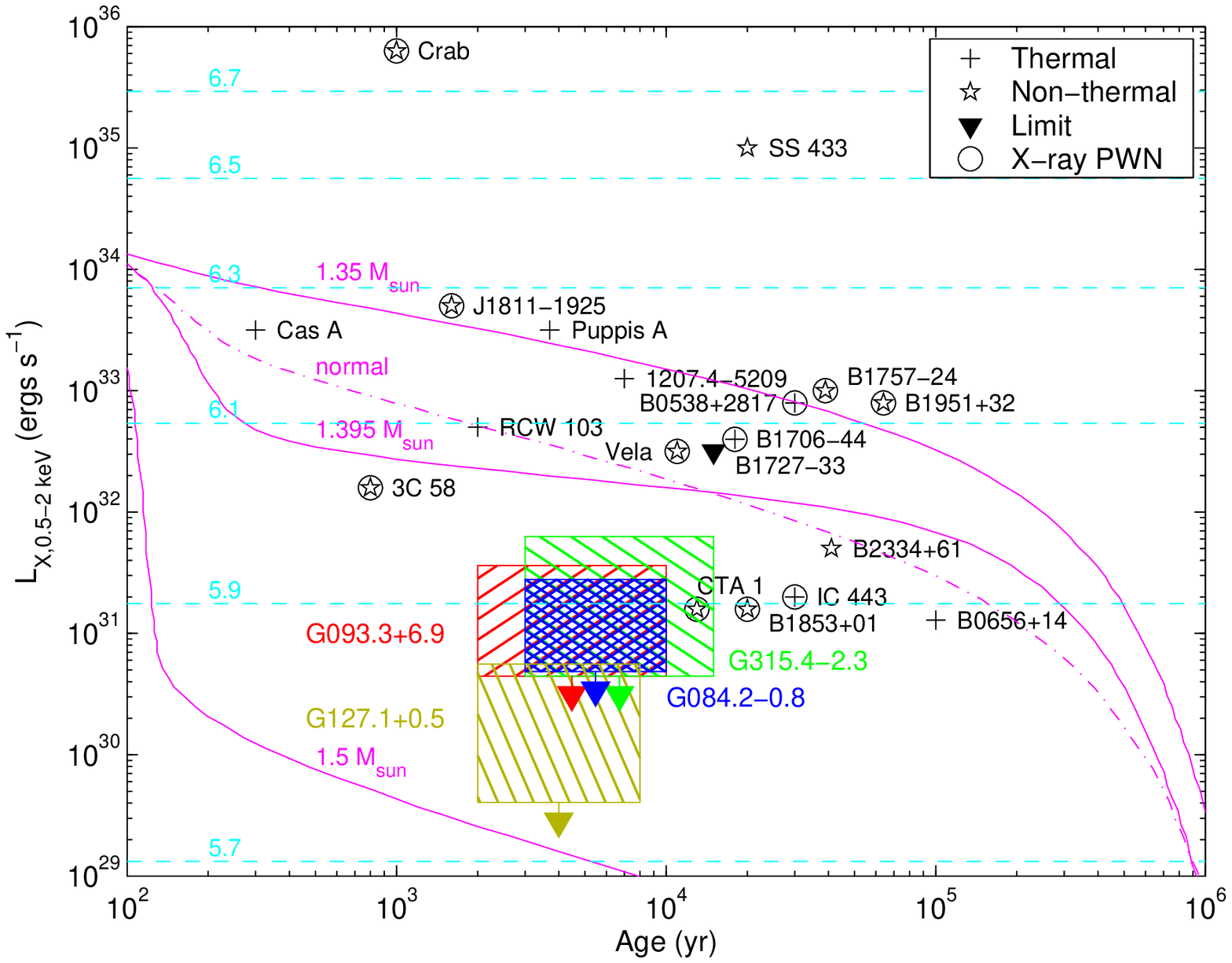}
\caption{\small X-ray luminosities (0.5--2~keV) as a function of age
for neutron stars in SNRs from Table~\ref{tab:psrs}.  Sources whose
emission is primarily thermal are indicated with plus symbols, those
whose emission is primarily non-thermal are indicated with stars, and
those with only limits are indicated with triangles.  The sources that
have X-ray PWNe, which are typically $>10$ times the X-ray luminosity
of the neutron stars themselves, are circled.  We also plot the limits
to blackbody emission from sources in SNRs \Ga\ (red hatched region),
\Gb\ (green hatched region), \Gc\ (blue cross-hatched region), and
\Gd\ (gold hatched region).  A 30\% uncertainty in the distance has
been added to the range of luminosities given in
Table~\ref{tab:limits} (i.e.\ we have taken the Type~I limits with a
30\% larger distance and the Type~II limits with a 30\% smaller
distance, to give the widest probable range of luminosities), and the
likely range of ages is also shown. The cooling curves are the 1p
proton superfluid models from \citet{ygk+03} (solid lines, with mass
as labeled) and the normal (i.e., non-superfluid) $M=1.35\,M_{\odot}$
model (dot-dashed line), assuming blackbody spectra and
$R_{\infty}=10$~km.  These curves are meant to be illustrative of
general cooling trends, and should not be interpreted as detailed
predictions.  The horizontal lines show the luminosity produced by
blackbodies with $R_{\infty}=10$~km and $\log T_{\infty}$ (K) as
indicated.  Faster cooling than the curves is possible, due either to
the presence of exotic particles in the NS core or to the full onset of
direct Urca cooling for a heavier NS \citep[e.g.,][]{ykhg02}. }
\label{fig:cool}
\end{figure*}

In Sections~\ref{sec:indva}, \ref{sec:indvb}, \ref{sec:indvc}, and
\ref{sec:indvd}, we showed that almost all of the X-ray sources from
Tables~\ref{tab:srcs}--\ref{tab:srcsd} can be identified either with
foreground or background sources.  Therefore, there does not appear to
be any detected neutron star in SNRs \Ga, \Gb, \Gc, or \Gd.  There are
a small number of cases where either the association or the type of
source (star versus galaxy; see Tab.~\ref{tab:class}) is uncertain,
either due to an optical/IR detection in only one band and/or a
detection at a somewhat large distance from the X-ray source, but
there are certainly no sources that scream out ``I am a neutron
star.''  If we accept this, we can then draw two limits to the flux of
any compact central source: a conservative limit (Limit I), and a
loose limit (Limit II).  The conservative limit will be the flux of
the brightest source for which the optical/IR counterpart is at all in
doubt.  For \snr, this would be source \snr:8: this counterpart is
faint, has unknown colors, is somewhat far from the X-ray source, and
is somewhat soft.  For \snrb, this would be source \snrb:1: again this
is faint, has unknown colors, and is in a crowded region.  For \snrc,
the limiting source would be \snrc:5: the multiple IR counterparts
make a firm association impossible, and the spectrum is not so hard as
to eliminate the possibility of a Crab-like pulsar.  Finally, for
\snrd, the limiting source is \snrd:18, where the crowding and
uncertain classification make a firm association impossible, and again
the spectrum is not so hard as to eliminate the possibility of a
Crab-like pulsar.  Here we have played devil's advocate, and
questioned all of the associations in Section~\ref{sec:results} as much
as possible.  We in fact believe that the associations are reasonably
secure, but we cannot be certain.  The looser limits in each case come
from the faintest sources in Tables~\ref{tab:srcs}--\ref{tab:srcsd},
and assume that all of the associations in
Tables~\ref{tab:match}--\ref{tab:matchd} are correct.  We present the
limits, along with approximate fluxes and luminosities for three
different source models, in Table~\ref{tab:limits}.  In
Figure~\ref{fig:cool}, we plot the blackbody limits along with the
luminosities of the sources in Table~\ref{tab:psrs}.  The limits for
\snrd\ are significantly below those of the other SNRs as it had never
been observed in the X-rays before, so we did not know what the level
of the diffuse background would be and therefore selected an exposure
time that would guarantee sufficient counts from a source above even
the most pessimistic background.

\begin{deluxetable}{l r r r c r r r }
\tablecaption{Limits on Central Sources in SNRs \Ga, \Gb, \Gc, and \Gd\label{tab:limits}}
\tablewidth{0pt}
\tablehead{\colhead{Model} & \mc{3}{c}{Limit I} & &\mc{3}{c}{Limit II}
  \\ \cline{2-4} \cline{6-8}
 & \colhead{Counts\tablenotemark{a}} & \colhead{$F_{\rm
      X}$\tablenotemark{a{\rm b}}} & \colhead{$L_{\rm X}$\tablenotemark{a{\rm c}}} & &\colhead{Counts\tablenotemark{a}} & \colhead{$F_{\rm X}$\tablenotemark{a{\rm b}}} & \colhead{$L_{\rm X}$\tablenotemark{a{\rm c}}} \\
}
\startdata
\mc{1}{l}{\bf \snr:}\\
%\sidehead{\snr:}
BB ($kT_\infty=0.25$~keV) & 25 & 9.1 & 2.6 && 11 & 4.0 & 1.1 \\
PL ($\Gamma=1.5$) & '' & 20.7 & 3.7 && '' & 9.1 & 1.6 \\
PL ($\Gamma=3.5$) & '' & 13.1 & 7.3 && '' & 5.8 & 3.2 \\
\tableline
\sidehead{\snrb:}
BB ($kT_\infty=0.25$~keV) & 45 & 24.7 & 4.5 && 11 & 6.0 & 1.1\\
PL ($\Gamma=1.5$) & '' & 55.8 & 6.4 && '' & 13.6 & 1.6\\
PL ($\Gamma=3.5$) & '' & 35.3 & 12.4 && '' & 8.6 & 3.0\\
\tableline
\sidehead{\snrc:}
BB ($kT_\infty=0.25$~keV) & 20 & 4.3 & 2.0 && 12 & 2.5 & 1.2\\
PL ($\Gamma=1.5$) & '' & 9.9 & 2.9 && '' & 5.9 & 1.7\\
PL ($\Gamma=3.5$) & '' & 6.3 & 5.7 && '' & 3.8 & 3.4\\
\tableline
\sidehead{\snrd:}
BB ($kT_\infty=0.25$~keV) & 20 & 5.4 & 0.1 && 12 & 3.2 & 0.1\\
PL ($\Gamma=1.5$) & '' & 12.5 & 0.3 && '' & 7.4 & 0.1\\
PL ($\Gamma=3.5$) & '' & 7.9 & 0.2 && '' & 4.8 & 0.1\\

% 0.5-2 keV: 
% \snr & 25 & 8.2 & 2.1 && 11 & 3.6 & 0.9 \\
% \snrb & 45 & 22.1 & 3.7 && 11 & 5.4 & 0.9\\
\enddata
\tablenotetext{a}{In the 0.3--8.0~keV band.}
\tablenotetext{b}{X-ray flux $\times 10^{-15}\mbox{ ergs s}^{-1}\mbox{
    cm}^{-2}$.}
\tablenotetext{c}{X-ray luminosity $\times 10^{31}\mbox{ ergs
    s}^{-1}$, corrected for absorption according to
  \S\S~\ref{sec:snra}--\ref{sec:snrd}, and using the distances in
  Table~\ref{tab:snrs}.}  
\tablecomments{Fluxes and luminosities were computed using W3PIMMS.  See
  \S~\ref{sec:limits} for definition of limit types.}
\end{deluxetable}

Below we discuss the implications of not detecting any sources in the
contexts of different models for what the sources could be
(\S~\ref{sec:class}). We also include discussions of limitations
imposed by our observing strategy, specifically the limited field of
view and the gaps in the ACIS-I detector.

\subsection{Instrumental Limitations}
\subsubsection{Field of View}
\label{sec:vel}
While we did not detect any bright X-ray sources that were obviously
compact objects, it may be that this was because the sources had
extremely high velocities that carried them beyond the ACIS-I
field-of-view.  This would imply $v_{\perp} > 1600\mbox{ km s}^{-1}$
for both \snr\ and \snrb, $v_{\perp} > 1700\mbox{ km s}^{-1}$ for
\snrc, and $v_{\perp} > 700\mbox{ km s}^{-1}$ for \snrd\ (assuming a
centered explosion).  $v_{\perp} = 1600\mbox{ km s}^{-1}$ is higher
than the velocities of 99\% of the radio pulsar population, while
$v_{\perp}=700\mbox{ km s}^{-1}$ is higher than 90\% of the population
\citep{acc02}. For one source, this might be acceptable, but for two
or more sources the chances become too low ($\sim 10^{-7}$ for all
four sources), requiring another explanation.

\citet{gv03} believe that \snrb\ was the result of an off-center
explosion in a cavity created by the SN's moving progenitor, and have
similar hypotheses about other SNRs \citep[e.g.,][]{bg02}.

\subsubsection{Chip Gaps}
The ACIS-I detector has gaps between the four CCDs where the
sensitivity falls to zero.  During normal
observations some compensation is made for this by the dithering of
the spacecraft.  For our observations, we ended up with approximately
5\% of the area having an effective exposure that was $\approx 50\%$
of the nominal exposure.  The \texttt{wavdetect} program used exposure
maps to account for this effect when detecting sources, so that
sources located in the chip gaps can have fewer detected counts but
end up with the same significance as a source in the middle of the
chips, so we should not be missing sources due to the chip gaps.  But
there is a 5\% chance of having a source in the gap region that is not
detected for which the flux/luminosity limits should be a factor of 2
higher than those stated in Table~\ref{tab:limits}.

\subsection{AXPs}
The properties of AXPs in Section~\ref{sec:axpdesc} allow us to state
that there are no such sources in the central $8\arcmin$ of SNRs \Ga, \Gb,
\Gc, or \Gd:
the luminosity limits for $\Gamma=3.5$ in Table~\ref{tab:limits} are
at most $10^{32}\mbox{ ergs s}^{-1}$, or two orders of magnitude below
those of most AXPs and a factor of 10 less than the ``quiescent''
states of the possibly variable AXPs.  This discrepancy cannot be solved by a
slight change in distance or absorption, and is therefore quite firm.

\subsection{Cooling Radio-Quiet Neutron Stars}
\label{sec:cool}
Using the standard cooling curve (modified Urca only) in
Figure~\ref{fig:cool}, we would estimate luminosities of (2--5)$\times
10^{32}\mbox{ ergs s}^{-1}$ for any compact sources.  Obviously, our
limits are below those values.  Our limits are also below  the
luminosities of most of the sources in Table~\ref{tab:psrs}: only
PSR~B2334+61, CXO~J061705.3+222127, PSR~B1853+01, and RX~J0007.0+7302
are comparable (while PSR~B1853+01 is dominated by non-thermal X-ray
emission, any cooling radiation would have to be below this level).
However, most of these objects are older ($\gsim 30$~kyr) than the
SNRs considered here ($\lsim 10$~kyr), and three of these sources have
substantial X-ray (and radio) PWNe that would make them detectable in
the absence of point-source emission.

 With our limits in Table~\ref{tab:limits}, any thermal emission from
sources in SNRs \Ga, \Gb, \Gc, or \Gd\ would have to have
$T_\infty\lsim \expnt{8}{5}$~K (or $kT_{\infty} \lsim 70$~eV, for
$R_{\infty}\approx 10$~km): see Figure~\ref{fig:cool}.  Even for radii
of 3~km (\S~\ref{sec:rqnsdesc}), the limits on $kT_\infty$ are
$\approx 100$~eV.  These are lower than expected from standard cooling
curves (\citealt{page98}, \citealt*{shm02}, \citealt{kyg02,ykhg02})
and would require some exotic physics (pion cooling, direct Urca
cooling, etc.) that may be related to a more massive neutron star
\citep{ygk+03}.

\subsection{Radio Pulsars}
\label{sec:psr}
If we compare the luminosity limits for the $\Gamma=2.0$ model for
SNRs \Ga, \Gb, \Gc, and \Gd\ with the luminosities of known radio
pulsars (Tab.~\ref{tab:radpsrs}), we see that our limits are below
most but not all of luminosities of pulsars found in SNRs.
Translating our limits to limits on $\dot E$ (roughly, $L_{{\rm
X},0.1-2.4\,{\rm keV}}\approx 0.5 L_{{\rm X},0.3-8\,{\rm keV}}$ and
using $L_{{\rm X},0.1-2.4\,{\rm keV}}\sim 10^{-3}\dot E$;
\citealt{bt97}) we find $\dot E \lsim \expnt{3}{34}\mbox{ ergs
s}^{-1}$ (for the type~I limit) or $\dot E \lsim \expnt{8}{33}\mbox{
ergs s}^{-1}$ (for the type~II limit) for \snr, \snrb, and \snrc, and
$\dot E \lsim \expnt{2}{33}\mbox{ ergs s}^{-1}$ (for the type~I limit)
or $\dot E \lsim \expnt{3}{32}\mbox{ ergs s}^{-1}$ (for the type~II
limit) for \snrd.  Comparing with the values of $\dot E$ given in
Section~\ref{sec:psrdesc}, these limits are below those of traditional
pulsars, but are compatible with the values for the low-$\dot E$
HBPSRs.  Our limits on $L_{X}$ are consistent with the one low-$\dot
E$ HBPSR that has detected X-ray emission --- PSR~J1718$-$37184
\citep{mks+03} --- but its age and distance are poorly known and its
luminosity of $\sim 10^{30}\mbox{ ergs s}^{-1}$ is roughly what is
expected from standard cooling curves for a source with $\tau \approx
30$~kyr in contrast to the sources discussed here (\S~\ref{sec:cool}).

Since we cannot constrain the existence of a low-$\dot E$ pulsar in
any of these four SNRs, we can ask what its period might be.  To do so
we must assume that initial period is much less than the current
period so that the characteristic age $\tau \equiv P/2\dot P$ is
similar to the actual age and that the braking index has the constant
value $n=3$.  We know that this is not always the case
\citep[e.g.,][]{mss+02,mgb+02,lyne04}, but it is the best guess that
one can make.  Under this assumption one finds $P\sim
\expnt{1.4}{23}\sqrt{I_{45}}(\tau \dot E)^{-0.5}$~s and $B\sim
\expnt{3.2}{42}\sqrt{I_{45}}(\tau^2 \dot E)^{-0.5}$~G, where
$I=10^{45} I_{45}\mbox{ g cm}^{2}$ is the moment of inertia.  With
$\tau\approx 4000$--6000~yrs, we find $P\gsim 3$~s for \snr, \snrb,
and \snrc, and $P\gsim 9$~s for \snrd.  These periods are larger than
those of most but not all radio pulsars (\citealt*{ymj99}, \citealt{ckl+00,msk+03}),
and may be high enough to take any source in \snr\ or \snrb\ beyond
the radio ``death line.''  The implied dipole magnetic fields are also
high, $\gsim \expnt{2}{14}$~G for \snr, \snrb, and \snrc\ and $\gsim
\expnt{5}{15}$~G for \snrd, similar to \psrckl\ and \psrmsk.  This may
be indicative of a growing population of such objects: young,
non-energetic long-period pulsars.  The lack of detected pulsed radio
emission in any of these four SNRs \citep{llc98,kmj+96} may be
intrinsic (i.e., there is no radio emission), it may be an orientation
effect, or it may just be that the SNRs have not been searched deeply
enough over enough of an area, as there is now a growing number of
radio pulsars with luminosities (defined here as $F_{\rm radio} d^2$)
below $1\mbox{ mJy kpc}^{2}$ \citep[e.g.,][]{csl+02,clb+02}, far lower
than typical for radio pulsars.

\subsubsection{Pulsar Wind Nebulae}
\label{sec:pwn}
Any PWNe in the SNRs discussed here are $\gsim 3000$~yrs old and may
have already interacted with the reverse shocks
(\S~\ref{sec:pwndesc}), so their sizes and brightnesses would be hard
to predict.  We therefore examine limits on PWNe for a range of sizes
(as in \S~\ref{sec:ext}).  We also scale to a fiducial size of
$1\mbox{ pc}\approx 1\arcmin$ --- we did not detect any sources with
those sizes in our images except for the known thermal emission from
RCW~86.

To quantify this, we take the limits on extended sources from
Section~\ref{sec:ext} and Table~\ref{tab:pwn}.  
We convert the count limits to luminosity limits using a photon index
of $\Gamma=1.5$, getting the limits in  Table~\ref{tab:pwn}.
These limits are below the
luminosities of virtually all young PWNe detected in X-rays
\citep{pccm02}, but are consistent with some older sources such
as the Vela PWN, CTB~80, and W44 \citep{pksg01}.  However, these PWNe
all have significant non-thermal radio emission, emission that is not
present in \snr, \snrb, \snrc, or \snrd\ since they are all shell-type
SNRs.

\subsection{Binary Systems}
As many as 50\% of massive stars originate in binary systems, which
presumably give rise to X-ray binaries and eventually millisecond
pulsars.  One might expect that we could see a binary system where the
more massive star has gone supernova but the less massive has not
evolved.  In this case, we would see X-ray emission that appears to be 
(but is not physically) associated with an optically-detected star that
might be hard to distinguish from the active stars that make up the
majority of Galactic X-ray sources (if the star did evolve and the
companion were close enough, it would donate matter to the compact
object and would appear as an X-ray binary and would have different
properties).

However, after only one supernova the binary system would have a small
space velocity, $< 100\mbox{ km s}^{-1}$ (\citealt*{prp02}, \citealt{prps02}).  It
would therefore be restricted to a smaller region than the full
search, which accommodates velocities up to $1500\mbox{ km s}^{-1}$
(\S~\ref{sec:vel}).  A velocity of $100\mbox{ km s}^{-1}$ is an
angular offset of $\approx 30\arcsec$ (for an average distance of
3~kpc and age of 5~kyr) --- much smaller than the ACIS-I
field-of-view.

As seen in Tables~\ref{tab:srcs}--\ref{tab:srcsd}, we have found only
one source within a radius of $30\arcsec$ of the center: \snrd:1,
which is known to be extragalactic.  In fact, none of the stellar
sources (as determined in \S\S~\ref{sec:indva}, \ref{sec:indvb},
\ref{sec:indvc}, and \ref{sec:indvd}) appear to be at distances beyond
1~kpc (comparing with Fig.~\ref{fig:color}).  This is likely a
selection effect of our X-ray flux limits: for stellar luminosities of
$10^{29}$--$10^{31}\mbox{ ergs s}^{-1}$, our flux limits translate
into distance limits of $\sim 1$~kpc.

If none of the stellar sources could be companions to a neutron star,
we must ask if any of the sources identified as galaxies could in fact
be stars, and their identifications as galaxies could be
coincidence. As seen from Figures~\ref{fig:color}--\ref{fig:colord},
a main-sequence star at the distances of SNRs \Ga, \Gb, \Gc, or \Gd\
would have $K_s\approx 15$--19, depending on stellar type (giant stars
at these distance would mostly be too bright, with $K \lsim 10$).  All
of the ``galaxies'' in Sections~\ref{sec:indva}, \ref{sec:indvb},
\ref{sec:indvc}, and \ref{sec:indvd} have $K>17$ except
\snrc:11 (and this is too red to be at the distance/reddening of
\snrc), so if they were main-sequence stars they would be type K5 or
later. However, even if one of these sources is a companion to a
neutron star, the Type~I limits still apply, and any neutron star
would be under-luminous (\S~\ref{sec:cool}).

\subsection{Accreting Black Holes}
If the SNe produced black holes (BHs) and not neutron stars, the black
holes themselves would be invisible.  X-ray
emission would only be detected if there were material accreting into
the BHs.  Models for such emission are not very well understood.
Thermal emission might be expected to come from the inner portion of
the accretion disk itself, with an area of several times $\pi R_{S}^2$
\citep{cph+01}, where $R_S$ is the Schwarzschild radius of the BH.
For a $10\,M_{\odot}$ BH (with $R_S=15$~km), the area would be $\gsim
1000\mbox{ km}^{2}$, or much larger than the limits on thermal
emission (\S~\ref{sec:cool}) for temperatures $\gsim 100$~eV.  It is
also possible that the X-ray emission arises from Compton scattering
in an optically thin corona over a thin disk or via optically thin
bremsstrahlung emission from a hot advection-dominated accretion flow
(ADAF, as in the model of \citealt*{nbmc97}), but these models are not
well enough developed to provide useful constraints.

\subsection{Type~Ia Supernovae}
Type Ia SNe are believe to result from carbon detonation/deflagration
of a white dwarf that has been pushed beyond its mass limit through
accretion from a companion star. The resulting explosion completely
disrupts the star, synthesizing nearly a solar mass of $^{56}$Ni which
ultimately decays to Fe. No compact core is left behind. The SNRs from
such events thus form a subsample in which we do not expect to find an
associated neutron star. The mean rate for Type Ia SNe is $\sim
20-25\%$ of that for core-collapse supernovae \citep*{cet99}, so we
expect less than 20\% of the observable SNRs to be the result of such
explosions.

The ejecta produced in Type Ia events differs considerably than that from
core-collapse SNe. The former are rich in Fe and Si while SNRs from
core-collapse events are richer in O and Ne. For young SNRs, the X-ray
spectra can be used to identify those of Type Ia origin (e.g., 
\citealt{hhh+95,hgrs03,lbh+03}). Once the
X-ray emission is dominated by swept-up circumstellar or interstellar
matter, such discrimination is much more difficult. SNRs originating
from core collapse are often found near the molecular clouds in which
the progenitors formed, while those from Type Ia events are from stars
with sufficiently long lifetimes that they can have traveled far from
their birthplaces. Thus, for SNRs located in the near vicinity of active
star-forming regions one can reasonably assume that they originated
from massive stars. However, the absence of nearby star formation or dense
molecular material is not necessarily a direct indicator of a Type Ia
progenitor.  There are three Galactic SNRs that are commonly thought
of as the products of Ia events --- Tycho, and SN~1006, and perhaps Kepler
\citep{baade45,fwlh88,apg01} ---   all reasonably young
``historic'' remnants \citep{sg02}.

If any of the SNRs discussed here were known to be the result of a
Type~Ia explosion (as suggested by some authors for both \snr\ and
\snrb), then we would not expect to see a compact remnant.  Observing
a known Ia remnant would then be a good test case for our methodology:
as we have not found any candidate compact remnant, it demonstrates
that our method is not inclined to find false positives.  However,
while knowing that a SNR is from a Type~Ia explosion precludes the
existence of compact objects, the lack of compact objects does not
require a SNR to be from a Type~Ia.  Therefore, we cannot conclude
that \snr\ or \snrb\ is the result of a Type~Ia explosion.

\section{Conclusions}
\label{sec:conclusions}
There are 45 known SNRs that are reliably within 5 kpc of the Sun.  Most of
these SNRs are expected to contain central neutron stars: $\lsim 20$\%
are expected to result from Type Ia SNe and thus not contain a central
compact source, while  $\sim 20$\% (dependent on the stellar
initial mass function, the limiting mass for black holes, and binary
evolution; \citealt{hfw+03}) are
expected to host a central black hole that may not be easily
identified as such.  Thanks to the persistent efforts by astronomers
over the past four decades, central sources have been detected in the
X-ray and/or radio bands in 18 of these SNRs, and three have been
identified as probable Type~Ia SNe. In some cases, only a centrally
located PWN is detected, but in those cases (i.e.\ IC~443) it is
reasonably assumed that the PWN is powered by a central compact
source.

We have begun a program of searching for compact sources in the
remaining 23 SNRs. The program has been motivated by the discovery of a
point-like X-ray source at the very center of the youngest known Galactic SNR,
Cas~A.  The program has two observational components: imaging with
\chandra\ or \xmm\ in the X-ray band followed by ground-based optical/IR
followup. The latter is essential because of the high incidence of
of interlopers (foreground stars and background galaxies).
Such a comprehensive program is possible thanks to the 
astrometric accuracy of the X-ray missions combined with high
sensitivity.

In this paper we report on \chandra\ ACIS-I observations of four shell
remnants (\Ga, \Gb, \Gc, \Gd).  We undertook optical and IR
observations of every X-ray source detected with more than 10
counts. For all detections, we found, within astrometric errors, a
counterpart in  one or more bands. These counterparts were
consistent with either foreground (stars) or background (galaxies)
sources.  In particular, the X-ray flux and/or the X-ray to optical
(IR) ratio of the detected X-ray sources were not as extreme as all
known classes of neutron stars: accreting neutron stars, radio
pulsars, AXPs and SGRs.

In \S~\ref{sec:limits} we discuss reasons why standard neutron stars
were not found in these SNRs: they could have fallen in the gap
between the chips, they could have escaped our field of view due to a
very high velocities, they could be undetectable black holes, or they
could not exist owing to the SNRs being the results of Type Ia
explosions.  All of these scenarios are unlikely for a single
source, and even more so for all four, but are technically possible.  If,
on the other hand, these scenarios do not apply, then four remnants
contain neutron stars that are fainter than our X-ray detection limit
(typically, $L_X \lsim 10^{31}\,$erg
s$^{-1}$ in the 0.5--10~keV band).

We now consider this last (and most interesting) possibility. In the
absence of other forms of energy generation (accretion, rotation
power, magnetic field decay) the minimum X-ray flux one expects is set
by the cooling of the neutron star.  From Section~\ref{sec:cool} and
Figure~\ref{fig:cool}  we immediately see that the central
neutron stars in these four remnants must be cooler than those present
for example in the similarly-aged Puppis~A, PKS~1205$-$51/52 and RCW~103.

Our knowledge of the physics of cooling is by no means firm. There is
considerable debate among theorists as to which of the multitude of
physical processes can significantly affect the cooling output and as
to which of the physical parameters (mass, rotation rate, magnetic
field) controls these processes \citep{ygkp02}.  Nonetheless, there is
agreement that more massive neutron stars (with their larger mean
densities) cool more rapidly than those with smaller mass; this
expectation is illustrated in Figure~\ref{fig:cool}.  Thus our upper
limits can be made consistent with the cooling possibility provided
the central neutron stars in these four SNRs are more massive than
known cooling neutron stars. Indeed, the known examples of radio-quiet
objects could well result from a strong selection effect, namely the
earlier X-ray observations by \textit{Einstein} and \rosat\ detected
the warmer cooling neutron stars (ignoring the neutron stars detected
because of non-thermal emission).  The existing data may already hint
at a parameter affecting cooling, as exemplified by PSR~J0205+6449
\citep{shm02} and the Vela pulsar \citep{pzs+01}, but also possibly by
PSR~B1853+01 \citep*{pks02} and RX~J0007.0+7302 \citep{szh+04}.

Of course, we also do not see rotation-powered pulsars such as the
majority of the objects in Table~\ref{tab:psrs}.  Recent observations
are finding pulsars with lower radio luminosities and values of $\dot
E$ than ever before, and our limits would only be consistent with
these newer sources.  One might then ask why we see neither a standard
cooling neutron star nor a standard active pulsar, assuming that there
is no  intrinsic correlation between these properties.  It is possible
that there truly are no neutron stars in these SNRs, allowing one to
speculate wildly about what actually is there.

In subsequent papers we plan to report the X-ray observations and
ground-based follow up of the remaining 19 SNRs.  If no central
sources are identified then the hypothesis proposed here, namely that
there is a parameter that determines cooling of neutron star will be
strengthened.  The simplest (and most reasonable) suggestion is that
this second parameter is the mass of neutron stars, so that cooling
observations could be used to ``weigh'' isolated neutron stars (as
discussed by \citealt*{khy01}). This hypothesis is at odds with the
strong clustering of binary neutron star masses \citep{tc99}. We note,
though, that mass determination is only possible for neutron stars in
compact binary systems where significant interaction with the
companion may have taken place. Thus it is possible that neutron stars
resulting from single stars are systematically more massive than those
which evolve in compact binary systems (and have gained mass through
accretion).

\acknowledgements

D.~L.~K.\ is supported by a fellowship from the Fannie and John Hertz
Foundation.  B.~M.~G., P.~O.~S., and A.~N.\ acknowledge support from
NASA Contract NAS8-39073 and Grant G02-3090.  E.~V.~G.\ is supported
by NASA LTSA grant NAG5-7935, and B.~M.~G.\ is supported by NASA LTSA
grant NAG5-13032. S.~R.~K.\ is supported by grants from NSF and NASA.
Support for this work was provided by the National Aeronautics and
Space Administration through Chandra award NAS8-39073 issued by the
\textit{Chandra X-Ray Observatory} Center, which is operated by the
Smithsonian Astrophysical Observatory for and on behalf of NASA under
contract NAS8-39073.  The NRAO is a facility of the National Science
Foundation operated under cooperative agreement by Associated
Universities, Inc.  We have made extensive use of the SIMBAD database,
and we are grateful to the astronomers at the Centre de Donn\'{e}es
Astronomiques de Strasbourg for maintaining this database.  We would
like to thank an anonymous referee for helpful comments, D.~Fox\ for
assistance with the X-ray analysis, E.~Persson for assistance with the
PANIC observing, M.~van~Kerkwijk for assistance with ESO and Magellan
observing, and P.~McCarthy for assistance with Las Campanas observing.
We would like to thank T.~Landecker and DRAO for supplying the radio
image of \snr.  The Dominion Radio Astrophysical Observatory (DRAO) is
operated as a national facility by the National Research Council of
Canada. The Canadian Galactic Plane Survey is a Canadian project with
international partners and is supported by a grant from the Natural
Sciences and Engineering Research Council of Canada.  Data presented
herein were collected at the European Southern Observatory, Chile
(program ESO 69.D-0072).  Data presented herein were also obtained at
the W.~M.~Keck Observatory, which is operated as a scientific
partnership among the California Institute of Technology, the
University of California, and the National Aeronautics and Space
Administration.  The Guide Star Catalog-II is a joint project of the
Space Telescope Science Institute and the Osservatorio Astronomico di
Torino.

%%%%%%%%%%%%%%%%%%%%%%%%%%%%%%%%%%%%%%%%%%%%%%%%%%
% END OF TEXT
%%%%%%%%%%%%%%%%%%%%%%%%%%%%%%%%%%%%%%%%%%%%%%%%%%

%\bibliography{magrefs,xray,casA,psrrefs,myrefs,snrs}

%%%%%%%%%%%%%%%%%%%%%%%%%%%%%%%%%%%%%%%%%%%%%%%%%%
% TABLES
%%%%%%%%%%%%%%%%%%%%%%%%%%%%%%%%%%%%%%%%%%%%%%%%%%

%%%%%%%%%%%%%%%%%%%%%%%%%%%%%%%%%%%%%%%%%%%%%%%%%%
% FIGURES
%%%%%%%%%%%%%%%%%%%%%%%%%%%%%%%%%%%%%%%%%%%%%%%%%%
%\clearpage

%\clearpage

\end{document}